\documentclass[referee,useAMS,usenatbib,usegraphicx]{mn2e}
\usepackage{color}

\title[Young open cluster Stock 8]{Stellar contents and star formation in the 
young open cluster Stock 8}
\author[Jessy Jose et al.]{Jessy Jose $^{1}$\thanks{E-mail: jessy@aries.ernet.in}, 
A.K. Pandey$^{2,3,1}$, D.K. Ojha$^3$, K. Ogura$^4$, W. P. Chen$^2$, B.C. Bhatt$^5$,
\newauthor
S.K. Ghosh$^3$, H. Mito $^6$, G. Maheswar $^{1,7}$, Saurabh Sharma$^1$\\
$^1$ Aryabhatta Research Institute of Observational Sciences (ARIES), Manora Peak, 
Naini Tal, 263129, India\\
$^2$ Institute of Astronomy, National Central University, Chung-Li, 32054, Taiwan\\
$^3$ Tata Institute of Fundamental Research, Mumbai (Bombay), 400 005, India\\
$^4$ Kokugakuin  University, Higashi, Shibuya-ku, Tokyo, 150-8440, Japan\\
$^5$ CREST, Indian Institute of Astrophysics, Koramangala, Bangalore, 560 034, india\\
$^6$ Kiso Observatory, School of Science,  University of Tokyo, Mitake-mura, Kiso-gun, 
Nagano 397-0101, Japan\\
$^7$ Korea Astronomy and Space Science Institute (KASI), 61-1, Hwaam-dong, Yuseong-gu, Daejeon, Republic of Korea 305-348\\}

\begin{document}

\date{}

\pubyear{2007}

\maketitle

\label{firstpage}

\begin{abstract}
We present $UBVI_c$ CCD photometry of the young open cluster Stock 8
with the aim of studying its basic properties such as the
amount of interstellar extinction, distance, age, stellar contents and
initial mass function (IMF). We also studied the star formation scenario
in this region. From optical data, the radius of the cluster  is found
to be $\sim 6^{\prime}$ ($\sim 3.6$ pc) and the reddening within the cluster
region varies from $E(B-V)=0.40$ to 0.60 mag. The cluster is located
at a distance of $2.05 \pm 0.10$ kpc. Using H$\alpha$ slitless
spectroscopy and 2MASS NIR data we identified H$\alpha$ emission
and  NIR excess young stellar objects (YSOs), respectively. 
From their locations in the  colour-magnitude diagrams,
majority of them seem to have ages between 1 to 5 Myr. The spread in
their ages indicate a possible non-coeval star formation in the
cluster. Massive stars in the cluster region reveal an average age of
$\le$ 2 Myr. In the cluster region ($r \le 6^\prime$) the slope of
the  mass function (MF),  $\Gamma$, in the mass range $\sim 1.0 \le
M/M_\odot < 13.4$ can be  represented by a power law having a slope of
$-1.38\pm0.12$, which agrees well with Salpeter value (-1.35).  In the
mass range $0.3 \le M/M_\odot < 1.0$, the MF is also found to follow a
power law with a shallower slope of $\Gamma  = -0.58\pm 0.23$
indicating a break in the slope of the IMF at $\sim 1 M_\odot$. The
slope of the $K$-band luminosity function for the cluster ($r \le 6^\prime$) 
is found to be $0.31\pm0.02$, which is smaller than the average value ($\sim$
0.4) obtained for  embedded star clusters.

A significant number of YSOs are distributed along a Nebulous Stream
towards the east side of the cluster. A small cluster is  embedded in
the Nebulous Stream. The YSOs lying in the Nebulous Stream and in the
embedded cluster are found to be younger than the stars in the cluster
Stock 8. The  radio continuum, MSX, IRAS mid- and far-infrared maps
and the ratio  of [S II]/H$\alpha$ intensities indicate that the
eastern region of Stock 8 is ionization bounded whereas the western
region is density bounded. The morphology seems to indicate that the
ionization/ shock front caused by the ionizing sources located in the
Stock 8 region and  westwards of Stock 8 has not reached the
Nebulous Stream. It appears that star formation activity in the
Nebulous Stream and embedded cluster may be independent from that of
Stock 8.

\end{abstract}

\begin{keywords}
stars: formation $-$ stars: luminosity function, mass function $­-$ stars:
pre$-$main$-$sequence $-­$ open clusters and associations: individual: Stock 8.

\end{keywords}

\section{Introduction}

The study of star formation process and stellar evolution
constitute one of the basic problems in astrophysics. Since
most of the stars tend to form in clusters or groups, star clusters
are useful objects to study the star formation process.  The
initial  mass function  (IMF),  defined  as  the distribution  of
stellar masses  at the  time of birth  is one  of the basic  tools for
understanding the formation and evolution of  stellar systems. Star
forming regions (SFRs)  and young star clusters have  proven to be the
ideal  laboratories for  studying the  form of  IMF and  its variation
through space  and time.  Since the young  clusters (age  $<$ 10 
Myr) are assumed to be less affected by  dynamical evolution, 
their mass function (MF) can be considered as the IMF. However,  a
recent study by Kroupa (2007) argues that even in the youngest clusters, it is 
difficult to trace the IMF, as clusters evolve rapidly and therefore eject
a fraction of their members even at a very young age. 

There  are many  unsolved issues  concerning the  universality  of the
stellar IMF.  The theoretical expectation  is that the  IMF of  a cluster
should depend on the location,  size, metallicity, density of the star
forming  environment  and  other  conditions such  as  temperature  or
pressure (Zinnecker 1986; Larson 1992; Price $\&$ Podsiadlowski 1995).
But, convincing  evidence for variation in the  stellar IMF has
not  yet   been  found  observationally  (Scalo   1998;  Kroupa  2001,
2002). Massey et al. (1995) concluded that the slope of the IMF at the
higher mass range is universal,  with a value of $\Gamma$= $-1.1\pm0.1$
irrespective of  the variations in  metallicity. Luhman et  al. (2000)
also found no systematic variations in IMF of the nearby SFRs (such as
$\rho$ Oph, IC 348, Trapezium) although there is a large difference in stellar
densities. On the other hand  Hillenbrand (1997) found a few SFRs with
unusual IMFs.

The  lack of  strong  evidence  for IMF  variations  suggests that
its fundamental form may be universal. Consequently, local conditions
may  not play any significant  role in the star formation
process. However, the influence of star  formation history of young
clusters  on the form of IMF is still an open issue.   High-mass stars
have strong influence on their  surroundings and  can
significantly affect the formation of low mass stars. Recently, a
relatively large number of low mass stars have been detected in a few
OB associations, e.g. Upper Scorpio, the $\sigma$ and $\lambda$ Ori
regions (Preibisch \& Zinnecker 1999, Dolan \& Mathieu 2002). Since
this realization, surveys have demonstrated that the IMFs must 
essentially be the same in all star forming regions. The apparent
difference is due mainly to the inherent low percentage of high mass
stars and the incomplete survey of low mass stars in high-mass star
forming regions (e.g. Preibisch \& Zinneker 1999, Hillenbrand 1997,
Massey et al. 1995).  Recently Parker \& Goodwin (2007)
have studied the origin of O-stars and their effect on  low-mass
star formation. Advancement in detectors along with various surveys
such as the 2MASS, DENIS and available archived data from {\it ISO} and {\it
Spitzer} space telescopes have permitted detailed studies of low-mass
stellar population in regions of high mass star formation.

With the aim of understanding the star formation process and IMF
in and around young star cluster regions containing OB stars, we selected
young cluster Stock 8 located within the HII region of IC 417
(Sh2-234) in the Auriga constellation of the Perseus arm. No CCD
photometric observations have been carried out for this
region so far. Mayer \& Macak (1971)  have carried out photoelectric
photometry  of 11 bright stars in the region and reported a distance
of 2.96 kpc for the cluster.  Using  the photographic  study  of stars
brighter  than V  = 16  mag, Malysheva (1990) found cluster age  as
$\sim$ 12  Myr with  an angular diameter of $20^\prime$ and a distance
of 1.9 kpc.

The present  study  is an  attempt  to  understand the structure, star
formation history, pre-main-sequence  (PMS) population and the form of
IMF of Stock 8  using the $UBVI_c$ CCD photometry data. In Sections 2 and
3, we describe the observations, data reductions and archived data
used in the present work. Sections 4 to 8 describe various results. The
star formation scenario in the IC 417 (Sh2-234) region is described in
Sections 9 to 12. The results are summarized in Section 13.

\section{OBSERVATIONS AND DATA REDUCTIONS}

The $30\times30$ arcmin$^2$ ($\sim 17.9\times17.9$ pc$^2$) area
containing the central region of Stock 8, reproduced from the DSS2-R band
image, is shown in  Fig. \ref{stock8}. As evident from the figure,
Stock 8 is embedded in an HII region. To the west of it, there are several
OB stars which appear to form an OB association. To the east of Stock 8, we
find a very peculiar chord-like nebulosity which we term as Nebulous Stream.
A small cluster is also embedded within the Nebulous Stream. 
The small embedded cluster and the Nebulous Stream are discussed in Sections 
9 and 11, respectively. We describe below the observations made to carry out
a detailed study of the region.

\subsection{Optical CCD Photometry}

The CCD $UBVI_c$ observations were carried out using the 2048 $\times$
2048 pixel$^2$ CCD camera on  the 105-cm Schmidt telescope of the KISO
Observatory, Japan on November 4, 2004. At the  Schmidt focus (f/3.1)
each pixel of  25 $\mu$m corresponds to 1.5 arcsec  and the entire
chip covers  a  field of  $\sim$  $50 \times  50$  arcmin$^2$ on the
sky. The read-out noise and gain of the CCD are 23.2   $\rm e^-$
and  3.4   $\rm e^-/ADU$, respectively.  The FWHM of  the star  images
was $\sim$ 3  arcsec.  A number  of  bias and  dome  flat frames  were
also taken during  the observing runs.  We  took short and long
exposures  in all the filters to avoid saturation of bright stars. SA
92 field of Landolt (1992) was observed to standardize the field on
the same night. The log of observations is tabulated in Table \ref{obslog}.

\begin{table}
\caption{Log of observations}
\label{obslog}
%\begin{minipage}{15mm}
%\caption{\label{Log of observations}}
%\centering
\begin{tabular}{|p{.40in}|p{.45in}|p{.48in}|p{.68in}|p{1.0in}|}
\hline
$\alpha_{(2000)}$ & $\delta_{(2000)}$ & Date of &Filter & Exp. time\\
(h:m:s) & (d:m:s)  &    observation  &          &  (sec)$\times$no. of frames \\
\hline
  {\it Kiso$^1$}                 &           &   &\\  
  
05:28:07& +34:25:42& 2004.11.04 & $U$ &  60$\times$3, 180$\times$4 \\
05:28:07& +34:25:42& 2004.11.04 & $B$ & 20$\times$4, 60$\times$4 \\
05:28:07& +34:25:42& 2004.11.04 & $V$ &  10$\times$4, 60$\times$4\\
05:28:07& +34:25:42& 2004.11.04 & $I_c$ &  10$\times$4, 60$\times$4\\
{\it ST$^2$} &                  &            &\\
05:28:07& +34:25:35&  2006.09.29 & $V$ &  5$\times$2, 300$\times$3, 600$\times$4\\
05:28:07& +34:25:35&  2006.09.29 & $I_c$ & 5$\times$3, 120$\times$5\\
05:28:54& +34:24:32& 2006.09.29 & $V$ &  5$\times$2, 300$\times$3, 600$\times$4\\
05:28:54& +34:24:32& 2006.09.29& $I_c$ & 5$\times$3, 120$\times$5\\
05:29:42& +34:42:13& 2006.09.29 & $V$ &  5$\times$2, 300$\times$3, 600$\times$4\\
05:29:42& +34:42:13& 2006.09.29& $I_c$ & 5$\times$3, 120$\times$5\\
{\it HCT$^3$} &                  &             &    \\
05:28:14& +34:24:52& 2005.09.26&H$\alpha$-Br &  60$\times$2\\
05:28:14& +34:24:52& 2005.09.26&Gr5/H$\alpha$-Br& 420$\times$3 \\
05:28:14&+34:15:55& 2005.09.27& H$\alpha$-Br &  60$\times$3\\
05:28:14&+34:15:55& 2005.09.27&Gr5/H$\alpha$-Br & 420$\times$3 \\
05:28:56& +34:16:43& 2005.10.10& H$\alpha$-Br &  60$\times$3 \\
05:28:56& +34:16:43& 2005.10.10& Gr5/H$\alpha$-Br & 420$\times$3 \\
05:28:32 &+34:29:55& 2005.10.10& H$\alpha$-Br &  60$\times$3\\
05:28:32 &+34:29:55& 2005.10.10& Gr5/H$\alpha$-Br & 420$\times$3 \\
05:28:07& +34:25:35& 2006.09.24&H$\alpha$-Br&450$\times$1\\
05:28:07& +34:25:35& 2006.09.24&[S II]&450$\times$1\\
05:28:07& +34:25:35& 2006.09.24&$R$&450$\times$1\\
05:28:47& +34:25:35& 2006.09.24&H$\alpha$-Br&450$\times$1\\
05:28:47& +34:25:35& 2006.09.24&[S II]&450$\times$1\\
05:28:47& +34:25:35& 2006.09.24&$R$&450$\times$1\\
\hline
\end{tabular} 

$^1$  105-cm Schmidt Telescope, Kiso, Japan\\
$^2$  104-cm Sampurnanand Telescope, ARIES, Naini Tal\\
$^3$  2-m Himalayan Chandra Telescope, IAO, Hanle\\

%\end{minipage}
\end{table}

%%%%%%%%%%%%%%%%%%%%%%%%%%%%%%%%%%%%%%%%%%%%%%%%%%%%%%%%%%%%%%%%%%%%%%%%%%

\begin{figure*}
\centering
\includegraphics[scale = 0.96, trim =00  55 0 170, clip]{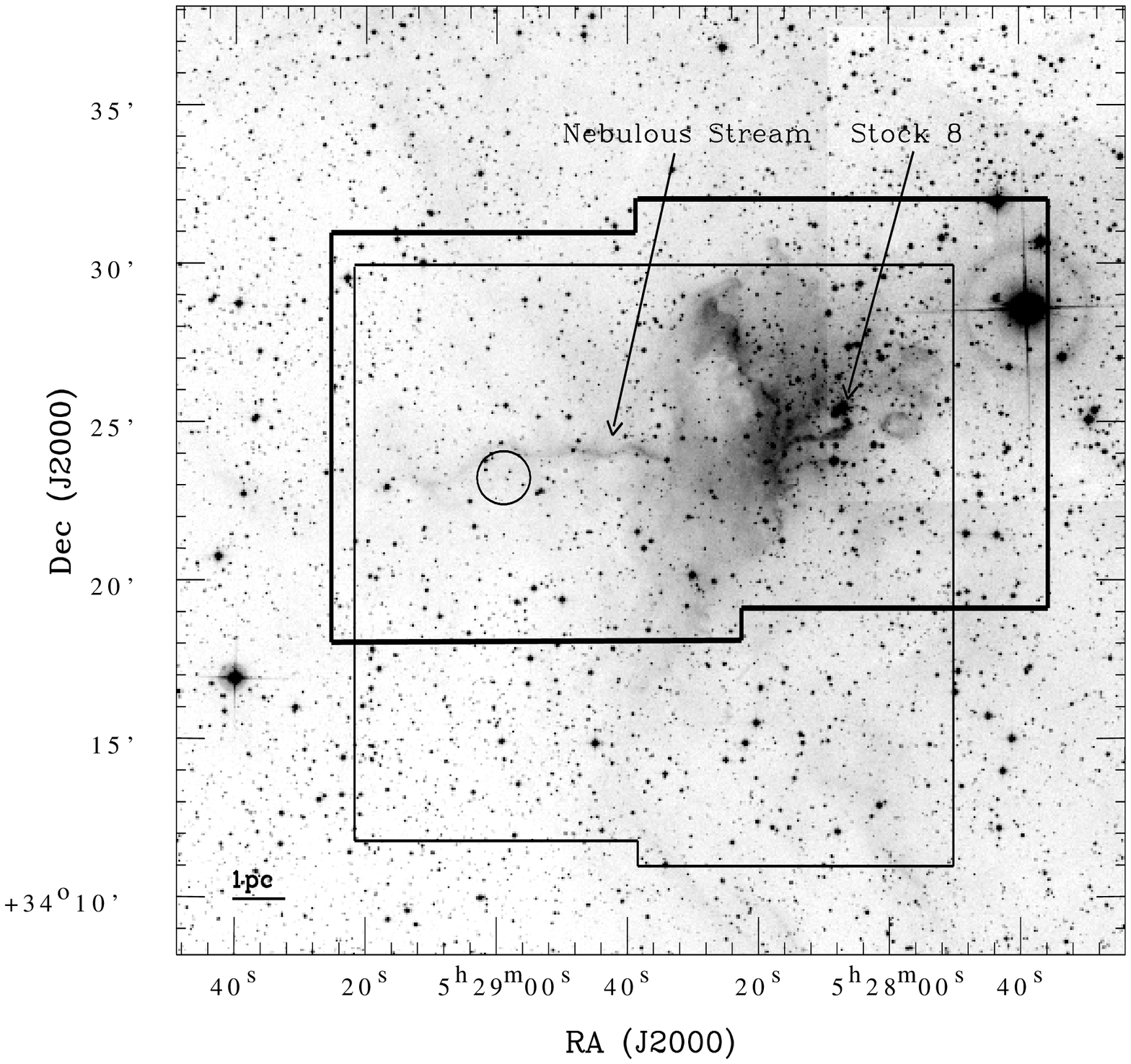}
\vspace{2mm}
\caption{The $30\times30$ arcmin$^2$ DSS2-R band image of the region around
Stock 8. The area  marked with  thin lines is the region covered by
slitless spectroscopic observations and the area marked with thick  lines
is the region for which deep images are taken in $V$ and $I_c$ bands. The
circle represents the location of the small embedded cluster (see Section 9). 
Locations of the Nebulous Stream (see Section 11) and Stock 8 are also marked.
The abscissa and the ordinates are for J2000 epoch. }
\label{stock8}
\end{figure*}
%%%%%%%%%%%%%%%%%%%%%%%%%%%%%%%%%%%%%%%%%%%%%%%%%%%%%%%%%%%%%%%%%%%%%%%%

The CCD  frames were bias-subtracted  and flat-field  corrected in the
standard manner using various tasks available under IRAF\footnote{IRAF
is distributed by National Optical Astronomy Observatories,
USA}. Aperture photometry was done for the standard stars of SA 92
field to estimate the atmospheric extinction and to calibrate the
field. Following calibration equations are derived using a
least-square linear regression:\\

\begin{figure}
\includegraphics[scale = 0.6, trim = 10 10 10 10, clip]{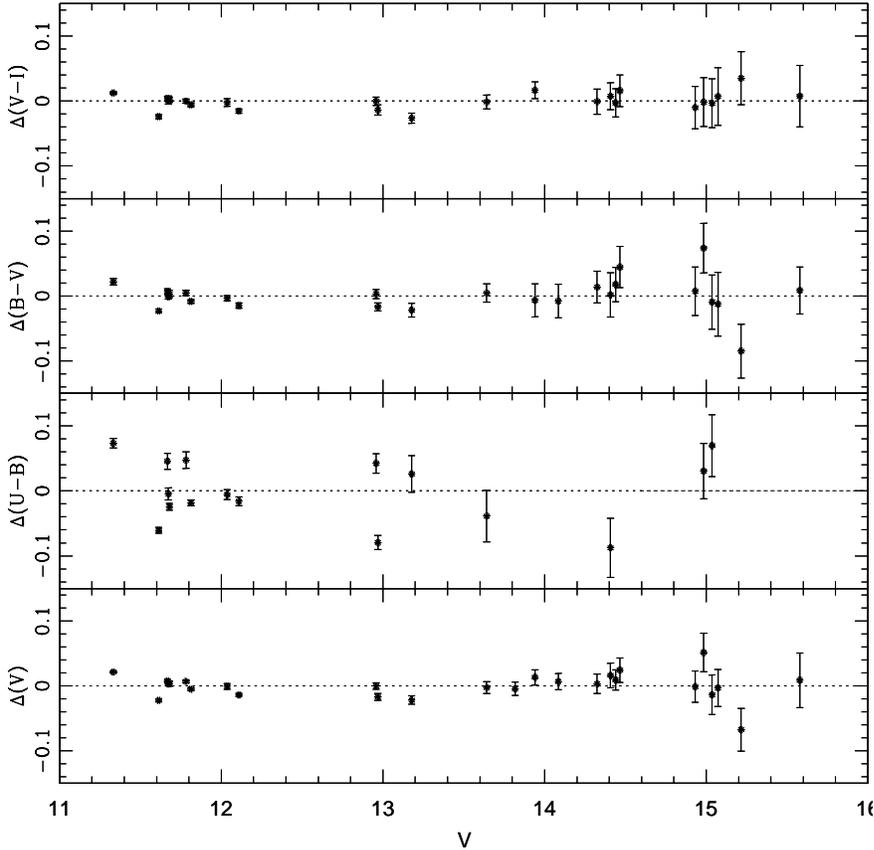}
\vspace{2mm}
\caption{Residuals  between  standard  and transformed  magnitude  and
colours  of standard stars  plotted against  the Landolt  standard $V$
magnitude. Theses observations are from Kiso Schmidt telescope. The error 
bars represent combined errors of Landolt (1992) and present measurements.}
\label{res}
\end{figure}

%\begin{equation}
\noindent
%\begin{center}
$(U-B) = (0.887\pm 0.025) (u-b) - (2.973\pm 0.010)$\\
%\end{equation}
%\begin{equation}
\noindent
$(B-V)=(1.233\pm 0.009) (b-v) + (0.482\pm0.010)$\\
%\end{equation}
%\begin{equation}
\noindent
$(V-I) = (0.847\pm 0.008) (v-i) + (0.758\pm0.006)$\\
%\end{equation}
%\begin{equation}
\noindent
$V = v-(0.095\pm 0.008) (V-I) - (2.578\pm0.007)$\\
%\end{equation}
%\end{center}

where  $u,b,v,i$ are the instrument magnitudes corrected for the
atmospheric extinctions and $U,B,V,I$ are the standard
magnitudes. Fig. \ref{res} shows the standardization  residuals,
$\Delta$,  between standard and  transformed $V$ magnitudes, $(U-B)$,
$(B-V)$ and $(V-I)$ colours of standard stars as a function of $V$
mag. The standard deviations in  $\Delta V,  \Delta (U-B),  \Delta
(B-V)$ and  $\Delta (V-I)$  are 0.021, 0.045, 0.023, 0.015 mag,
respectively.

\begin{figure}
\includegraphics[scale = .6, trim = 10 10 10 10, clip]{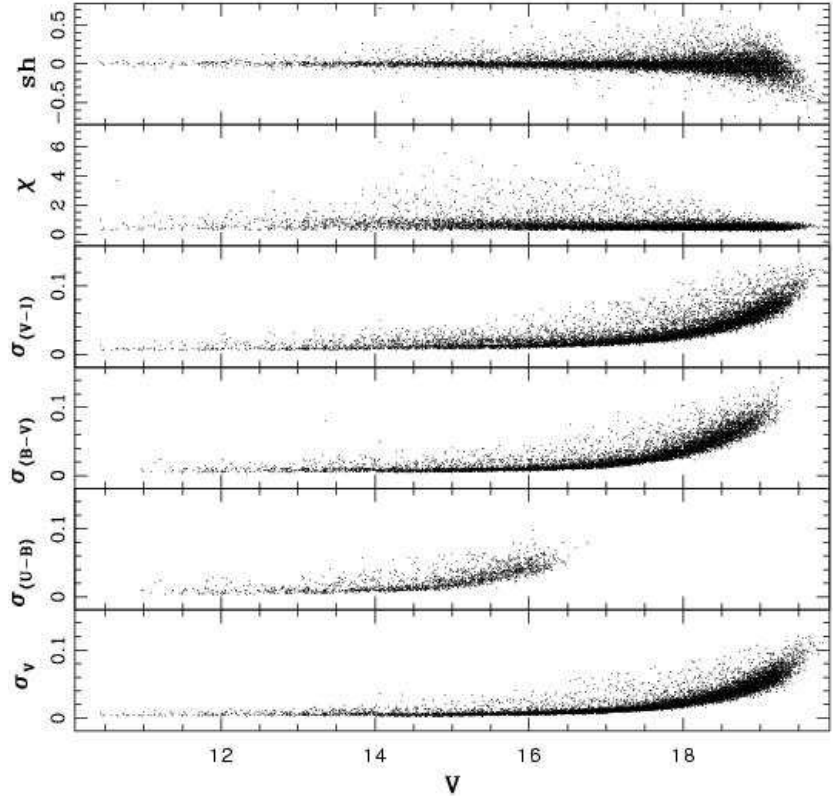}
\vspace{2mm}
\caption{The  photometric errors  in  $V,(U-B),(B-V),(V-I)$, image
parameters $\chi$ and Sharpness as a function of $V$ magnitude  for Kiso 
observations. }
\label{error}
\end{figure}

Different frames of cluster region of same exposure  time and filters
were averaged. Photometry of cleaned frames was carried out using the
DAOPHOT-II (Stetson 1987) profile fitting software. The magnitudes of
bright stars were taken from short exposure frames whereas that of
faint stars were taken from deep exposures.   Profile-fitting
photometry gives the error in magnitude determination, the goodness of
the fit parameter, $\chi$, which is a measure of the average rms
deviation to the PSF fit normalized to the expected errors. It also
gives a shape parameter, Sharpness, which measures how well the
PSF fits the object. The photometric errors in magnitude, colours
and image parameters as a function of $V$ magnitude are shown in
Fig. \ref{error}. These parameters were used to reject poor measurements.

%%%%%%%%%%%%%%%%%%%%%%%%%%%%%%%%%%%%%%%%%%%%%%%%%%%%%%%%%%%%%%%%%%%%%%%%
\noindent

%%%%%%%%%%%%%%%%%%%%%%%%%%%%%%%%%%%%%%%%%%%%%%%%%%%%%%%%%%%%%%%%%%%%%%%%%%
%%%%%%%%%%%%%%%%%%%%%%%%%%%%%%%%%%%%%%%%%%%%%%%%%%%%%%%%%%%%%%%%%%%%%%%%%%
%%%%%%%%%%%%%%%%%%%%%%%%%%%%%%%%%%%%%%%%%%%%%%%%%%%%%%%%%%%%%%%%%%%%%%%%

Deep images of two  central regions of the cluster and a nearby field  
($\alpha_{2000}$ = $05^{h}29^{m}42^{s}$; $\delta_{2000}$ =
$+34^{\circ}42^{\prime}13^{\prime\prime}$) were also obtained in $V$
and $I_c$ filters on September 29, 2006 using  a 1024  $\times$ 1024
pixel$^2$ CCD camera mounted  at the f/13 Cassegrain  focus of  the 104-cm
Sampurnanand Telescope  (ST)  of the Aryabhatta Research Institute of
Observational sciencES (ARIES), Naini Tal, India. Log of
the observations is given in Table \ref{obslog}. In this set  up, each pixel  
of the
CCD corresponds to $0.37$ arcsec and the entire chip covers a field of
$\sim 13\times13$ arcmin$^2$ on the sky. To improve the  signal to
noise  ratio, the observations were carried out in binning mode of
$2\times2$ pixels. The FWHM of the star images was $\sim 1.8$
arcsec. The  observed region is  marked by thick lines in
Fig. \ref{stock8}. The secondary standards from  the KISO
observations  were  used to  calibrate  the  data taken at ARIES. The
typical DAOPHOT photometric errors at brighter end ($V\sim$ 15)  are
of the order of $\sim$  0.01 mag, whereas the errors  increase towards
the fainter end ($\sim$ 0.05 at $V$ $\sim$ 21). The combined photometric
data of KISO Schmidt telescope and ST in an area of $40^\prime \times
40^\prime$ along  with positions of  the stars  are given  in a table,
which is only available in electronic form  as  part of the online material from
{\it http://www.blackwell-synergy.com}. Format of the table  is
shown in Table \ref{photdata}.

\begin{table}
%\begin{minipage}{80mm}
\caption{$UBVI_c$ photometric data of the sample stars. The complete 
table is available in electronic form only.}
\label{photdata}
%\medskip
%\scriptsize
\begin{tabular}{ccccccc} \hline
star& $\alpha_{(2000)}$& $\delta_{(2000)}$& $V$ &$(U-B)$ &$(B-V)$ & $(V-I)$\\
no & (h:m:s) &  (d:m:s)    & & & &  \\
\hline
1 & 05:27:48.31  & +34:21:26.6 & 11.210 & 0.032 & 0.746 & 0.724\\
2 & 05:28:01.22  & +34:27:00.7 & 11.498 & 0.129 & 0.174 & 0.281\\
3 & 05:27:43.61  & +34:21:24.8 & 11.271 & -0.481& 0.417 & 0.617\\
4 & 05:28:50.26   &+34:19:23.2  & 11.507 & 0.015 & 0.174 & 0.298 \\
5 & 05:28:04.46   &+34:29:20.8  & 11.710 &-0.474 & 0.298 & 0.494 \\
\hline
\end{tabular}
%\end{minipage}
\end{table}

To study  the luminosity function (LF)/ MF it  is  necessary to  take
into  account  the incompleteness  in  the observed data that may
occur for various reasons (e.g. crowding of the stars).   A
quantitative evaluation of the completeness of the photometric data
with respect to the brightness and the position on a given frame is
necessary to convert the observed luminosity function (LF) to a true
LF. We used the ADDSTAR routine of DAOPHOT II to determine the
completeness factor (CF). The procedure has been outlined in detail in
our earlier work (Pandey et al. 2001). Briefly, we randomly added
artificial stars to both $V$ and $I$ images in such a way that they
have similar geometrical locations but differ in I brightness
according to mean $(V-I)$ colour ($\sim 1.5$) of the data sample. 
The luminosity distribution of artificial stars is chosen in
such a way that more stars are inserted towards the fainter magnitude
bins. The frames are reduced using the same procedure used for the
original frame. The ratio of the number of stars recovered to those
added in each magnitude interval gives the CF as a function of
magnitude. The minimum value of the CF of the pair (i.e. $V$ and $I$
band  observations) for the two sub-regions, given in Table \ref{cf_opt}, is
used to correct the data for incompleteness. The incompleteness of the
data increases with increasing magnitude as expected, however it does
not depend on the area significantly.

\begin{table}
%\begin{minipage}{80mm}
\caption{Completeness Factor (CF) of photometric data in
the cluster and field regions.}
\label{cf_opt}
%\medskip
%\scriptsize
\begin{tabular}{cccc} \hline
V range&    Stock8 &  & Field region\\
 (mag)& $r\le 3^\prime$ & $3^\prime <r \le6^\prime$&  \\
\hline
10 - 11&1.00&1.00&1.00\\
11 - 12&1.00&1.00&1.00\\
13 - 14&1.00&1.00&1.00\\
14 - 15&1.00&1.00&1.00\\
15 - 16&1.00&1.00&1.00\\
16 - 17&0.98&0.97&0.98\\
17 - 18&0.95&0.95&0.97\\
18 - 19&0.93&0.93&0.96\\
19 - 20&0.88&0.90&0.92\\
20 - 21&0.88&0.86&0.86\\
21 - 22&0.74&0.75&0.76\\

\hline
\end{tabular}
%\end{minipage}
\end{table}

\subsection {Slitless Spectroscopy, H$\alpha$ and [S II] imaging}

The cluster was observed in slitless mode with grism as the dispersing
element using  the  Himalaya Faint Object Spectrograph
Camera (HFOSC) on the 2-m Himalayan Chandra Telescope (HCT) of Indian
Astronomical Observatory (IAO), Hanle, India. The  central $2K \times
2K$ pixels of  $2K \times 4K$ CCD were used for data acquisition. The
pixel size is 15 $\mu$m with an image scale of
$0^{\prime\prime}$.296/pixel and it covers an area of 10$\times$10
arcmin$^2$ on the sky. A combination of H$\alpha$ broad-band filter
(H$\alpha$-Br; 6100 - 6740 \AA) and Grism 5 were used without any slit.
The resolution of  Grism 5 is 870.  The 
regions where slitless spectroscopic observations were
carried out are identified with thin lines in Fig. \ref{stock8}.
 Emission-line stars with enhancement over the
continuum at H$\alpha$ wavelength are visually identified. Multiple
frames were taken to confirm the presence of H$\alpha$ emitting
sources. Positions of the H$\alpha$ emission stars are given in Table
\ref{halphalog}.  H$\alpha$-Br  and [S II] ($\lambda$ = 6724 \AA, $\Delta \lambda
\sim$ 100 \AA) images of two central regions of the  cluster were also
acquired on September 24, 2006 using  the HFOSC at the 2-m HCT.
Log of the observations is given in Table \ref{obslog}.

\begin{table}
%\begin{minipage}{80mm}
\caption{Coordinates and photometric data of detected H$\alpha$ emission stars}
\label{halphalog}
%\medskip
%\scriptsize
\begin{tabular}{cccccc} \hline
ID & $\alpha_{(2000)}$& $\delta_{(2000)}$&V& $B-V$&$V-I$ \\
 & (h:m:s) &  (d:m:s)    & (mag)& (mag) &  (mag)      \\
\hline

1 & 05:28:04.65 &  +34:21:51.0 & 15.642 &     0.961   &    1.205  \\
2 & 05:28:00.00 &  +34:24:40.5 & 19.522 &        -    &    1.728  \\
3 & 05:28:00.45 &  +34:25:57.8 & 18.493 &     1.594   &     2.030 \\
4 & 05:28:06.09 &  +34:24:55.4 & 18.084 &        -    &    1.460  \\
5 & 05:28:06.90 &  +34:24:49.6 & 14.800 &      0.970  &     1.285 \\
6 & 05:28:05.72 &  +34:25:28.1 & 17.108 &        -    &    1.778  \\
7 & 05:28:08.65 &  +34:25:38.9 & 16.700 &     1.218   &    1.812  \\
8 & 05:28:16.07 &  +34:27:28.8 & 18.284 &     1.379   &    2.125  \\
9 & 05:28:17.12 &  +34:28:04.1 & 15.565 &     1.249   &    1.568  \\
10& 05:28:25.90 &  +34:23:10.2 & 16.467 &     0.852   &    1.161  \\
11& 05:28:35.83 &  +34:24:32.8 & 15.267 &     1.256   &    1.694  \\
12& 05:28:18.22 &  +34:16:52.7 & 16.486 &     1.144   &    1.337  \\
13& 05:28:34.91 &  +34:26:01.9 & 20.352 &        -    &    2.240 \\
14& 05:28:39.38 &  +34:24:33.2 & 18.892 &        -    &    1.980 \\
15& 05:28:39.85 &  +34:24:39.3 & 19.423 &        -    &   1.848 	\\
16& 05:28:46.04 &  +34:24:07.5 & 21.147 &        -    &   2.419 	\\
17& 05:28:48.08 &  +34:24:09.6 & 17.777 &        -    &   1.423 	\\
18& 05:28:46.68 &  +34:22:18.9 & 19.751 &        -    &   1.828 	\\
19& 05:28:49.55 &  +34:23:26.8 & 17.658 &        -    &   2.248 	\\
20& 05:28:53.23 &  +34:23:35.2 & 19.581 &        -    &   2.587 	\\
21& 05:28:56.05 &  +34:23:00.3 & 16.876 &	 -    &   1.955         \\ 
22& 05:28:58.48 &  +34:23:10.2 & 18.579 &        -    &   2.439 	\\
23& 05:29:03.31 &  +34:24:13.6 & 19.792 &        -    &   2.576 	\\
24& 05:29:09.69 &  +34:23:32.2 & 18.000 &        -    &   2.287 	\\
25& 05:29:19.14 &  +34:17:47.1 &  -     &	     &	  -  	 \\         
		   
\hline		   
\end{tabular}	   
%\end{minipage}
\end{table}

\section {archival data}

Near-infrared (NIR) $JHK_s$ data for  point sources within a radius of
$20^\prime$ around Stock  8 have been obtained from  the Two Micron All
Sky   Survey  (2MASS)   Point  Source   Catalogue  (PSC)   (Cutri  et
al. 2003).  To improve the photometric accuracy, we used photometric
quality flag (ph$\_$qual = AAA) which gives a S/N $\ge$ 10  and a photometric
uncertainty $ <$ 0.10 mag. This selection criterian ensures best-quality 
detection in terms of photometry and astrometry (cf. Lee et al. 2005). 
 The $JHK_s$
data  were transformed  from 2MASS  system  to CIT  system using  the
relations given on the 2MASS web 
site\footnote{http://www.astro.caltech.edu/$\sim$jmc/2mass/v3/transformations/}.

The {\it Midcourse  Space experiment} ({\it MSX})  surveyed the Galactic  plane in
four  mid-infrared bands  - A  ($8.28~  \mu$m), C  ($12.13~ \mu$m),  D
($14.65~  \mu$m) and  E ($21.34~  \mu$m)  at a  spatial resolution  of
$\sim18^{\prime \prime}$  (Price et al.  2001). Two of these  bands (A
and   C)  with  ${\it   \lambda(\Delta  \lambda)}$   corresponding  to
8.28 $\mu$m (3.36 $\mu$m)  and  12.13 $\mu$m (1.71 $\mu$m)  include  several 
polycyclic aromatic  hydrocarbons  (PAHs) features  at  6.2, 7.7, 8.7, 11.3, 
and 12.7 $\mu$m. MSX images  in these four  bands around  the cluster  region 
were  used to study  the  emission  from  the  PAHs.

The data  from the {\it IRAS} survey  around the cluster region  in the four
bands  (12,  25, 60,  100  $\mu$m)  were  HIRES processed  (Aumann  et
al. 1990) to obtain high angular resolution maps. These maps were  used
to study   the spatial distribution of  dust colour temperature  and optical depth.

\section{Radial stellar surface density and cluster size}\label{rd}

The radius of a cluster is one of the important parameters used to study the
dynamical  state of  the cluster. We used star count  technique  to
determine the  statistical properties of clusters with  respect to the
surrounding stellar background, to study
the surface density distribution of stars in the cluster region and to
derive the radius of the cluster.

To  determine   the  cluster  center,  we  used   the  stellar  density
distribution of  stars having $V  \leq 18$ mag  in a $\pm$  75 pixel
wide  strip along  both X  and Y  directions around  an   eye
estimated center. The  point of maximum density obtained  by fitting a
Gaussian  curve is  considered as  the  center of  the cluster.   The
coordinates of  the cluster center  are found to be  $\alpha_{2000}$ =
$05^{h}28^{m}07^{s}.5 \pm 1^{s}.0$; $\delta_{2000}$ =
$+34^{\circ}25^{\prime}42^{\prime\prime} \pm 15^{\prime\prime}$.  The
2MASS NIR data yields the cluster center  $\sim7^{\prime\prime}$ away
southwards of the optical center. However, this difference 
is within the uncertainty and therefore can be
considered identical. Henceforth, we have used the optical center
as the center of the cluster.

\begin{figure}
%\centering
\includegraphics[scale =.6,trim=10 10 10 10,clip]{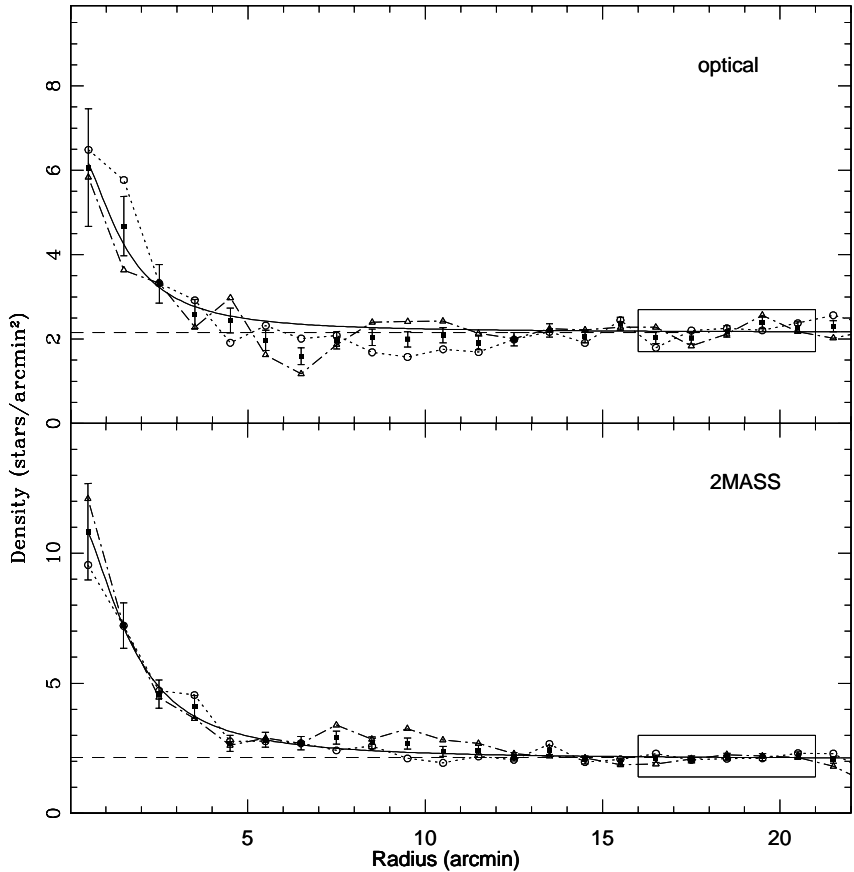}

\caption{Stellar  density (filled squares with error bars) as  a
function  of radius  from  the adopted cluster  center for  the
optical  and 2MASS  data.  The dashed  line represents the  mean
density  level of the  field stars and  the continuous  curve shows
the least square  fit of the  King (1962) profile  to the observed
data points from Kiso Schmidt observations. Open circles connected by
dotted line and open triangles connected by dot-dashed line represent
RDPs for the east and west region of the cluster respectively.  The
error bars  represent $\pm$  $\sqrt{N}$ errors. Box represents the
radial range considered as a field region. }
\label{rad_18}
\end{figure}

To investigate the radial structure of the cluster, we derived the radial density
profile (RDP)  using the Kiso Schmidt observations for stars
brighter than  $V$ = 18 mag  by dividing star counts  inside the
concentric  annuli  of $1^{\prime}$  wide  around  the cluster  center
by  the respective  annulus area.  The  densities thus obtained are
plotted as a function of radius in Fig. \ref{rad_18},  where one
arcmin at the distance of cluster (2.05 kpc, cf. Sec 6) corresponds to
$\sim$ 0.6 pc. The upper and the  lower panels  show the RDPs
obtained from optical  and 2MASS NIR data respectively. The error
bars are derived assuming that the number of stars in each annulus
follows Poisson statistics.

The radius of the cluster $(r_{cl})$ is defined as the point where the
cluster stellar density merges  with the  field stellar  density. The
contribution of the  field stars has been estimated  by counting stars
in  the annular region  16$^\prime$ - 21$^\prime$ from the  cluster
center. The horizontal  dashed line  in Fig. \ref{rad_18} shows  the
field star density. Optical RDP indicates a  dip in  stellar counts
at $r\sim 6^{\prime}$.5, whereas this dip is  absent in NIR RDP. The
NIR RDP indicates an enhancement in stellar density at $r\sim
7^{\prime}.5$.  To find out the cause of dip in the optical RDP and
enhancement in stellar density in the NIR RDP, in Fig. \ref{rad_18},  
we also show RDPs for the eastern and western regions of Stock 8.
 The dip in the optical and
enhancement in NIR RDP is visible only towards the western region of the
cluster. A comparison of the RDPs with spatial distribution of
$E(B-V)$, shown in Fig. \ref{red}, indicates that the dip in the optical RDP
at $r\sim 6^{\prime}$.5 may be an artifact of relatively higher
extinction around $r\sim 6^{\prime}.5 - 7^{\prime}$. The NIR
observations can probe deeper in the embedded region, hence enhancement
in stellar density in NIR RDP around $r\sim 7^{\prime}.5$ may be due
to embedded population.  
Since star formation activity in the
Nebulous Stream, as discussed in Sec. 11, is found to be independent
of the star formation activity in the cluster region, we used radius
of the cluster  as $r_{cl}\sim  6^{\prime}$ (as obtained from optical
data) to study the properties of the cluster.

To  parametrize the RDP,  we fitted  the observed RDP with the
empirical model of King (1962) which is given by,
\begin{equation}
\hspace{20mm}{\rho (r) = {{f_0} \over \displaystyle {1+\left({r\over
r_c}\right)^2}}}
\end{equation}
where  $r_c$ is  the core  radius at  which surface  density $\rho(r)$
becomes  half of  the central  density, $f_0$  . The  best fit  to the
observed RDPs, obtained by a $\chi^2$ minimization technique, is shown
in Fig. \ref{rad_18}. The core  radii thus estimated 
from optical and NIR RDPs are $1^\prime.4 \pm 0^\prime.2$  and 
$1^\prime.6 \pm 0^\prime.1$ respectively.
 
%%%%%%%%%%%%%%%%%%%%%%%%%%%%%%%%%%%%%%%%%%%%%%%%%%%%%%%%%%%%%%%%%%%%%%%%%%
%%%%%%%%%%%%%%%%%%%%%%%%%%%%%%%%%%%%%%%%%%%%%%%%%%%%%%%%%%%%%%%%%%%%%%%%
%%%%%%%%%%%%%%%%%%%%%%%%%%%%%%%%%%%%%%%%%%%%%%%%%%%%%%%%%%%%%%%%%%%%%%%%%%
\begin{figure}
%\centering
\includegraphics[scale = .6, trim = 10 10 10 10, clip]{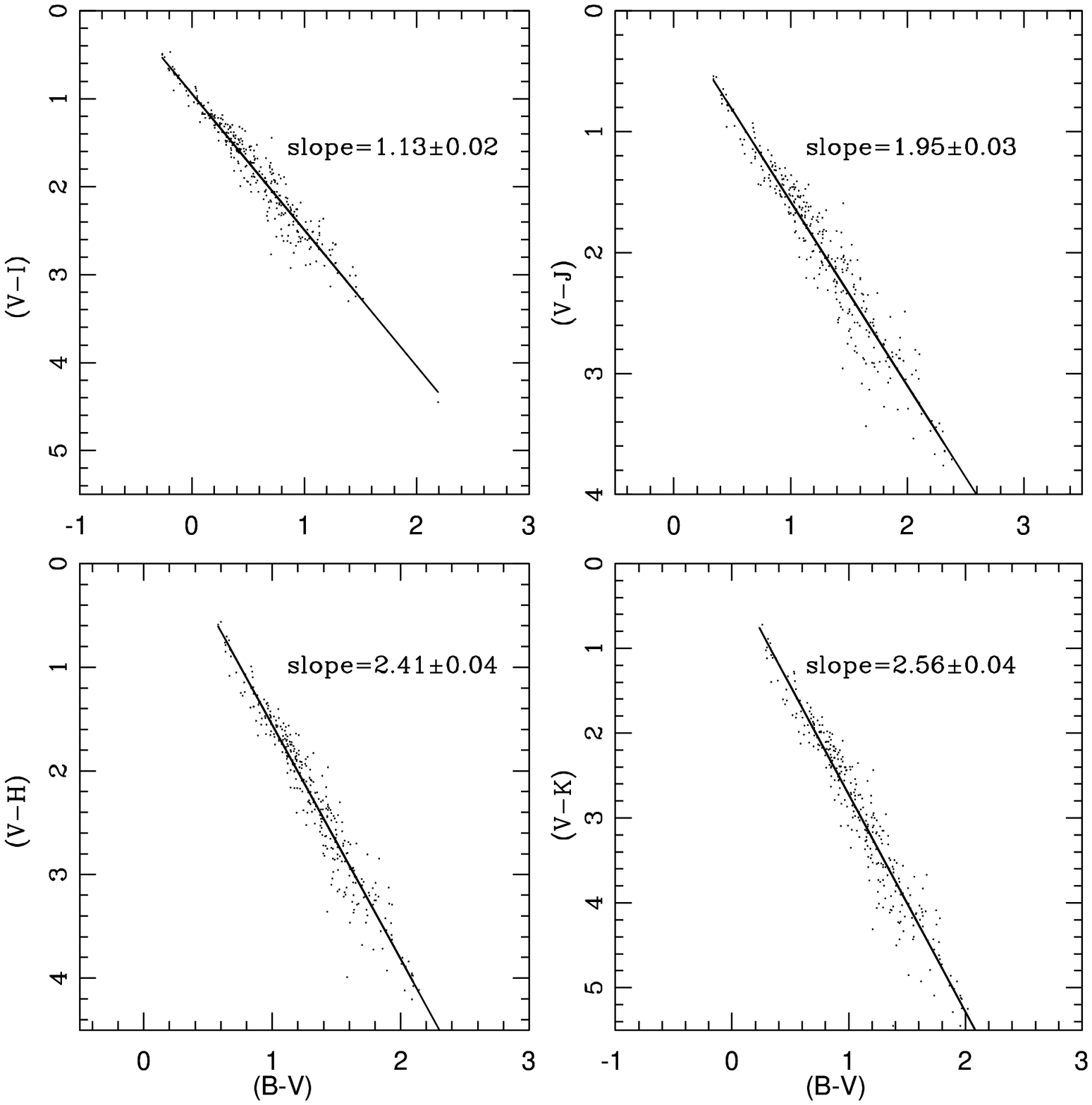}

\caption{Two - colour diagrams  for stars within  the radius  $r \le
12^{\prime}$ from the center of Stock 8. The  continuous line represents 
the  least square fit to the  data. }
\label{Rvalue}
\end{figure}
%%%%%%%%%%%%%%%%%%%%%%%%%%%%%%%%%%%%%%%%%%%%%%%%%%%%%%%%%%%%%%%%%%%%%%%%
\section{Reddening in the cluster}
\label{reddening}

 The extinction in a star cluster arises due to two distinct sources: 
(1) the general 
ISM in the foreground of the cluster [$E(B-−V)_{min}$], and (2) the 
localized cloud associated with the cluster 
[$E(B−-V) = E(B-−V)_* -− E(B-−V)_{min}$], where $E(B−-V )_*$ is
the reddening of the star embedded in the parent cloud. The former
component is characterized by the ratio of the total-to-selective
extinction $R_V$ $[= A_V /E (B -− V )]$= 3.1 (Wegner 1993; He et al. 1995;
Winkler 1997), whereas for the intracluster regions of young clusters
embedded in dust and gas cloud the value of $R_V$ may vary
significantly (Chini \& Wargau 1990; Tapia et al. 1991; Hillenbrand et
al. 1993; Pandey et al. 2000).

The ratio of total-to-selective extinction $R$ is
an important quantity, that must be known,  to get the dereddened
magnitudes of the stars. Consequently, the value of $R$ affects the
distance determination significantly. Several studies have already
pointed out anomalous reddening law with high $R$ values in the
vicinity of star forming regions (see. e.g. Neckel \& Chini 1981,
Chini \& Krugel 1983, Chini \& Wargau 1990, Hillenbrand et al. 1993,
Pandey et al. 2000, Samal et al. 2007). Since the cluster is embedded
in the parent molecular cloud, it will be useful to understand the
reddening law in and around the Stock 8 region.

To study  the nature of the  extinction law in the  cluster region, we
used two - colour diagrams (TCDs) as  described by Pandey et
al. (2003). The  TCDs of the form of  ($V-\lambda$)  vs. ($B-V$),
where  $\lambda$ is  one   of  the  broad-band
filters ($R,I,J,H,K,L$),  provide  an effective method for separating
the influence of the normal extinction produced by the diffuse
interstellar  medium from that of the abnormal extinction arising
within regions  having a peculiar  distribution of dust sizes
(cf. Chini \& Wargau 1990, Pandey et al. 2000).

The   TCDs    for   the   region   $r<12^{\prime}$    are   shown   in
Fig. \ref{Rvalue}.  The  ${E(V-\lambda)}\over  {E(B-V)}$ values  in  the
cluster region are estimated 
using the procedure as described in Pandey et al. (2003).
The slopes of the distributions $m_{cluster}$  (cf. Pandey et al. 2003) are
found to  be $1.13\pm0.01, 1.95\pm0.03,  2.41\pm0.03,  2.56\pm0.03$ for
$(V-I),    (V-J),    (V-H),    (V-K)$    vs.    $(B-V)$    diagrams
respectively. Identical values for the  slopes are found for the region
$r<6^{\prime}$.  The  ratio of  total-to-selective  extinction in  the
cluster region, $R_{cluster}$, is also derived using the procedure given by
Pandey  et  al.  (2003).  The  ratios  ${E(V-\lambda)}\over  {E(B-V)}$
$(\lambda  \ge \lambda_I)$  yield $R_{cluster}  = 3.1  \pm  0.1$ which
indicates a normal reddening law in the cluster region.

\begin{figure}
%\centering
\includegraphics[scale = .6, trim = 20 20 100 100, clip]{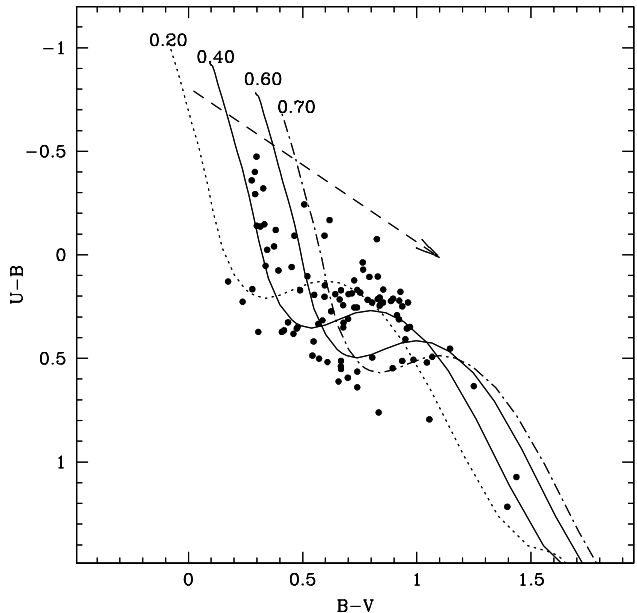}

\caption{$(U  - B)/(B  - V)$ colour-colour  diagram for the stars
within  the region  $r \le 6^\prime$. The dotted curve represents the
ZAMS by Girardi et al. (2002) shifted along the reddening slope of
0.72 (shown as a dashed line) for $E(B-V)$=0.20 mag (probable
foreground stars). The continuous curves  are the ZAMS shifted by $E(B
-  V)$ = 0.40 and 0.60 mag respectively to match the observations of probable
cluster members. The dot-dashed curve represents the ZAMS reddened by
$E(B-V)$=0.70 mag to match the probable background population (see
text).}
\label{ubbv}
\end{figure}

 In the absence of spectroscopic observations, 
the interstellar extinction $E(B -− V)$ 
toward the cluster region can be estimated using
the $(U −- B )/(B -− V )$ colour-colour (CC) diagram. The CC diagram of
the cluster region ($r \le 6^\prime$) is presented in
Fig. \ref{ubbv}. Since the cluster  is very young, a variable reddening  
is expected  within  the cluster
region. The figure also indicates a large amount of contamination due
to field stars. The probable foreground stars follow the intrinsic
zero-age-main-sequence (ZAMS) by Girardi et al. (2002) reddened by
$E(B-V)$=0.20 mag along a normal reddening vector (i.e., $E(U - B)
/E(B - V )$ = 0.72). In Fig. \ref{ubbv}, the continuous line represents
the theoretical ZAMS which is  shifted by $E(B  -V)$ = 0.40 and 0.60  mag
respectively, along the normal reddening vector to  match  the
observations of probable cluster members. Fig. \ref{ubbv}  yields  a
variable reddening with $E(B - V)_{min}$ = 0.40 mag to $E (B -
V)_{max}$ = 0.60 mag in the cluster region. A careful inspection of CC
diagram indicates the  presence of further reddened population. The
theoretical ZAMS, shown by dot-dashed line, is further shifted to
match the reddened sequence. This population may belong to blue plume
(BP) of Norma-Cygnus arm (cf. Carraro et al. 2005, Pandey et al.
2006). The $E(B  -V)$ value for the background population comes out to be
$\sim$0.70 mag, which is comparable to the $E(B  -V)$ value of BP population 
around $l \sim 170^\circ$ (cf. Pandey et al. 2006).

\begin{figure}
%\centering
\includegraphics[scale = .6, trim = 10 10 10 10, clip]{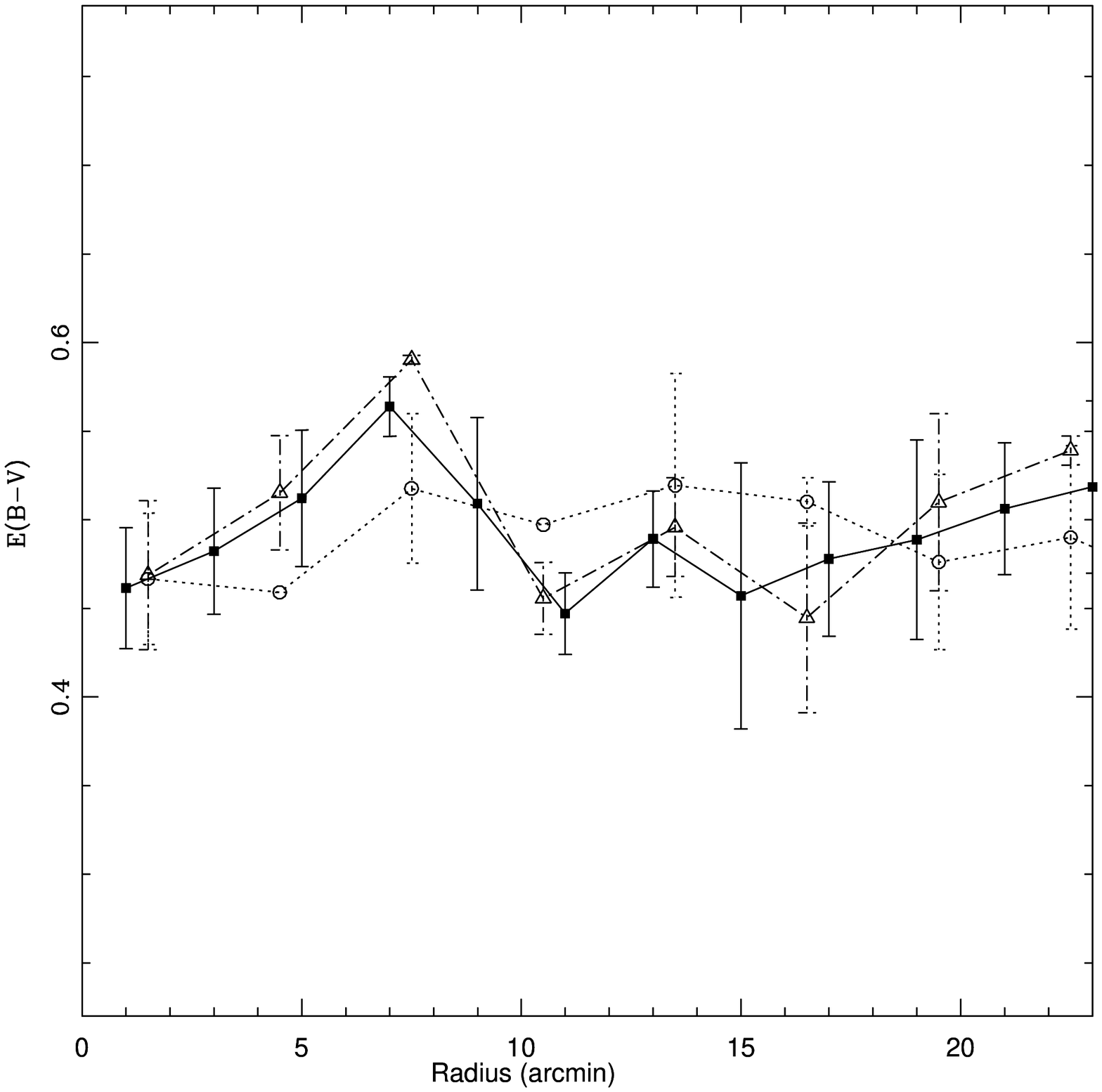}

\caption{Distribution of average reddening (filled squares) as a
function of radial distance from the cluster center. Error bars
represent standard errors.  Open circles connected by dotted line and
open triangles connected by dot-dashed line represent distribution of
average reddening in the east and west region of the cluster respectively. }
\label{red}
\end{figure}

Reddening of  individual stars having spectral types  earlier than $A0$
have  also been computed  by means  of  the reddening  free index  $Q$
(Johnson $\&$ Morgan 1953).  Assuming a normal reddening law we
can construct a reddening-free parameter index $Q = (U-B) - 0.72\times
E(B-V)$. For the MS stars, the intrinsic $(B-V)_0$ colour and
colour-excess can be obtained from the relation $(B-V)_0 = 0.332\times
Q$ (Johnson 1966, Hillenbrand et al. 1993) and $E(B-V) = (B-V) -
(B-V)_0$, respectively.  The  distribution of  reddening as  a function
of radial  distance from the cluster center is shown in
Fig. \ref{red}, which indicates a slight enhancement of $E(B-V)$ at
$r\sim$ $7^\prime$.  In Fig. \ref{red} we also show variation of
reddening towards eastern and western directions of the cluster.
From the figure it can be noted that the reddening towards eastern region
of the cluster is rather uniform whereas the reddening towards the western region
of the cluster shows relatively higher value (although statistically weak) 
at $r\sim$ $7^\prime$.

%%%%%%%%%%%%%%%%%%%%%%%%%%%%%%%%%%%%%%%%%%%%%%%%%%%%%%%%%%%%%%%%%%%%%%%%%%
%%%%%%%%%%%%%%%%%%%%%%%%%%%%%%%%%%%%%%%%%%%%%%%%%%%%%%%%%%%%%%%%%%%%%%%%
%%%%%%%%%%%%%%%%%%%%%%%%%%%%%%%%%%%%%%%%%%%%%%%%%%%%%%%%%%%%%%%%%%%%%%%%%%
%%%%%%%%%%%%%%%%%%%%%%%%%%%%%%%%%%%%%%%%%%%%%%%%%%%%%%%%%%%%%%%%%%%%%%%%

\section{\bf Optical Colour Magnitude Diagrams }

\begin{figure}
\centering
\includegraphics[scale = .82, trim = 10 10 10 160, clip]{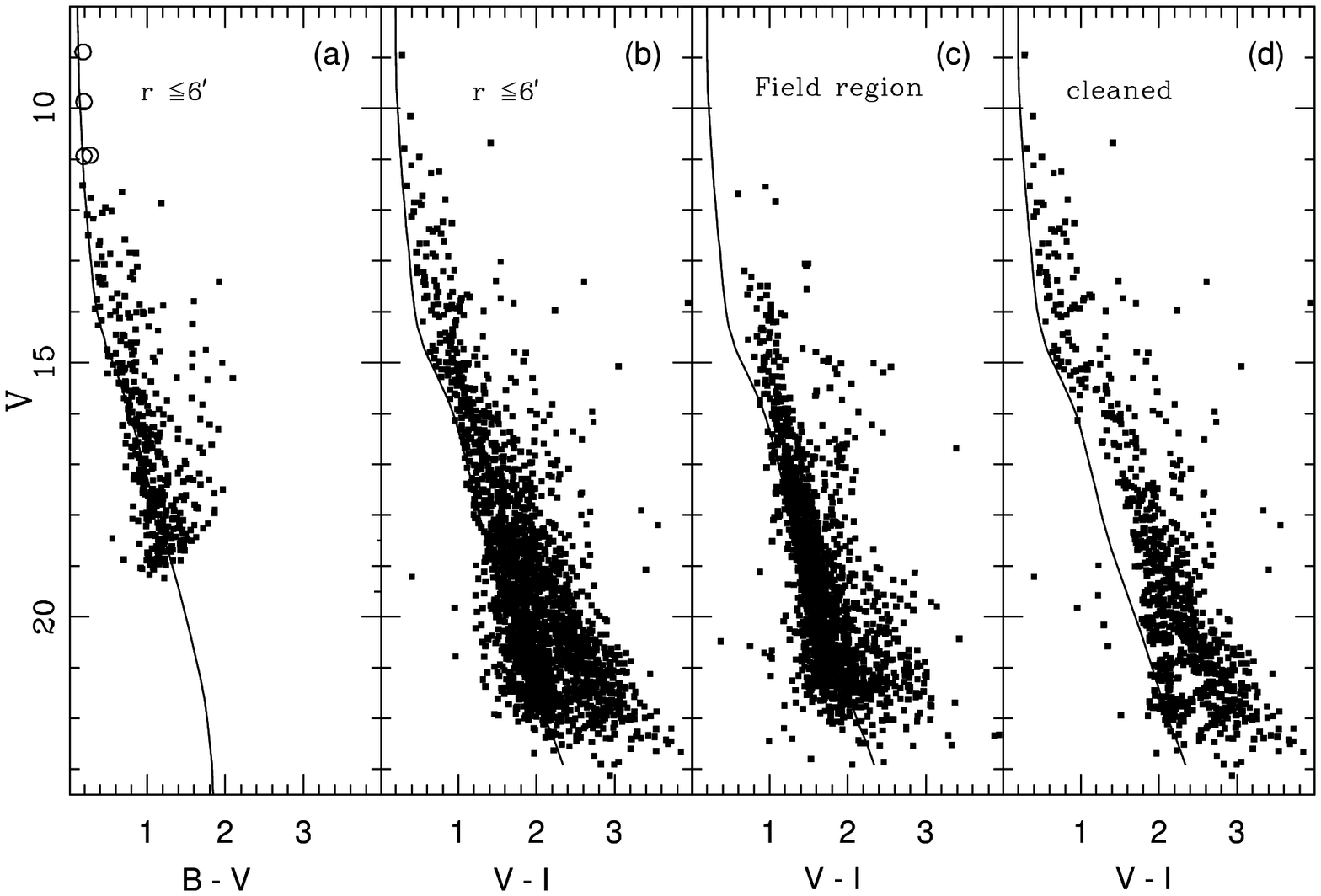}
\caption{(a) and (b): $V/(B-V)$ and $V/(V-I)$ CMDs for stars within
the region  $r \le 6^\prime$ of the cluster Stock 8, (c): $V/(V-I)$
CMD for stars in the nearby field  region having same area as in
Figs. \ref{cmd}a and \ref{cmd}b, (d): statistically cleaned $V/(V-I)$
CMD for the cluster region. The  continuous line is the isochrone of 2
Myr from Girardi et al. (2002) corrected for the cluster distance and
reddening. The stars in $V/(B-V)$ CMD shown by open circles are taken
from  Mayer \& Macak (1971). }
\label{cmd}
\end{figure}

The optical colour- magnitude diagrams (CMDs) were used  to derive the
cluster fundamental parameters such as age, distance etc. The $V/(B -
V)$ and $V/(V - I)$ CMDs of stars within radial distance $r \le
6^\prime$ of Stock 8 and $V/(V - I)$  CMD for the nearby field region
are shown in Fig. \ref{cmd}. The contamination due to background field
population is apparent in the CMDs of the cluster region. The CMDs  of
the cluster region also show a significant number of stars towards
the right of the ZAMS. Stars brighter than $V \sim$ 11.5 mag observed in B-band
using  Kiso Schmidt telescope are saturated even in frames with short
exposure.

%%%%%%%%%%%%%%%%%%%%%%%%%%%%%%%%%%%%%%%%%%%%%%%%%%%%%%%%%%%%%%%%%%%%%%%%%%
%%%%%%%%%%%%%%%%%%%%%%%%%%%%%%%%%%%%%%%%%%%%%%%%%%%%%%%%%%%%%%%%%%%%%%%%

To study the LF/MF, it is necessary to remove field star contamination
from the sample of stars in the cluster region. 
Membership determination is also crucial for assessing the presence of PMS stars
because  both PMS and  dwarf foreground  stars occupy  similar
positions above the ZAMS in  the CMDs. In the absence  of proper
motion  study, we  used  statistical criterion  to estimate the number
of probable member stars in the cluster region. To remove contamination 
of field  stars from the  MS and PMS  sample, we
statistically subtracted the contribution  of field stars from the
observed CMD of  the  cluster region  using  the following  procedure.
For any star in  the $V/(V-I)$ CMD of the  field region, the nearest
star in the cluster's  $V/(V-I)$   CMD  within   $V\pm0.125$  and
$(V-I)\pm0.065$ of  the field star  was removed.  While removing
stars from the cluster CMD, necessary corrections for incompleteness
of the data sample were taken into account. The  statistically cleaned
$V/(V-I)$ CMD  of the cluster region  is shown in  Fig. \ref{cmd}d
which clearly shows a sequence towards red side of the MS. The
contamination due to field stars at $V > 20$ mag and $(V-I) \sim 2.1$
mag can still be seen in Fig. \ref{cmd}d.

%%%%%%%%%%%%%%%%%%%%%%%%%%%%%%%%%%%%%%%%%%%%%%%%%%%%%%%%%%%%%%%%%%%%%%%%%%
\begin{figure}
\centering
\includegraphics[scale = .6, trim = 10 10 10 10, clip]{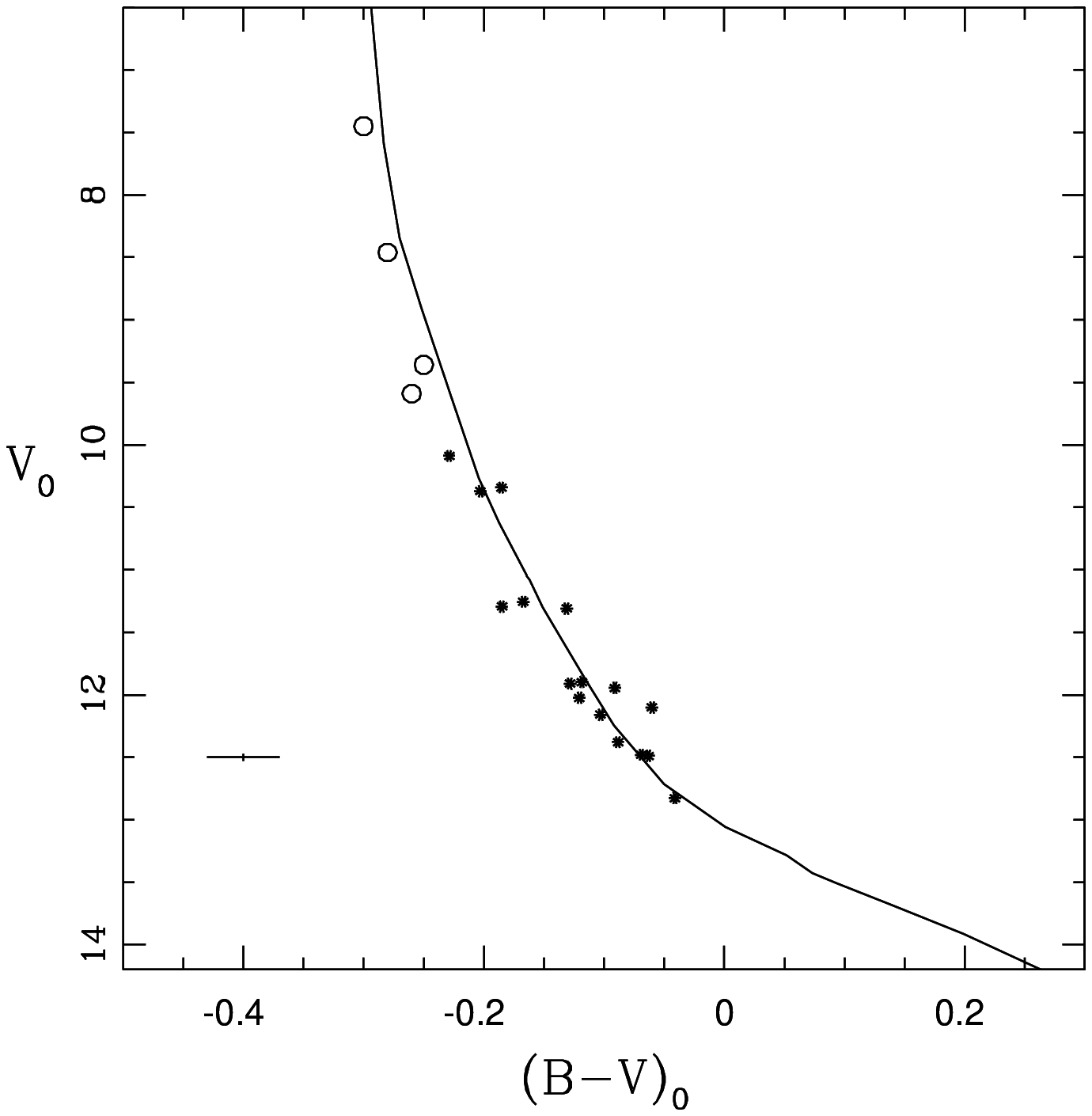}

\caption{$V_0/(B-V)_0$ CMD for stars lying within the region 
$r \le  6^{\prime}$ of the cluster 
Stock 8. The data for stars shown by open circles are taken
from Mayer \& Macak (1971). The isochrone (continuous curve) for 2 
Myr  by Girardi et al. (2002) is also shown. The average photometric
errors in magnitude and colour are shown in the lower left of the figure.}
\label{q}
\end{figure}
%%%%%%%%%%%%%%%%%%%%%%%%%%%%%%%%%%%%%%%%%%%%%%%%%%%%%%%%%%%%%%%%%%%%%%%%
\begin{figure}
\centering
\includegraphics[scale = .6, trim = 10 10 10 10, clip]{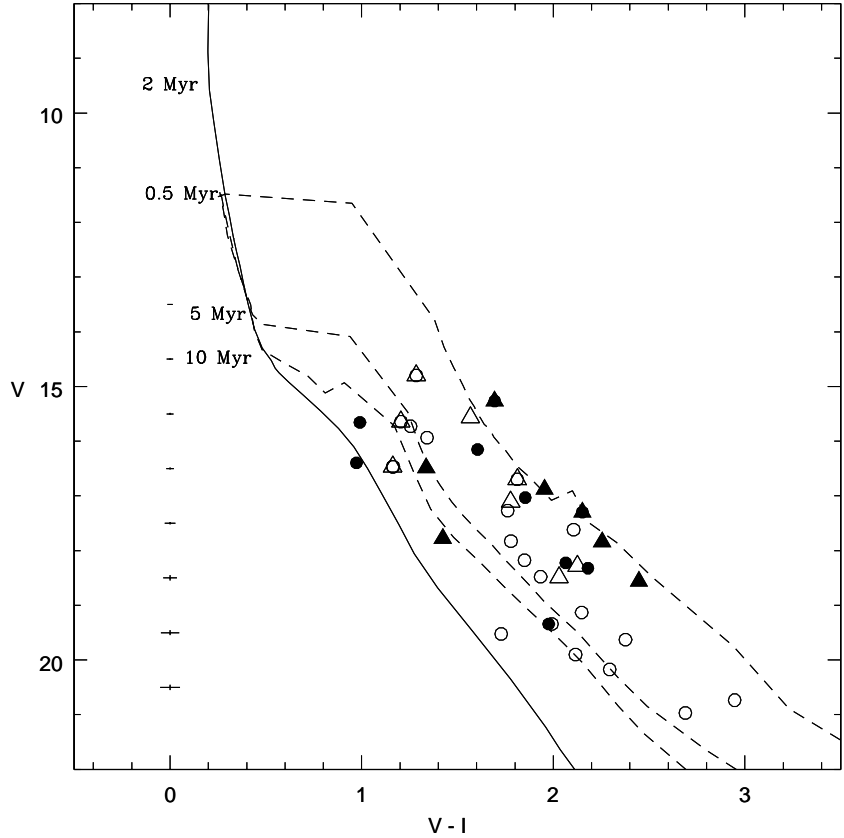}

\caption{The $V/(V-I)$ CMD  for H$\alpha$ emission (open and filled triangles) 
and NIR excess  stars (open and  filled circles),
lying  within  the regions  $r  \le  6^{\prime}$ and  $6^{\prime} < $  r  $\le
12^{\prime}$  respectively.   Isochrone  for  2  Myr  by  Girardi  et
al. (2002)  (continuous curve) and PMS  isochrones for 0.5, 5,  10 Myr by
Siess et al. (2000) (dashed  curves) are also shown. All the isochrones
are corrected  for the distance  of 2.05  kpc and reddening  $E(B -  V)$ =
0.40 mag. Average photometric errors in $(V-I)$ colour as a function 
of magnitudes are shown  in the left side of the figure.}
\label{excess}
\end{figure}
%%%%%%%%%%%%%%%%%%%%%%%%%%%%%%%%%%%%%%%%%%%%%%%%%%%%%%%%%%%%%%%%%%%%%%%%

\begin{figure}
%\centering
\includegraphics[scale = .6, trim = 10 10 10 10, clip]{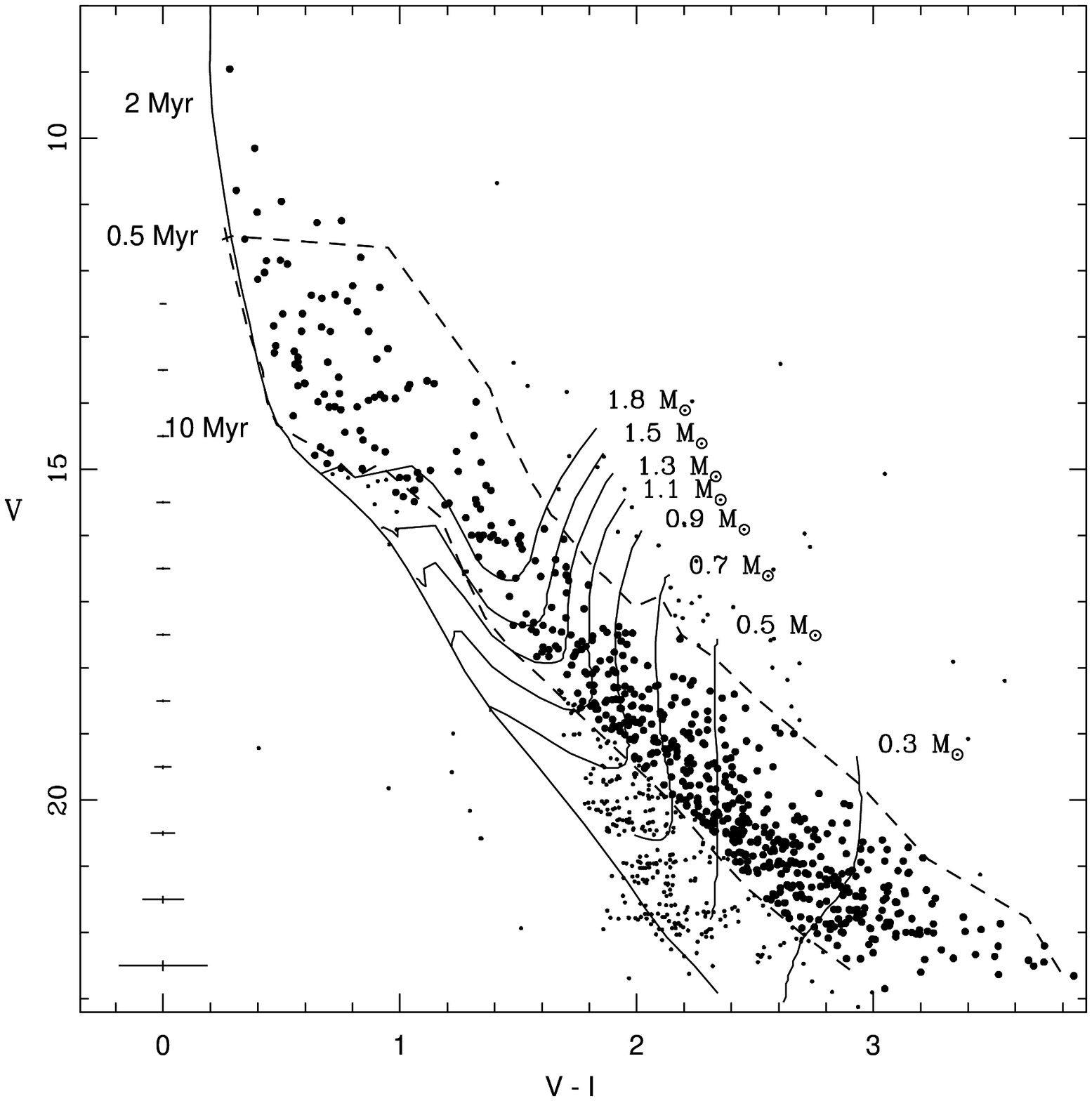}
\caption{Statistically cleaned $V/(V -  I)$ CMD for stars lying 
within the region $r \le  6^{\prime}$ of Stock 8 and between PMS  
isochrones of age 0.5  Myr and 10
Myr. The  isochrone for 2 Myr age  by Girardi et al.   (2002) and PMS
isochrones of 0.5, 10 Myr  along with evolutionary tracks of different
mass stars  by Siess et al.  (2000) are also shown. All  the isochrones are
corrected for  the cluster  distance and reddening.  The corresponding
values of  masses in solar mass  are given at the  right side of
each track. Points shown by small dots are considered as non-members.
Average photometric errors in $(V-I)$ colour as a function of magnitudes 
are shown  in the left side of the figure.}
 \label{calone}
\end{figure}
%%%%%%%%%%%%%%%%%%%%%%%%%%%%%%%%%%%%%%%%%%%%%%%%%%%%%%%%%%%%%%%%%%%%%%%%

In Fig. \ref{cmd}, using $E(B-V)_{min} =0.40$ mag
and  relations  $A_{V}=3.1\times  E(B-V)$;
$E(V-I)=1.25\times E(B-V)$, we visually fitted  theoretical isochrone
of log age  = 6.3  (2  Myr) and $Z=0.02$  by  Girardi et  al.  (2002)
to  the  blue  envelope of  the observed MS. We estimated  a distance
modulus of $(m-M)_V = 12.8  \pm 0.15$ mag which corresponds to a distance of
$2.05 \pm  0.10$  kpc. Fig. \ref{q} shows dereddened $V_0/(B-V)_0$ CMD
of the stars lying within $r \le 6^{\prime}$. The stars having
spectral type earlier than $A0$ were dereddened individually using the
$Q$  method as discussed in Section \ref{reddening}. 
Because bright stars in our observations got saturated even in short exposure
frames, we obtained their B-band data from from Mayer \& Macak (1971).
The isochrone for 2 Myr by Girardi et al. (2002) are also
plotted in the figure. This figure yields an average
post-main-sequence age of the massive stars of the cluster as $\leq$ 2
Myr.

In  Fig. \ref{excess} we have  plotted $V/(V-I)$ CMD  for  the young
stellar objects (YSOs) i.e. H$\alpha$ emission  and  NIR  excess
sources (cf. Section \ref{nircc}) lying  within  regions $r  \le
6^{\prime}$ and  $6^{\prime} < r$  $\le 12^{\prime}$.  The H$\alpha$
sources  are represented  by open  and filled triangles  and the  NIR
excess sources  by open and filled circles.  The PMS isochrones by
Siess et  al. (2000) for 0.5, 5,  10 Myr (dashed lines)  and isochrone
for  2 Myr  by Girardi et al. 2002  (continuous  line) are also
shown. Fig. \ref{excess} reveals that majority of the YSOs (i.e.,
H$\alpha$ and NIR excess stars) have ages $\le $ 5 Myr. The age spread
may indicate a non-coeval star formation in the
cluster. Fig. \ref{excess} also indicates a tendency that stars  lying
at $r > 6^\prime$ are  relatively younger than those in the cluster
region.

In Fig. \ref{calone} we have shown statistically cleaned CMD along
with PMS  isochrones  for  ages  0.5 Myr  and 10  Myr by Siess et
al. (2000). The 2 Myr  isochrone by Girardi et al. (2002) and
evolutionary tracks by Siess et al. (2000)  for different  masses are
also shown. A comparison of Figs. \ref{excess} and  \ref{calone}
suggests that the stars lying on the right side of the MS in
Fig. \ref{calone} may be  probable PMS stars. The stars lying between
the 0.5 - 10 Myr isochrones are considered  as PMS stars and this  set
of data  is used for calculation of the IMF (see Sec. 8).

\section{Near Infrared colour-colour and colour-magnitude diagrams}

The NIR data are very useful  to study the nature of young stellar
population  within the star forming regions. From 2MASS point source
catalogue, we obtained $JHK_s$ data (with photometric errors $\le$ 0.1 mag 
in all the three bands) for $445$ and $898$  NIR sources
lying  within  $r \le 6^{\prime}$ and $6^{\prime} < r$ $\le
12^{\prime}$ respectively. In the following Section, we discuss NIR
colour-colour diagram and CMDs.

\subsection {Colour-Colour Diagram}
\label{nircc}

%%%%%%%%%%%%%%%%%%%%%%%%%%%%%%%%%%%%%%%%%%%%%%%%%%%%%%%%%%%%%%%%%%%%%%%%%%
\begin{figure}
\centering
\includegraphics[scale = .8, trim = 0 10 0 150, clip]{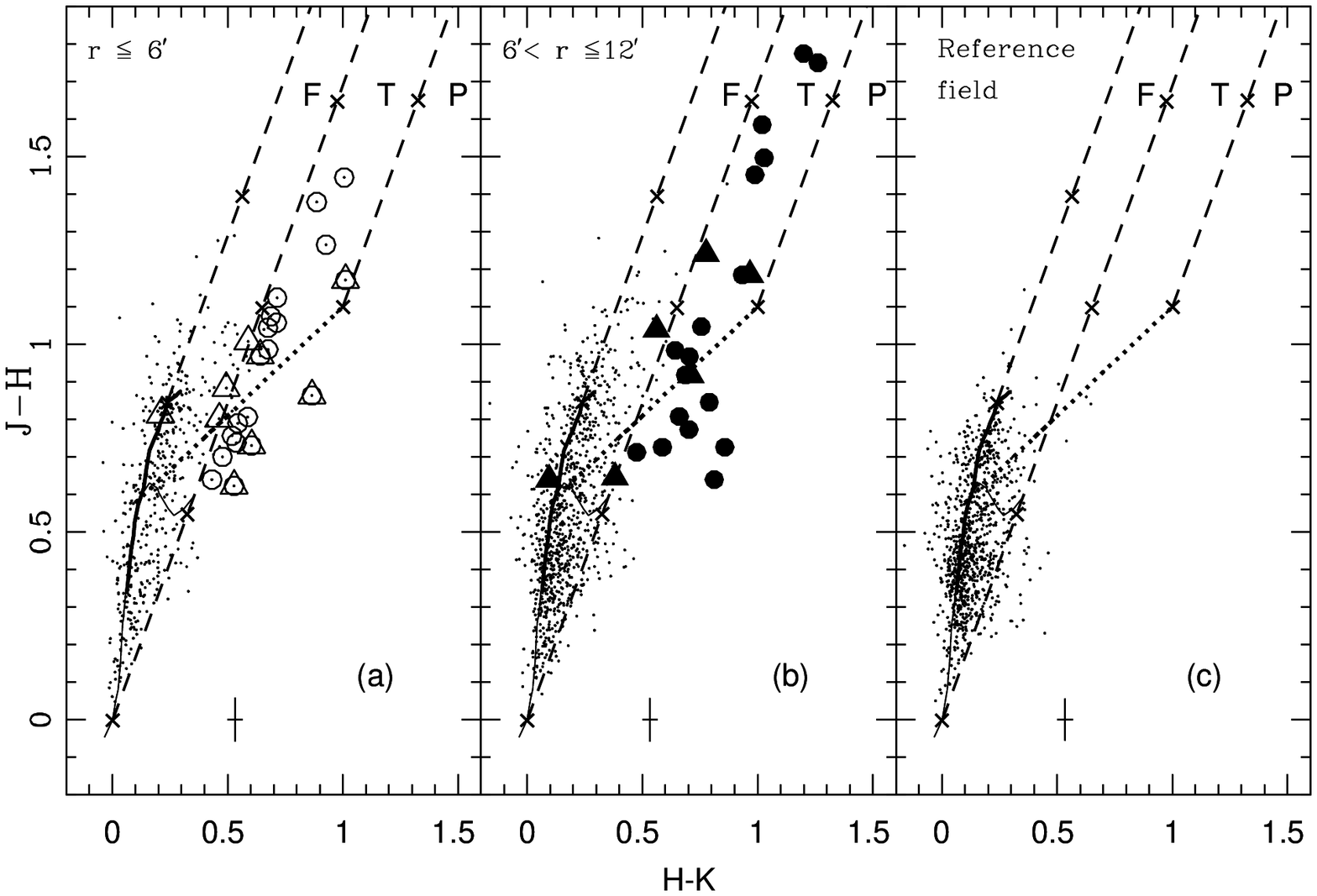}

\caption{$(J - H)/(H -  K)$ colour-colour diagrams of sources detected
in the  $J H K_s$ bands with  photometric errors less than  0.1 mag in
(a)  Stock 8 region  ($r \le  6^{\prime}$)  (b)  outside the Stock 8  region
$6^{\prime} < r$  $\le  12^{\prime}$ (c) the nearby reference field.  
The locus for dwarfs  (thin solid curve) and
giants (thick  solid curve) are  from Bessell $\&$ Brett  (1988). The
dotted  line  represents  the  locus   of  T  Tauri  stars  (Meyer  et
al.  1997).  Dashed straight  lines  represent  the reddening  vectors
(Cohen et al. 1981). The crosses  on the dashed lines are separated by
$A_V$  =  5  mag.  The  open  and filled  triangles are H$\alpha$ emission
sources while open and filled circles are NIR excess sources for the inner
and outer regions, respectively. The average photometric errors are shown 
in the lower left of the panel. }

\label{jhhk}
\end{figure}
%%%%%%%%%%%%%%%%%%%%%%%%%%%%%%%%%%%%%%%%%%%%%%%%%%%%%%%%%%%%%%%%%%%%%%%%

The $(J-H)/(H-K)$  colour-colour diagrams for the cluster region $r
\le 6^{\prime}$, outside the cluster region  $6^{\prime} <  r$ $\le
12^{\prime}$,  and a nearby field region are  shown in
Fig. \ref{jhhk}.  The thin and  thick solid lines are the locations of
unreddened main-sequence and giant stars (Bessel $\&$ Brett 1988),
respectively. The dotted line represents the locus of T Tauri stars
(Meyer et al. 1997). The parallel dashed lines are the reddening
vectors for the early MS and giant type  stars (drawn from the base
and tip of the two branches).  The crosses on the dashed lines are
separated by  an $A_{V}$ value of  5 mag. The extinction ratios,
$A_J/A_V = 0.265, A_H/A_V = 0.155$ and $A_K/A_V=0.090$, 
are adopted from Cohen et al. (1981). All the 
2MASS magnitudes and colours
have been converted into the CIT system. The curves are also in the
CIT system.

 Presently young stellar objects are classified  as an evolutionary
sequence spanning a few million years as:  Class 0/ Class I - the
youngest embedded protostars surrounded by infalling envelopes and
growing accretion  disk; Class II - PMS stars with less active
accretion disks and Class III - PMS stars with no disks or optically
thin remnant dust (Adams et al. 1987). 
Following Ojha et al. (2004a), we classified sources
according to their locations in $(J-H)/(H-K)$  colour-colour diagrams.
The `F' sources are those located between 
the reddening vectors projected from the
intrinsic colour of main-sequence and giant. These are considered
to be field stars (main-sequence, giants) or Class III /Class II
sources with small NIR excesses. The `T' sources are located redward of
`F' but blueward of the reddening line projected from the red
end of the T Tauri locus of Meyer et al. (1997). These sources are
considered to be mostly classical T Tauri stars (Class II objects)
with large NIR excesses. There may be an overlap in NIR colours of
Herbig Ae/Be stars and T Tauri stars in the `T' region (Hillenbrand et
al. 1992). The `P' sources are those located in the region redward of
region `T' and are most likely Class I objects (protostellar objects).

\begin{figure}
%\centering
\includegraphics[scale = .6, trim = 10 10 10 10, clip]{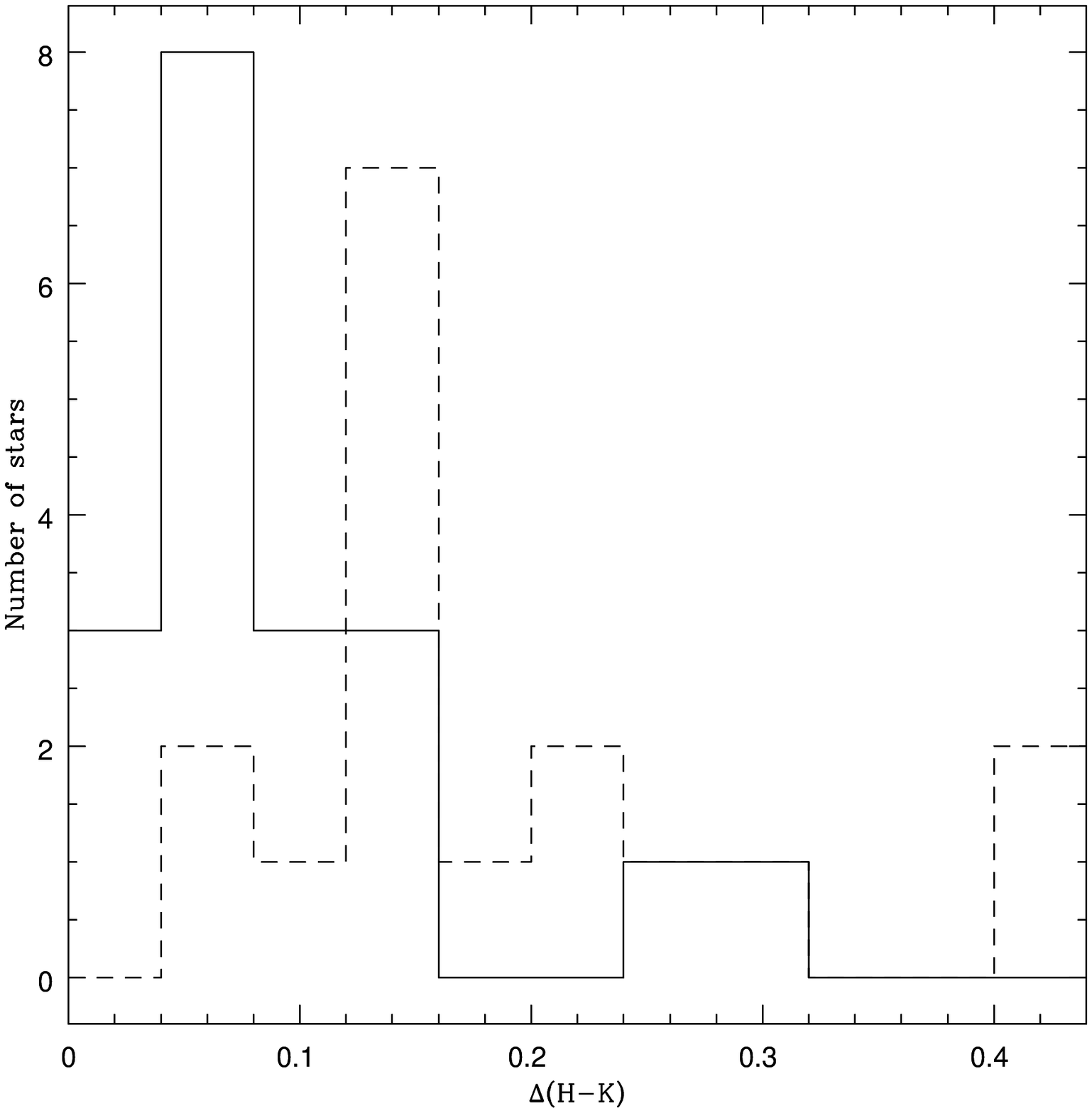}

\caption{Distribution of NIR excess in two regions. Continuous and dashed 
line histograms show the distribution in $r\le 6^{\prime}$ and 
$6^{\prime} < r$ $\le 12^{\prime}$ regions respectively. }

\label{hist}
\end{figure}

In Fig. \ref{jhhk}, a significant number of stars show NIR excess
indicating that the Stock 8 region is  populated by YSOs. A comparison
of the  colour-colour diagrams of  the cluster region  with the  field
region (Fig. \ref{jhhk}c) indicates  that stars in  the cluster region
having  $J-H>$ 0.6 and lying towards the right side of the reddening
vector at the boundary of  `F' and `T' regions can  be  safely
considered  as NIR excess stars.  This criterion yields 19 and 17 NIR
excess sources in the Stock 8 ($r\le 6^{\prime}$) and outside the
cluster  ($6^{\prime} < r$ $\le 12^{\prime}$) regions, respectively,
which are shown as  open and filled circles in
Fig. \ref{jhhk}. Distribution of H${\alpha}$ emission stars  is also
shown  as open ($r\le 6^{\prime}$) and  filled   ($6^{\prime} < r$
$\le 12^{\prime}$) triangles. Fig. \ref{jhhk} indicates that the YSOs
located outside the cluster region have relatively higher extinction
and  NIR excess as compared to those located in the inner region. 
The distribution of NIR excess, $\Delta (H-K)$, defined as horizontal
displacement  from the reddening vector at the boundary of `F' and `T'
regions (cf. Fig. \ref{jhhk}), for the two regions is shown in Fig. \ref{hist}
which manifests that YSOs located in the region  $6^{\prime} < r$
$\le 12^{\prime}$ have relatively higher NIR excess.

%%%%%%%%%%%%%%%%%%%%%%%%%%%%%%%%%%%%%%%%%%%%%%%%%%%%%%%%%%%%%%%%%%%%%%%%%%%%%%%%%%%%%%%%%%%%%%%%%
%%%%%%%%%%%%%%%%%%%%%%%%%%%%%%%%%%%%%%%%%%%%%%%%%%%%%%%%%%%%%%%%%%%%%%%%%%%%%%%%%%%%%%%%%%%%

\subsection{Colour-Magnitude Diagrams}

%%%%%%%%%%%%%%%%%%%%%%%%%%%%%%%%%%%%%%%%%%%%%%%%%%%%%%%%%%%%%%%%%%%%%%%%%%%%%%%%%%%%%%%%%%%%%%%%%
\begin{figure}
%\centering
\includegraphics[scale = .6, trim = 10 10 10 10, clip]{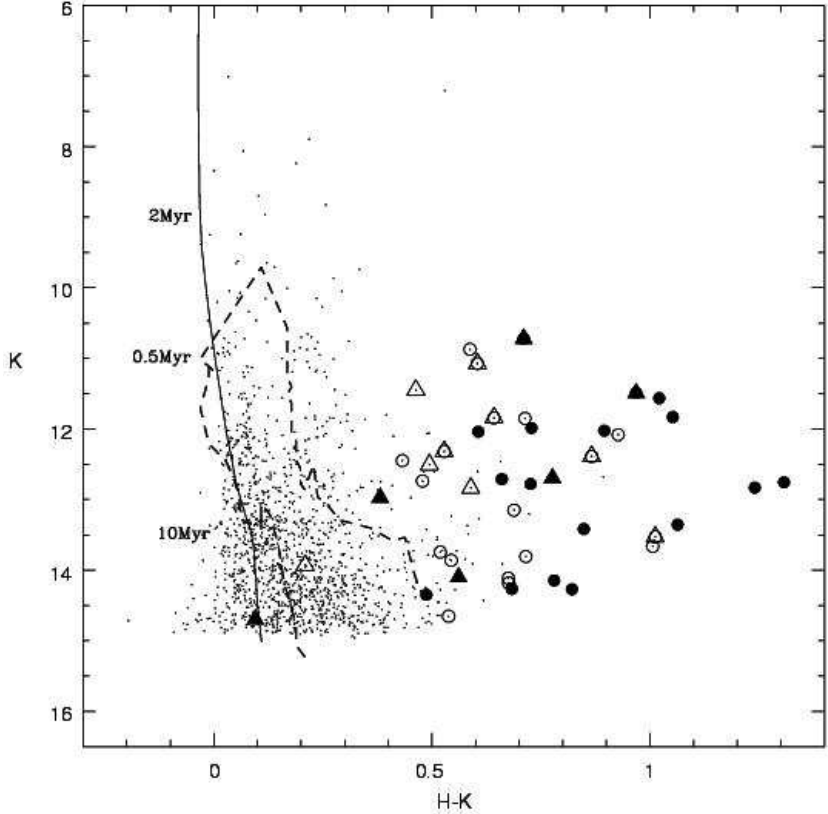}

\caption{$K/(H - K)$ CMD for stars within the region $r \le
12^{\prime}$. The symbols are same as in Fig. \ref{excess}. Isochrone
for 2 Myr by Girardi et al. (2002) (continuous curve) and PMS
isochrones from 0.5 and 10 Myr by Siess et al. (2000) (dashed curves),
corrected for cluster distance and reddening are also shown. }

\label{hkk}
\end{figure}
%%%%%%%%%%%%%%%%%%%%%%%%%%%%%%%%%%%%%%%%%%%%%%%%%%%%%%%%%%%%%%%%%%%%%%%%%%%%%%%%%%%%%%%%%%%%
\begin{figure}
%\centering
\includegraphics[scale = .6, trim = 10 10 10 10, clip]{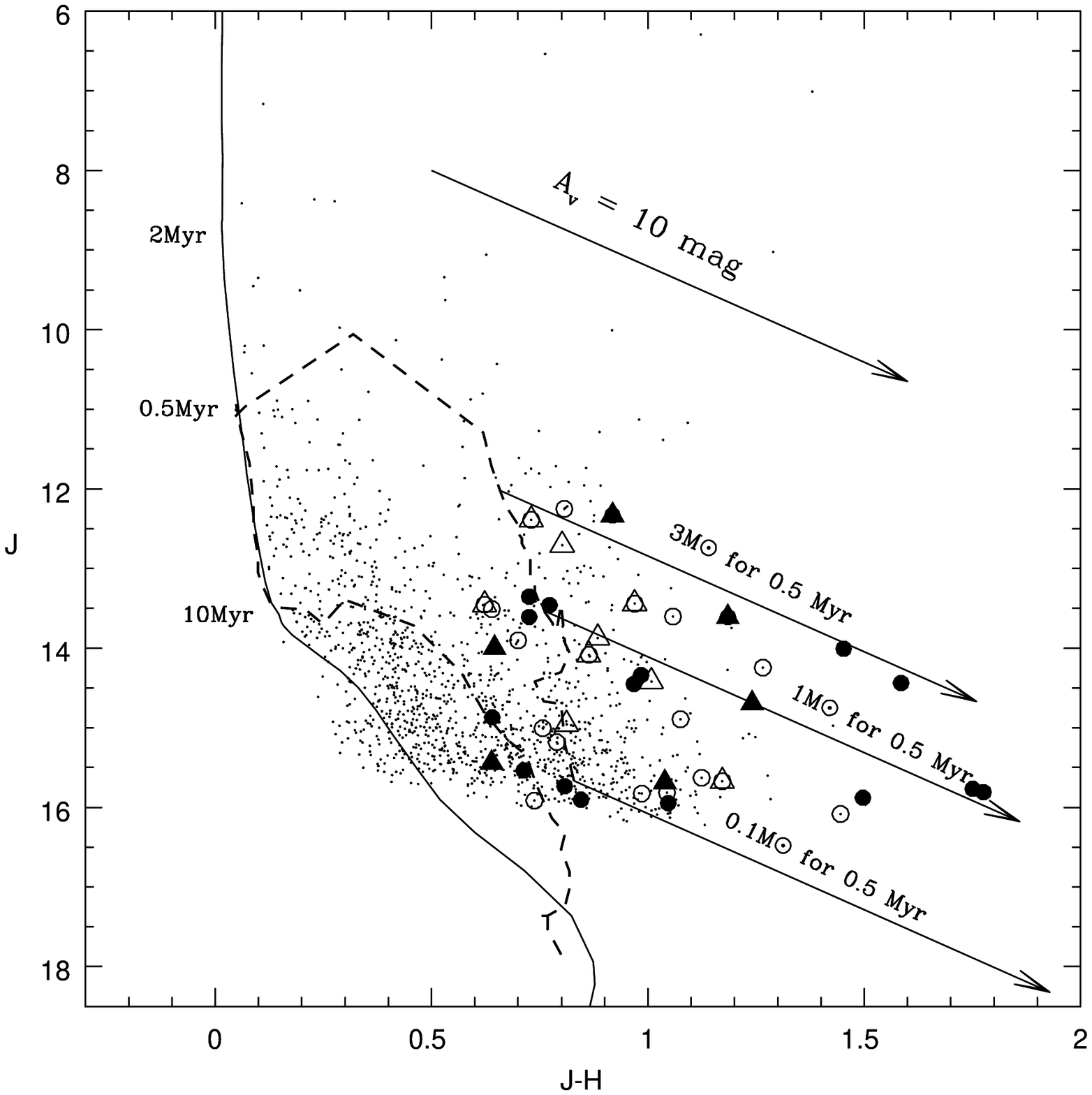}

\caption{$J/(J - H)$ CMD  for stars within the region $r \le
12^\prime$ . The symbols are same as in Fig. \ref{excess}. The
isochrone of 2 Myr (continuous curve) by Girardi  et al. (2002) and
PMS isochrones of age 0.5  and 10  Myr (dashed curves) by Siess  et
al.  (2000), corrected  for cluster distance and  reddening are also
shown. The continuous oblique reddening lines  denote  the  positions
of   PMS  stars  of  0.1,  1.0 and  3.0 $M_{\odot}$ for 0.5 Myr.}

\label{jhj}
\end{figure}

%%%%%%%%%%%%%%%%%%%%%%%%%%%%%%%%%%%%%%%%%%%%%%%%%%%%%%%%%%%%%%%%%%%%%%%%%%%%%%%%%%%%%%%%%%%%%%%%%

The $K/(H-K)$ and $J/(J-H)$ CMDs for the  NIR excess and H$\alpha$
sources  detected  within $r \le 12^\prime$  region  are  shown  in
Figs. \ref{hkk}  and \ref{jhj}. The isochrone for age 2 Myr by
Girardi et al.  (2002)  and PMS isochrones for ages 0.5, 10 Myr by
Siess et al. (2000) respectively, have been  plotted assuming 
a distance of  2.05 kpc  and an extinction of $E(B-V)_{min}=0.40$ mag as
obtained   from   the  optical   data.  Fig. \ref{hkk}  also indicates
that sources lying  outside the cluster region i.e.  $6^\prime  < r$
$\le  12^\prime$  have  relatively higher  NIR excess.  
Therefore, from Figs. \ref{excess}, \ref{jhhk}, \ref{hist}  and  \ref{hkk}, we infer
that sources lying in the region   $6^\prime  < r$  $\le  12^\prime$
 may possibly be  younger than those located in the
cluster region ($r \le  6^\prime$).

The mass  of the  probable YSO candidates can  be estimated  by
comparing their  locations on  the CMD with the  evolutionary models
of PMS stars.  To estimate  the stellar masses, the $J$  luminosity is
recommended rather than  that of $H$ or $K$,  as  the  $J$  band   is
less  affected  by  the  emission  from circumstellar material
(Bertout et al. 1988). In  Fig. \ref{jhj}, the continuous oblique
reddening lines denote  the positions of PMS  stars of 0.5 Myr age
having masses 0.1, 1.0 and 3.0 $M_\odot$. The  YSOs, 
in general, have masses in the range  0.1 to 3.0  $M_\odot$.

\section{Initial Mass Function}

Young  clusters are important  tools to study the IMF since their
MF can be considered  as the IMF  as they  are too young  to loose
significant number   of  members  either  by   dynamical  or  stellar
evolution.  To study  the IMF  of Stock 8  we  used  the data within
$r \le 6^\prime$.

The mass function (MF) is often expressed by the power law,  $N (\log
m) \propto m^{\Gamma}$  and the slope of the MF is given as:
    $$ \Gamma = d \log N (\log m)/d \log m $$

\noindent
where $N  (\log m)$ is the  number of stars per  unit logarithmic mass
interval.  For the mass range $0.4  < M/M_{\odot} \le 10$,  the classical  
value derived  by Salpeter  (1955)  for the slope of IMF is $\Gamma = -1.35$ .

%%%%%%%%%%%%%%%%%%%%%%%%%%%%%%%%%%%%%%%%%%%%%%%%%%%%%%%%%%%%%%%%%%%%%%%%%%%
\begin{figure}
\centering
\includegraphics[scale = .6, trim = 10 10 10 10, clip]{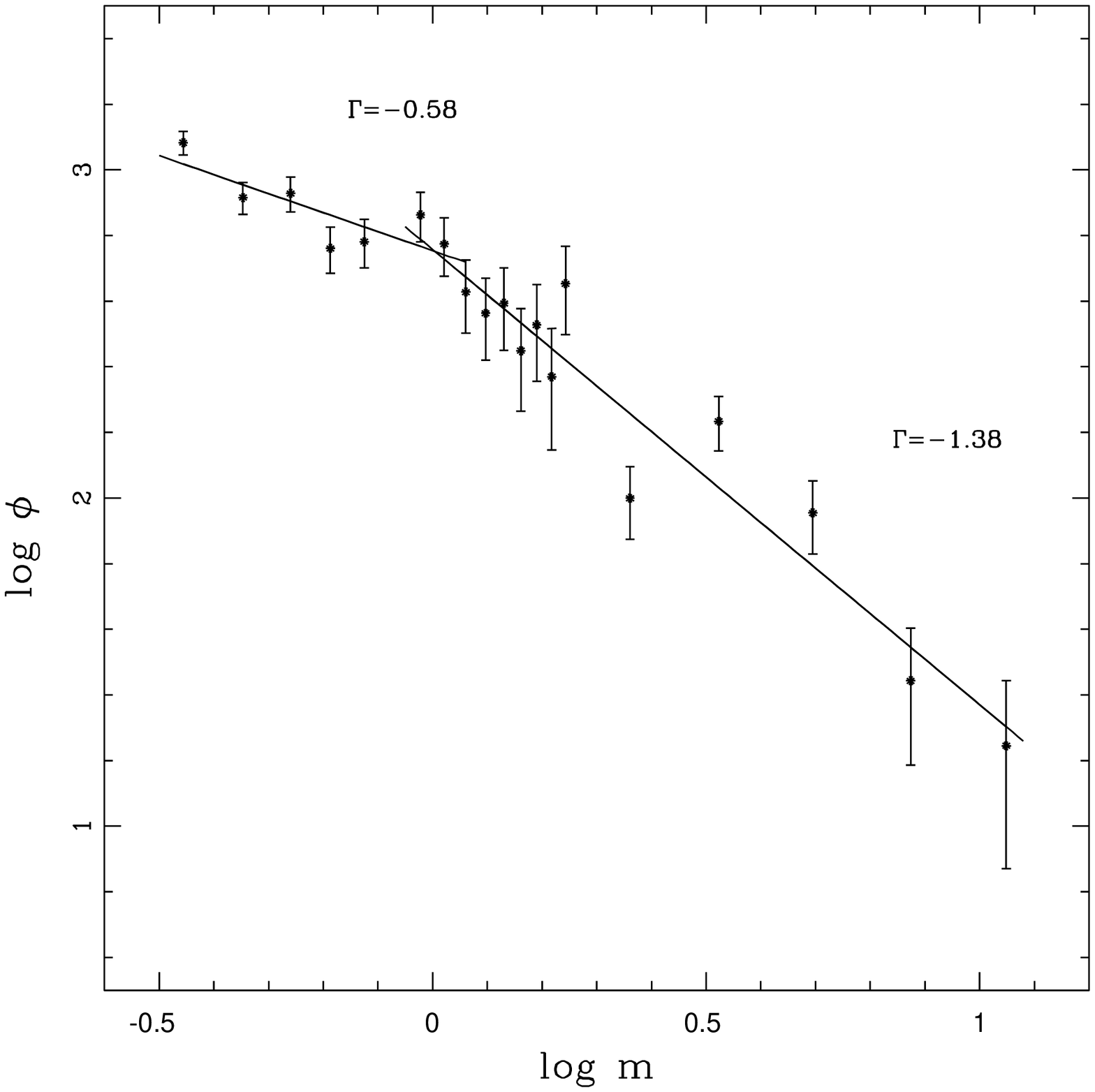}

\caption{A plot of the mass function for the cluster region $r \le
6^\prime$. The  log $\phi$ represents $N$/dlog $m$.  The error bars
represent  $\pm$$\sqrt{N}$ errors. The continuous lines show
least-squares fit  to the mass ranges described in the text. The
values of the slopes obtained are also mentioned in the figure.}
\label{mf}
\end{figure}
%%%%%%%%%%%%%%%%%%%%%%%%%%%%%%%%%%%%%%%%%%%%%%%%%%%%%%%%%%%%%%%%%%%%%%%%%

With the  help of statistically cleaned  CMD shown in
Fig. \ref{calone}, we can derive the MF using  theoretical
evolutionary models. Since the post-main-sequence age of the cluster is
$\le$ 2 Myr,   stars  having  $V \le 15$ mag  have  been considered
to be on the main sequence. For  the MS stars LF was converted to MF using
the theoretical model by  Girardi et al. (2002) (cf. Pandey et
al. 2001, 2005).  The MF for PMS stars was  obtained by counting the
number of stars in various mass bins (shown as  evolutionary tracks in
Fig. \ref{calone}). Necessary corrections for incompleteness of the
data sample  were taken into account for each magnitude bin to
calculate  the MF. The  MF of  the cluster is  plotted in
Fig. \ref{mf}. The slope  of  the  mass function  $\Gamma$,  in  the
mass range $1.0 \le M/M_{\odot}<13.4$, can be  represented by a power
law having a slope of $\Gamma$ = $-1.38\pm0.12$, which agrees well
with Salpeter value (-1.35). In the mass range $0.3 \le M/M_\odot <
1.0$, the mass function also follows a power law but 
with a shallower slope $\Gamma = -0.58\pm 0.23$ indicating a 
break in the slope of the
MF at $\sim$ $1 M_{\odot}$.  The break in the power  law has already
been reported in  the case of a  few young clusters e.g. Trapezium
and IC 348 (Muench  et al.  2002, 2003). They showed a flattening 
in slope of the IMF  at $\sim 0.6 M_\odot$.  Recently, in the case of
young clusters NGC 1893 (Sharma et al. 2007) and Be 59 (Pandey et
al. 2007), a break in the power law was reported at $\sim$ $2 M_{\odot}$
and $\sim$ $2.5 M_{\odot}$, respectively. However, 
$\Gamma = -0.58\pm 0.23$ obtained for Stock 8 below $1 M_{\odot}$ is 
shallower than that obtained for NGC 1893 ($-0.88\pm 0.09$) and is 
steeper than that obtained for Be 59 (approximately flat). 
In the case of NGC 2362 the MF flattens at $\sim$ $3 - 1 M_{\odot}$ 
(Damiani et al. 2006) which is at a
much higher mass than observed in some young clusters. Prinsinzano et
al. (2005) found that the MF of young open cluster NGC 6530, in the
mass range   $0.4 \le M/M_\odot < 4.0$, can be represented by a power
law having slope ($ -1.22 \pm 0.17$). However the MF does not flatten
below  $\sim 0.6 M_{\odot}$ as in the case of some young clusters
e.g. Trapezium, Taurus and IC 348, instead it decreases for masses
lower than $\sim 0.4 M_{\odot}$. Here it is worthwhile to mention
that exclusion of the lowest mass bin in Fig. \ref{mf} will yield a shallower slope
$\Gamma  =-0.3 \pm 0.3$, although with large error, for masses below $1 M_{\odot}$,
which is comparable with the slope obtained by Muench et al. (2002, 2003) for 
Trapezium and IC 348 clusters. However a deeper photometry is required 
to study the nature of mass function of Stock 8 below  $1 M_{\odot}$.

%%%%%%%%%%%%%%%%%%%%%%%%%%%%%%%%%%%%%%%%%%%%%%%%%%%%%%%%%%%%%%%%%%%%%%%%%
\begin{figure}
\centering
\includegraphics[scale = .6, trim = 10 10 10 10, clip]{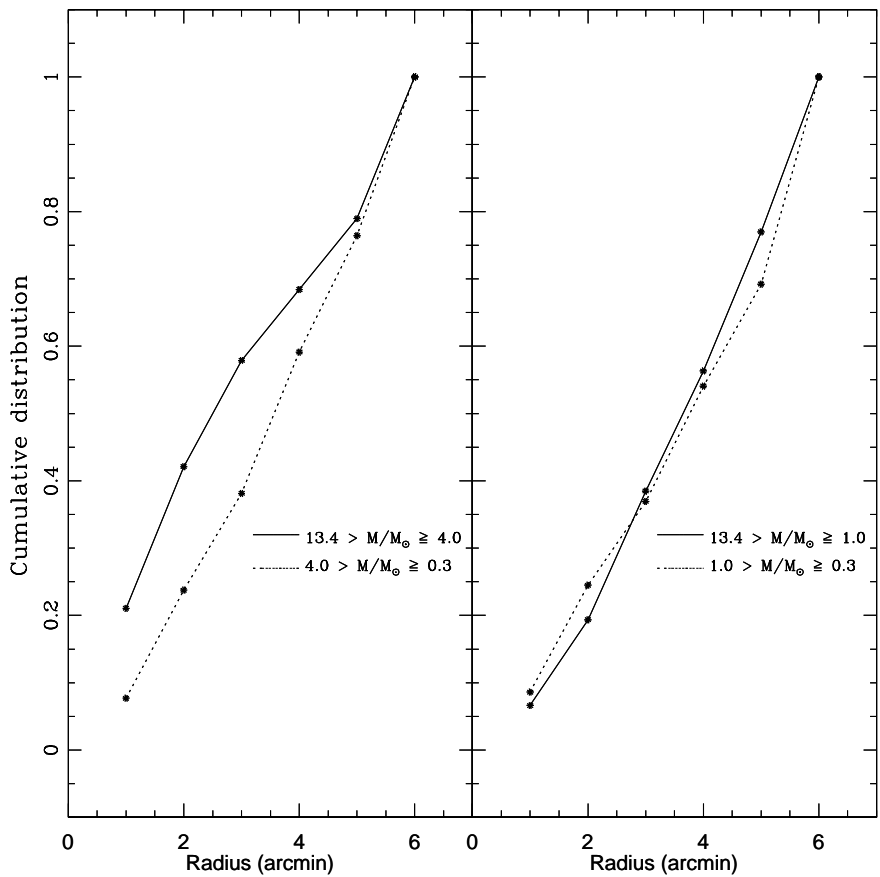}

\caption{Cumulative distributions of stars of different mass intervals as a 
function of radial  distance.} 

\label{masegn}
\end{figure}
%%%%%%%%%%%%%%%%%%%%%%%%%%%%%%%%%%%%%%%%%%%%%%%%%%%%%%%%%%%%%%%%%%%%%%%%%%

The  mass segregation,  in the  sense  that massive  stars being  more
centrally concentrated than the lower mass stars, has been reported in
several LMC and Milky Way star clusters (see  Chen et al. 2007; de
Grijs et al. 2002 a, b, c; Fischer et al. 1998; Pandey et al. 1992,
2001, 2005). In the case of intermediate or  old clusters, the  mass
segregation is mainly  due to the effect of  dynamical evolution. But
in the  case of young clusters the  mass segregation  may be the
imprint of  star formation  process. To characterize  the degree  of
mass  segregation in  Stock 8,  we plotted cumulative distribution of
stars as a function of distance from the  cluster center in two mass
groups as shown in Fig. \ref{masegn}. The left panel shows the  distribution for
mass groups $4\le M/M_\odot<13.4$ (high mass group) and $0.3\le
M/M_\odot<4$ (low mass group),  whereas the  right panel  shows the 
distribution  for  $1.0\le M/M_\odot<13.4$ (high mass group) and
$0.3\le M/M_\odot<1.0$ (low mass group).  The Kolmogorov-Smirnov test
indicates that the left panel shows mass  segregation at a confidence
level  of $\sim  75\%$, whereas the right panel does not show any mass 
segregation. This  indicates  that  mass  segregation,
although weak, is effective  only towards the  higher mass end. Since
variation of reddening within the cluster is not significant
(cf. Fig. \ref{red}), and incompleteness of the data does not depend on
the radial distance (cf. Table \ref{cf_opt}), we believe that the mass
segregation may be the imprint of star formation process.

%%%%%%%%%%%%%%%%%%%%%%%%%%%%%%%%%%%%%%%%%%%%%%%%%%%%%%%%%%%%%%%%%%%%%%%%%
\begin{figure}
%\centering
\includegraphics[scale = .6, trim = 10 10 10 10, clip]{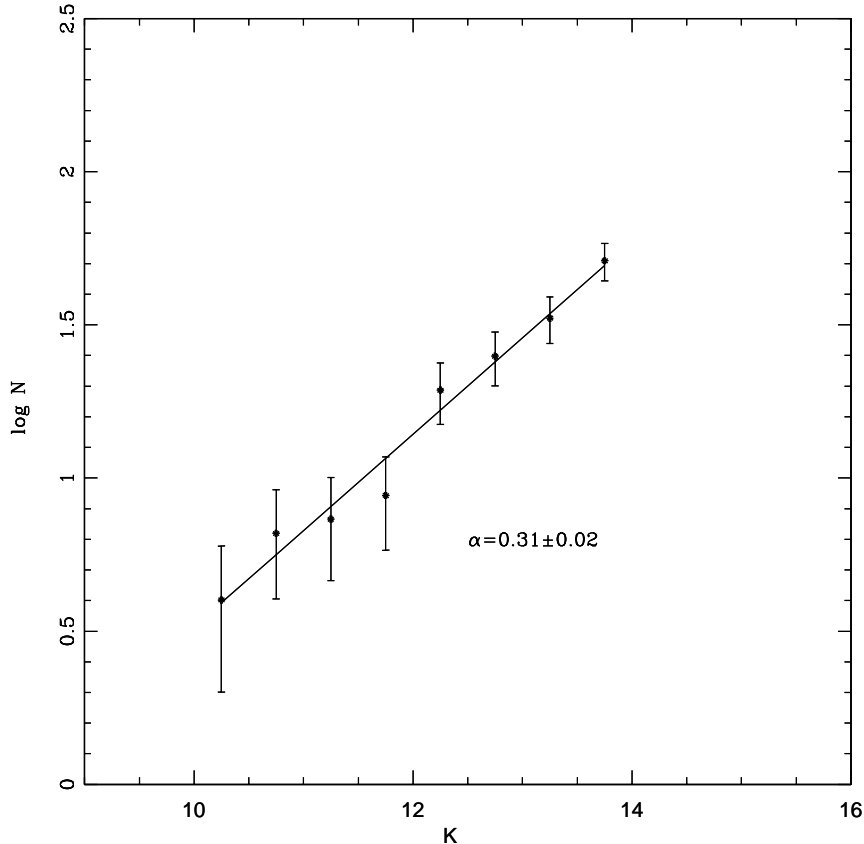}
\caption {KLF derived  after subtracting  the model corrected field star 
contamination (see the text). The linear fit is  represented by the continuous line.}
\label{klf}
\end{figure}

%%%%%%%%%%%%%%%%%%%%%%%%%%%%%%%%%%%%%%%%%%%%%%%%%%%%%%%%%%%%%%%%%%%%%%%%

The $K$-band luminosity function (KLF) is a powerful tool to investigate
the IMF of young embedded star clusters.
During the last decade several studies have been carried out 
with the aim of determining the KLF of
young open clusters  (e.g. Lada \&  Lada 2003,  Ojha et  al. 2004b,
Sanchawala et al. 2007). We have used 2MASS $K_s$-band data to 
study the KLF in Stock 8 region. The completeness of the 2MASS data is
obtained by using ADDSTAR routine as discussed in Sec. 2.1. The CFs
obtained for two sub-regions are given in Table \ref{cf_2mass}.    The KLF  in
Stock  8  region  is  studied  by using the Besan\c con Galactic model
of stellar population  synthesis (Robin  et al.  2003) 
and  stars from a nearby reference field to take  into account  
foreground/background
field star contamination (cf. Sharma et al. 2007).  The reference
field, also taken from 2MASS, is located at a radial
distance of $\sim 30^{\prime}$ away from the cluster center.  Star
counts are predicted using the  Besan\c con model towards the
direction of the control  field. An advantage of using this model is
that we can simulate foreground ($d<2.05$ kpc) and  the
background  ($d>2.05$   kpc)  field star populations separately.  
The foreground population was simulated using the model
with $A_V$ = 1.24 mag ($E(B-V) = 0.40$ mag (cf. Section \ref{reddening})
and $d < 2.05$ kpc. The background
population ($d>2.05$ kpc) was simulated with $A_V$ = 1.86 mag.  
Thus we  determined  the  fraction  of  the contaminating  stars
(foreground + background)  over  the  total  model counts. This fraction
was used to scale the nearby reference field and subsequently  the
star counts  of  the  modified  control field  were subtracted from
the  KLF of the cluster to  obtain the final corrected KLF.  The
scale factor in different magnitude bins was found to be in the range of 0.8 - 1.0
with a mean value of $\sim 0.9$.

\begin{table}
%\begin{minipage}{80mm}
\caption{Completeness Factor (CF) of 2MASS data in
the cluster and field regions.}
\label{cf_2mass}
%\medskip
%\scriptsize
\begin{tabular}{cccc} \hline
K range&    Stock8 &  & Field region\\
 (mag)& $r\le 3^\prime$ & $3^\prime <r \le6^\prime$&  \\
\hline

10.0 - 10.5  &  1.00   &    1.00     &       1.00  \\
10.5 - 11.0  &  1.00   &    1.00     &       1.00\\
11.0 - 11.5  &  1.00   &    1.00     &       0.98\\
11.5 - 12.0  &  1.00   &    1.00     &       0.97\\
12.0 - 12.5  &  0.95&  0.96    &       0.91\\
12.5 - 13.0  &  0.99&    1.00     &       0.96\\
13.0 - 13.5  &  0.94&  0.94    &       0.97\\
13.5 - 14.0  &  0.96&  0.96    &       0.97\\

\hline
\end{tabular}
%\end{minipage}
\end{table}

The KLF is expressed by following power-law:

%\midskip
%\hspace{5mm}
\begin{center}
${{ \rm {d} N(K) } \over {\rm{d} K }} \propto 10^{\alpha K}$
\end{center}
%\midskip

where ${ \rm {d} N(K) } \over  {\rm{d} K }$ is the number of stars per
0.5 magnitude   bin  and   $\alpha$  is  the   slope  of   the  power
law. The KLF for the cluster region $r \le 6^{\prime}$ shown in
Fig. \ref{klf}, yields a slope of  $0.31\pm0.02$, which is smaller
than the average value  of  slopes  ($\alpha  \sim 0.4$) for  young
clusters (Lada  et al. 1991;  Lada \& Lada  1995; Lada \& Lada
2003).  Smaller values of KLF slope ($\sim 0.3 - 0.2$) have been
  reported for various young embedded clusters (Megeath et
  al. 1996, Chen et al. 1997, Brandl et al. 1999, Ojha et al. 2004b,
  Leistra et al. 2005, Sanchawala et al. 2007, Pandey et al. 2007).

The KLF of Stock 8 is worth comparing with those of
NGC 1893 and Be 59 since  all the KLFs are obtained using a similar
technique (NGC 1893, Sharma et al. 2007; Be59, Pandey et
al. 2007). The slope  ($\alpha = 0.31\pm0.02$) obtained for Stock 8 is
in agreement with those obtained for NGC 1893  ($\alpha =
0.34\pm0.07$) and Be 59 ($\alpha = 0.27\pm0.02$). 

\section {Embedded Cluster in the Nebulous Stream}

%%%%%%%%%%%%%%%%%%%%%%%%%%%%%%%%%%%%%%%%%%%%%%%%%%%%%%%%%%%%%%%%%%%%%%%%
\begin{figure*}
\centering
\includegraphics[scale = .95, trim = 0 50 10 160, clip]{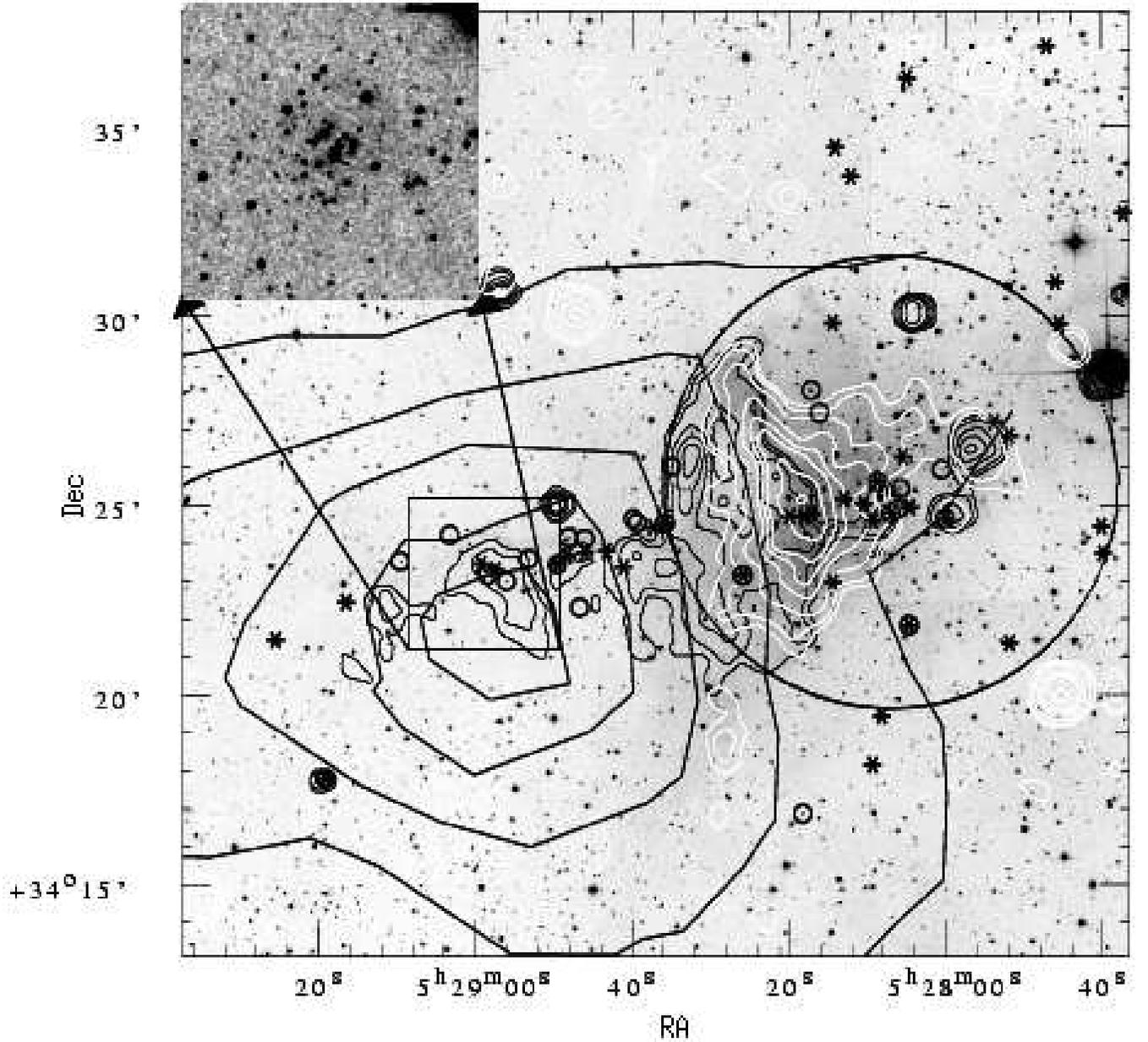}
\caption{Spatial distributions of $H\alpha$ emitters (open circles) and 
IR-excess sources (asterisk symbols) overlaid on DSS-2 $R$ band image. 
The thick contours represent  $^{12}$CO emission from 
Leisawitz et al. (1989). NVSS (1.4 GHz) radio contours  (white) and MSX A-band 
intensity contours (thin black) have also been shown. The MSX A-band 
contours are 1.5, 2, 3, 4, 5, 10, 20, 40, 60, 80 $\%$ of the peak value 
$9.666 \times 10^{-5}$ $W m^{-2} Sr^{-1}$ and the NVSS radio contours are  
3, 5, 10, 20, 40, 60, 80, 90 $\%$ of the peak value 29.5 mJy/beam. 
The circle having radius, $r= 6^{\prime}$ represents optical extent 
of the cluster (see Fig. \ref{rad_18}). The inset box shows the enlarged 
2MASS $K_s$-band image of the embedded cluster CC14. The abscissa and the 
ordinates are for the J2000 epoch. }
\label{kimage}
\end{figure*}
%%%%%%%%%%%%%%%%%%%%%%%%%%%%%%%%%%%%%%%%%%%%%%%%%%%%%%%%%%%%%%%%%%%%%%%%55

\begin{figure}
%\centering
\includegraphics[scale = .6, trim = 10 10 10 10, clip]{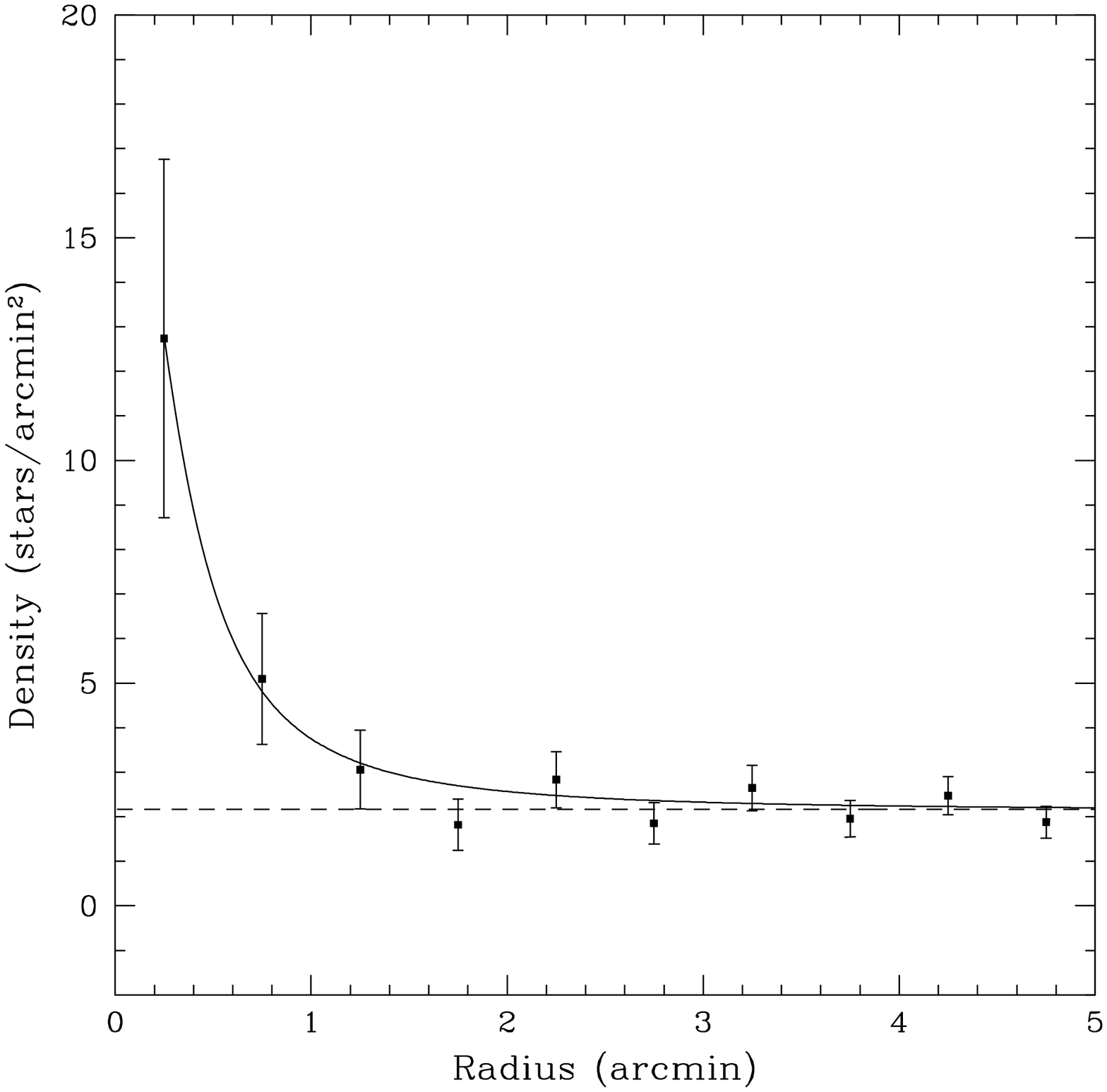}

\caption{Radial density profile for the embedded cluster CC14 obtained 
from the 2MASS data. The continuous curve is the least square fit of 
the King (1962) profile to the observed data points and the error
bars represent  $\pm$  $\sqrt{N}$ errors. The dashed line represents 
the mean density level of the field stars.}
\label{radens_nircluster}
\end{figure}

A Nebulous Stream, as identified in Fig. \ref{stock8}, is
clearly seen towards the eastern region of the cluster, Stock 8.
An embedded cluster, at a radial distance of $\sim$ $13^\prime$ from 
the center of Stock 8, can be noticed in the Nebulous Stream 
at $\alpha_{2000}$ = $05^{h}29^{m}00^{s}$; $\delta_{2000}$ = 
$+34^{\circ}23^{\prime}12^{\prime\prime}$. Its 2MASS $K_s$ band image 
together with the distribution of NIR excess stars and 
H$\alpha$ emitters is shown in Fig. \ref{kimage}. Borissova et al. (2003) 
detected this cluster  in their search 
for 2MASS PSC and designated it  as CC 14.
Recently, Ivanov et al. (2005)  have studied 
CC 14 using deep NIR observations. They estimated a distance modulus 
$(m  - M)_0  = 16.6\pm0.8$  mag, corresponding  to a  distance
of  $\sim 21$  kpc which indicates that  CC 14 is not associated with
the IC 417 region. However, the  distribution of NIR-excess and  
H$\alpha$ emission stars shown in Fig. \ref{kimage} seems to indicate a 
physical  relationship between CC 14 and the  H {II}  region IC 417, 
which is discussed later. In Fig. \ref{radens_nircluster} we show the 
RDP for CC 14 obtained from the 2MASS data around the  position  mentioned above, 
which indeed manifests a clustering with a radius of $\sim 1^\prime.5$.

%%%%%%%%%%%%%%%%%%%%%%%%%%%%%%%%%%%%%%%%%%%%%%%%%%%%%%%%%%%%%%%%%%%%%%%%%
%%%%%%%%%%%%%%%%%%%%%%%%%%%%%%%%%%%%%%%%%%%%%%%%%%%%%%%%%%%%%%%%%%%%%%%%%%

%%%%%%%%%%%%%%%%%%%%%%%%%%%%%%%%%%%%%%%%%%%%%%%%%%%%%%%%%%%%%%%%%%%%%%%%%
\begin{figure}
%\centering
\includegraphics[scale = 0.9, trim = 10 00 10 320, clip]{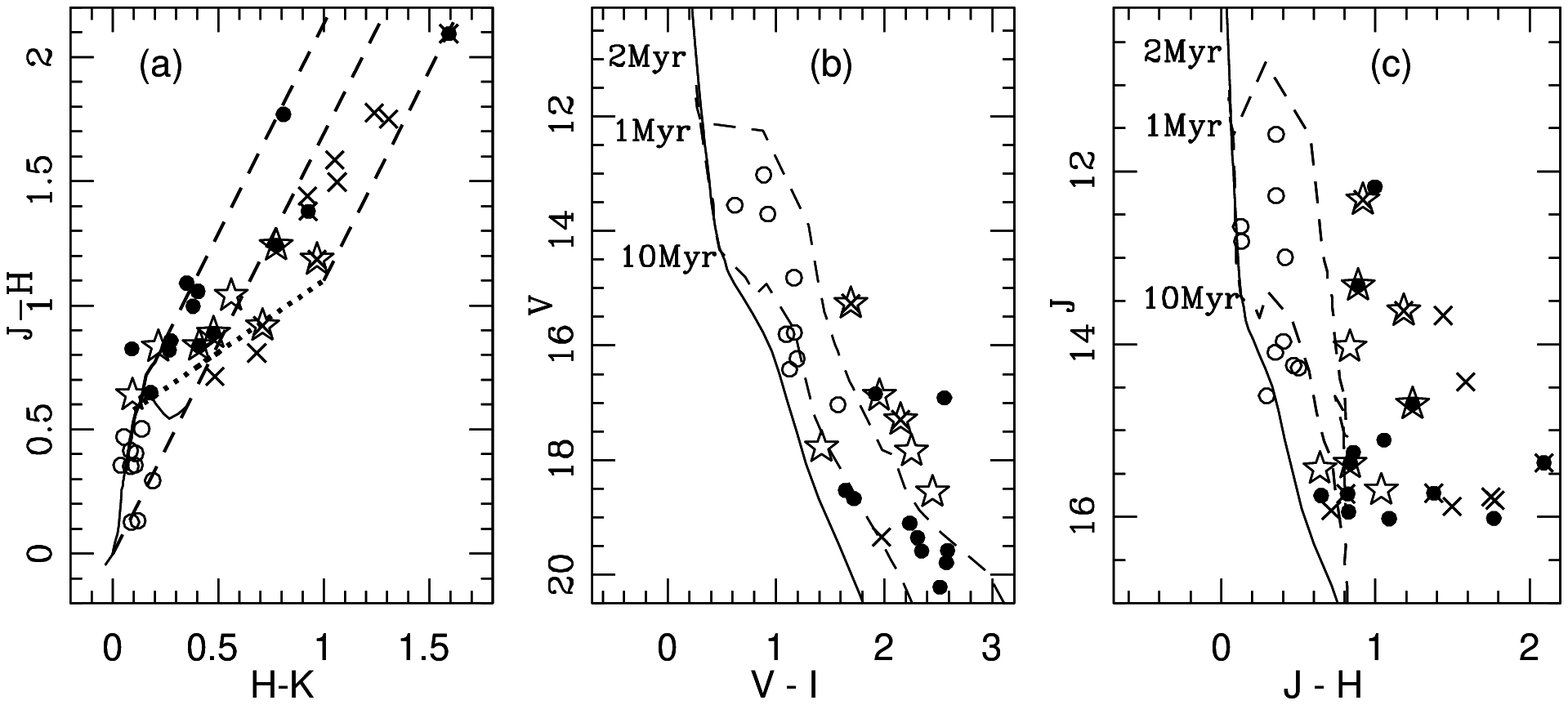}

\caption{Colour - colour diagram and CMDs for the stars lying within $1^\prime.5$ 
radius of the embedded cluster CC14 as well as for YSOs lying along the Nebulous Stream. 
$(a)$: $(J - H)/(H - K)$ colour - colour 
diagram with probable cluster members ($J-H > 0.6$) as filled circles and field 
stars as open circles. Crosses and star symbols are the NIR excess  and H$\alpha$ 
emission sources respectively, lying along the Nebulous Stream. $(b)$ and $(c)$ are 
$V/(V - I)$ and $J/(J-H)$ CMDs respectively. The symbols are same as 
in Fig. \ref{nir_dust}a. In Figs. \ref{nir_dust}b and \ref{nir_dust}c the isochrone
of 2 Myr (continuous curve) by Girardi  et al. (2002) and PMS isochrones of age
1  and 10  Myr (dashed curves) by Siess  et al.  (2000), corrected  for cluster
distance and  reddening, are also  shown.}
\label{nir_dust}
\end{figure}
%%%%%%%%%%%%%%%%%%%%%%%%%%%%%%%%%%%%%%%%%%%%%%%%%%%%%%%%%%%%%%%%%%%%%%%%%%

In Fig. \ref{nir_dust}a we show $(J-H)/ (H-K)$ colour-colour diagram for
stars in CC 14 cluster as well as for the YSOs distributed along
the Nebulous Stream. A comparison of Fig. \ref{nir_dust}a with
Fig. \ref{jhhk}  indicates that the stars  having $(J-H) > 0.6$ can be
considered as probable YSOs associated with CC 14 and the Nebulous
Stream. Assuming that the YSOs have average mass of $ \sim 1
M_\odot$ (i.e. $ M_V \sim  4.5$ at 1 Myr), mean $V \sim$ 18.5 mag and
average $A_V \sim$ 2.5 mag (cf. Fig. \ref{nir_dust}), we estimate a 
true distance modulus of $\sim$ 11.5, which clearly indicates that CC 14 and 
YSOs associated with the Nebulous Stream can not be located at the
distance ($\sim$ 21 kpc) suggested by Ivanov et al. (2005).

Of the 23  stars located within CC 14 region, the location of 10 stars
(open circles) on  the $(J-H)/ (H-K)$  colour-colour diagram indicates
that they  may be  probable field stars. The RDP
(Fig. \ref{radens_nircluster}) for   CC 14 also indicates 
the presence of $\sim$ 14 field stars in the  region. Probable members of  CC
14 are shown by filled circles  in Fig. \ref{nir_dust}a. Figures
\ref{nir_dust}b and \ref{nir_dust}c show  $V/(V-I)$  and $J/(J-H)$
CMDs for the CC 14 region where  open and filled circles represent
field and probable cluster populations respectively.  Comparison of
$V/(V-I)$  CMD (Fig. \ref{nir_dust}b) with the field region  $V/(V-I)$
CMD (Fig. \ref{cmd}c) clearly indicates that the field population
(open  circles) in the  embedded cluster region belongs to the same
population  as that of    the  nearby field  region.  The probable
cluster members (filled circles) follow  $\sim$ 1 Myr isochrone. In
Figs. \ref{nir_dust}a-c we have also plotted the  NIR-excess and
H$\alpha$ emission  stars lying along the Nebulous Stream 
using  crosses and star symbols, respectively. The $V/(V-I)$ CMD indicates that the
stars lying on the Nebulous Stream also have ages of $\sim$ 1 Myr.

\section {Distribution of Gas and Dust around Stock 8}

%%%%%%%%%%%%%%%%%%%%%%%%%%%%%%%%%%%%%%%%%%%%%%%%%%%%%%%%%%%%%%%%%%%%%%%%%
\begin{figure*}
%\centering
\includegraphics[scale = .94, trim = 20 120 0 120, clip]{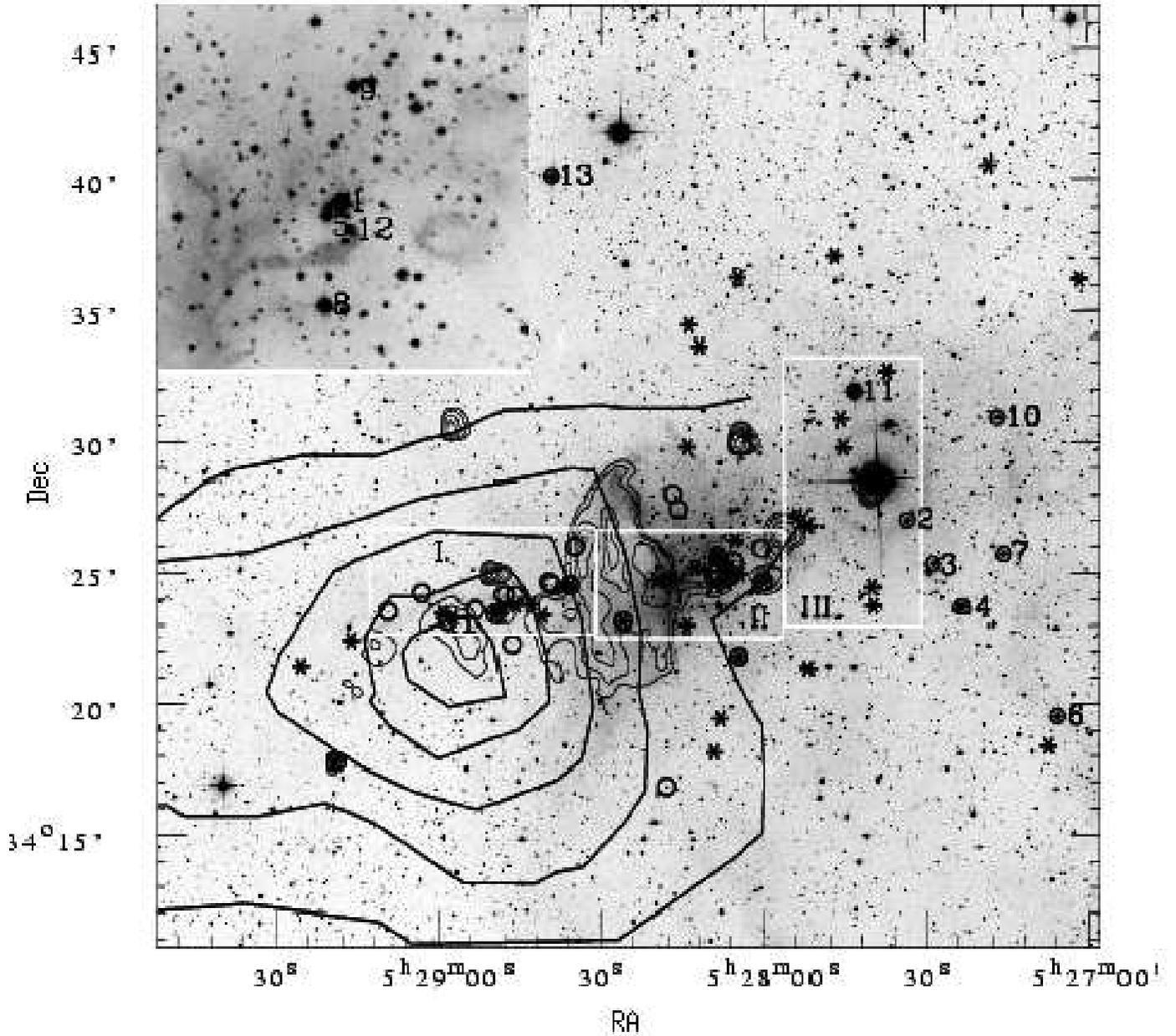}

\caption{Spatial distributions of OB stars, H$\alpha$ emitters (open circles)
and IR-excess sources (asterisks) overlaid on DSS-2 $R$ band image of Stock 8 
region. The inset box shows an enlarged view of $6^\prime$ central region 
of Stock 8. The thick contours represent  $^{12}$CO emission from 
Leisawitz et al. (1989). MSX A-band intensity contours
(thin lines) have also been shown. The contour levels are the  same as in Fig. \ref{kimage}. 
The star numbers from Table 3  and three sub regions (I - III, see the text) are also shown.}
\label{comap}
\end{figure*}
%%%%%%%%%%%%%%%%%%%%%%%%%%%%%%%%%%%%%%%%%%%%%%%%%%%%%%%%%%%%%%%%%%%%%%%%%%

No detailed observations have  been made so far, to probe the molecular cloud 
associated with Stock 8. Only available is the $^{12}$CO contour map 
of IC 417 (Sh2-234) region traced from figure 27c of Leisawitz et al. (1989), 
which is shown in Fig. \ref{kimage}. Although the resolution is low, a moderate-sized 
molecular cloud of $\sim 3.4 \times 10^{3} M_{\odot}$ is found 
adjacent to the east of Stock 8. 

In Fig. \ref{kimage}, NVSS (1.4 GHz) radio contours (white) and MSX
A-band  MIR contours (black) are overlaid on DSS-2 $R$ band
image. The  radio   continuum emission peaks around
$\alpha_{2000}\sim$ $05^{h}28^{m}18^{s}$  and $\delta_{2000}\sim$
$+34^{\circ}25^{\prime}30^{\prime\prime}$ and shows  a  sharp cutoff
at  $\alpha_{2000}\sim$ $05^{h}28^{m}30^{s}$  indicating an interface
between the ionized gas and molecular cloud.  The integrated flux
density  of the radio continuum above $3\sigma$ level is  estimated to
be 2.474 Jy. Assuming a spherical symmetry for the ionized region  and
neglecting absorption of ultraviolet radiation by dust inside the H II
region,  the above flux density together with assumed distance,  allow
us to estimate the number of Lyman continuum photons ($N_{Lyc}$)
emitted per second, and hence the spectral type of the exciting star.
Using the relation given by Mart\'{i}n-Hern\'{a}ndez et al. (2003)
and assuming  an electron temperature of 8000 K, we 
estimated log$N_{Lyc} = 47.67$,  which corresponds to a MS spectral type of
$\sim$ B0 (Panagia 1973; Schaerer \& de Koter 1997). This seems to be
consistent with the B0/O9.5 star (cf. Table 3, star no. 1), at the
center of Stock 8, as the ionizing source of the region.  Major uncertainty 
in the above calculation may arise due to neglect of 
the absorption of Lyman continuum photons by the dust.
We estimated the effect of dust using the following two  methods; 1) the dust continuum
density was estimated by averaging FIR optical depth $\tau_{100}$
(shown in Fig. \ref{tau_temp}a) over the radio emission 
region, and 2) attributing the spread in $E(B-V)$ in the cluster 
region (cf. Sec. 5) to the local
dust distribution. In both the cases, we assumed a uniform dust
distribution and  the dust composition to be an equal mixture of
Silicate and Graphite with properties identical to 
those given by Draine \& Lee (1984). The calculation yields log$N_{Lyc} \sim$  
48.15 (from method 1) and 48.32 (from method 2) implying the ZAMS spectral type 
of the ionizing star to be O9/O8.5 and O8 from methods 1 and 2, respectively.
Keeping in mind the uncertainties in estimating the spectral type of the
star and the assumptions made to calculate the total number of Lyman continuum
flux, we feel that the above estimates are comparable.

However star no. 1 is located almost at the western edge of the
ionized gas, so roughly half of its UV emission may not be used in
creating the HII region. The H II region IC 417 contains several OB
stars  (cf. Mayer \& Macak, 1971), whose distribution is shown in
Fig.  \ref{comap}.  Presumably these stars are the first generation of
stars in the IC 417 region.   The details of the OB stars are given in
Table 3. We suspect that the O8  and O9 stars (star Nos. 2 and 3,
respectively), which jointly emit  Lyman continuum photons of log
$N_{Lyc}$ = 48.78, also contribute  to ionize the IC 417 HII
region to a similar extent of that by B0/O9.5  star in Stock 8, although
they are located relatively far away.

\begin{table}
%\begin{minipage}{80mm}
\caption{List of OB stars in IC 417 region }
\label{oblog}
%\medskip
\scriptsize
\begin{tabular}{llllll} \hline
Star & $\alpha_{(2000)}$& $\delta_{(2000)}$& Spectral& $E(B-V)$&References\\
no.      & (h:m:s) &      (d:m:s)    & type & mag&  \\

\hline
1 & 05:28:07&+34:25:27 & O9.5 V & - & Georgelin, et al., 1973\\
& &      &       B0 IV & 0.48   &Mayer \& Macak, 1971   \\
2 &05:27:35 &+34:27:00 & O8 V   &1.25 &Mayer \& Macak, 1971   \\
3& 05:27:29& +34:25:03 & O9 IV-V&0.98    & Mayer \& Macak, 1971 \\
4&05:27:23& +34:23:41  & B0 V &0.55  & Mayer \& Macak, 1971 \\
5& 05:28:08 &+34:25:14 & B1 V &0.46 & Mayer \& Macak, 1971 \\
6&05:27:06 &+34:19:32  & B0.5 V & 0.44& Mayer \& Macak, 1971 \\
7&05:27:16& +34:25:44  & B0.5 V &0.45& Mayer \& Macak, 1971 \\
8&05:28:08& +34:23:45  & B0.5 V&0.47 & Mayer \& Macak, 1971 \\
9& 05:28:06 &+34:27:22 & B1.5 V & 0.52& Mayer \& Macak, 1971  \\
10& 05:27:17 &+34:30:57& B2 V &0.54  & Mayer \& Macak, 1971  \\
11& 05:27:43 &+34:31:56& B0.5 IV&0.60&Savage, 1985 \\
12& 05:28:06 & +34:25:00&B1 V &0.45   &Mayer \& Macak, 1971  \\
13& 05:28:39& +34:40:09& O8 V& 0.57   & Savage, 1985\\
\hline
\end{tabular}
%\end{minipage}
\end{table}

%%%%%%%%%%%%%%%%%%%%%%%%%%%%%%%%%%%%%%%%%%%%%%%%%%%%%%%%%%%%%%%%%%%%%%%%%
\begin{figure*}
\centering
\includegraphics[scale = .4, trim = 10 10 10 150, clip]{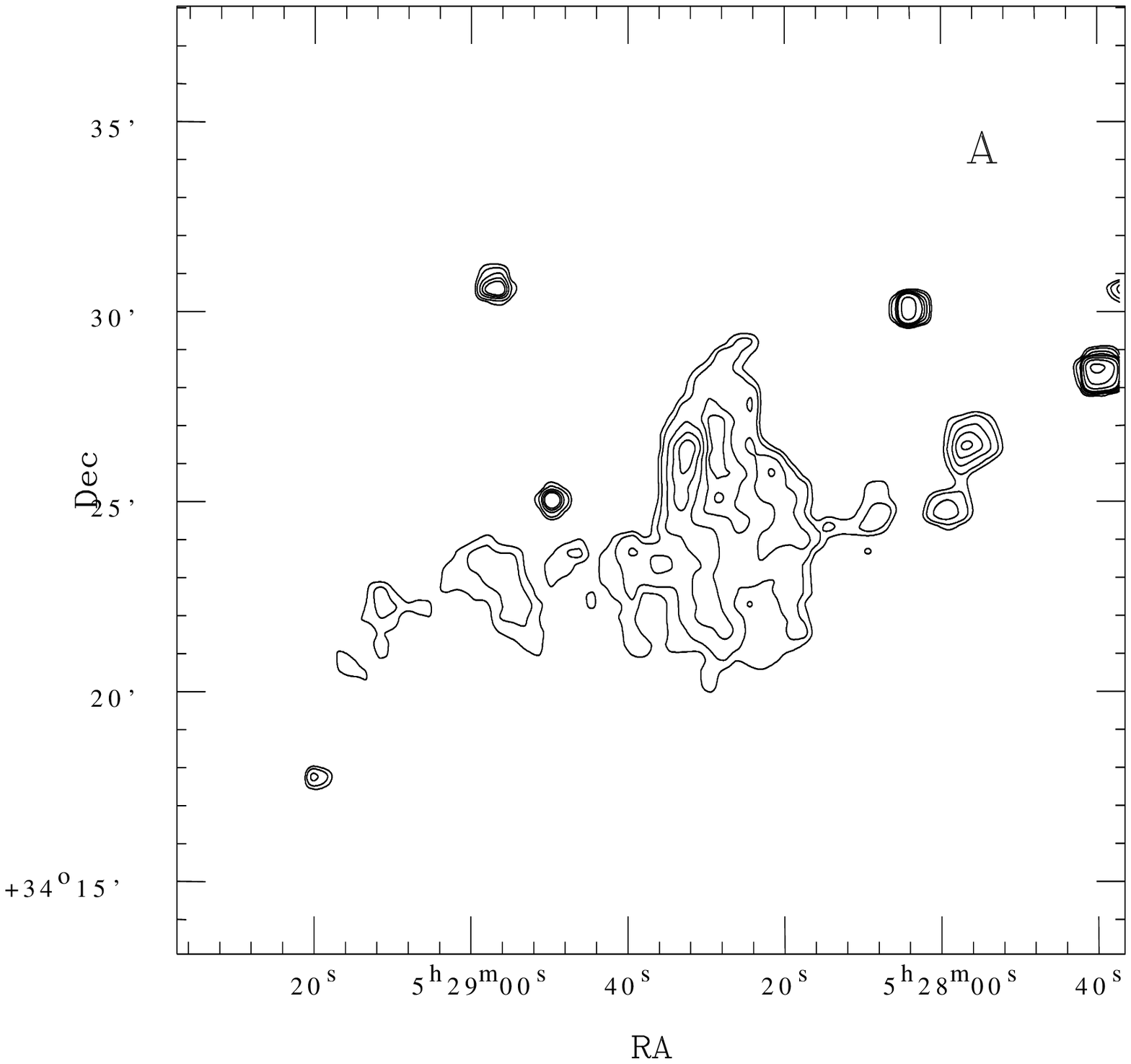}
\includegraphics[scale = .4, trim = 10 10 10 150, clip]{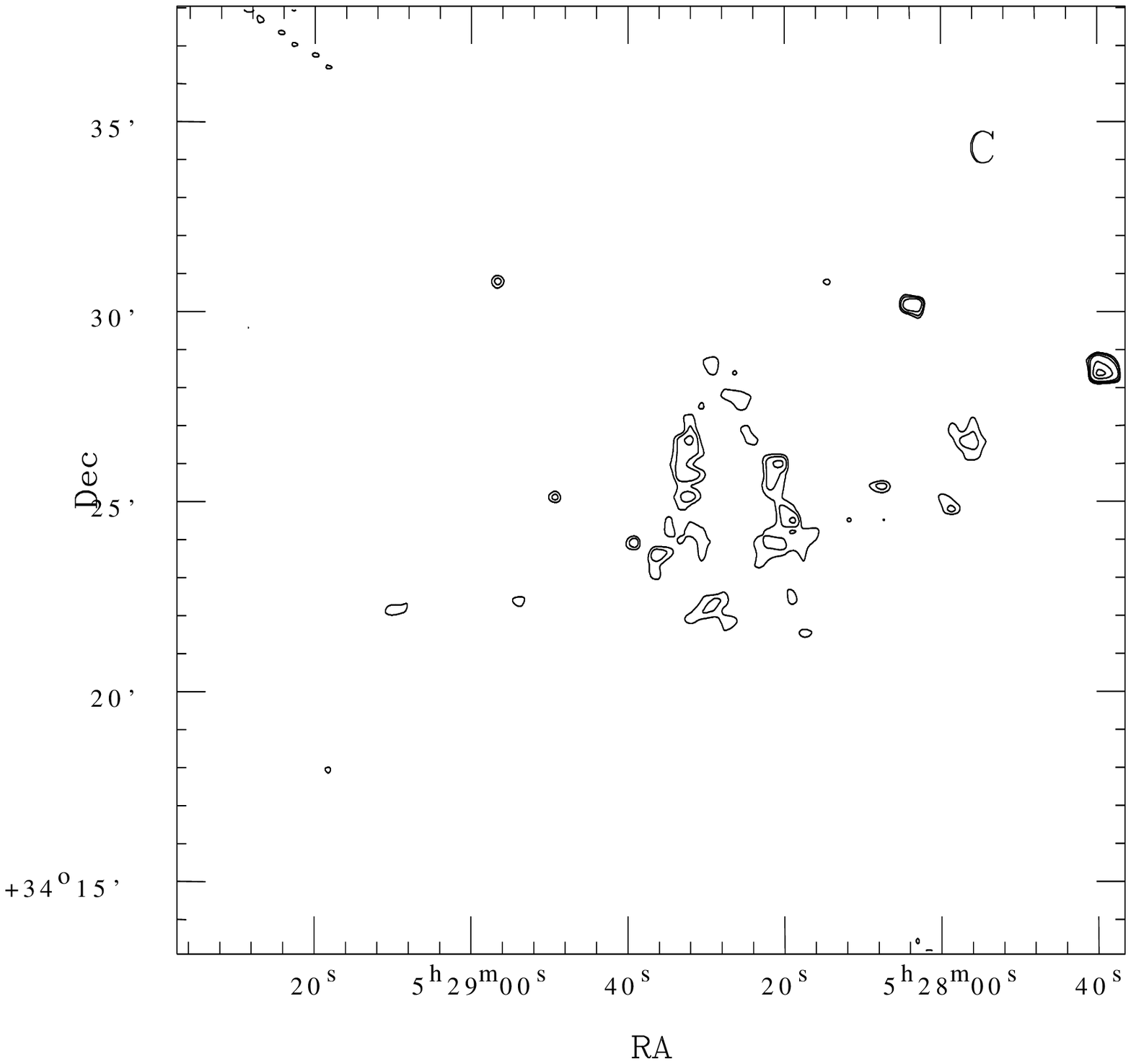}
\hspace{-.01cm}
\includegraphics[scale = .4, trim = 10 10 10 150, clip]{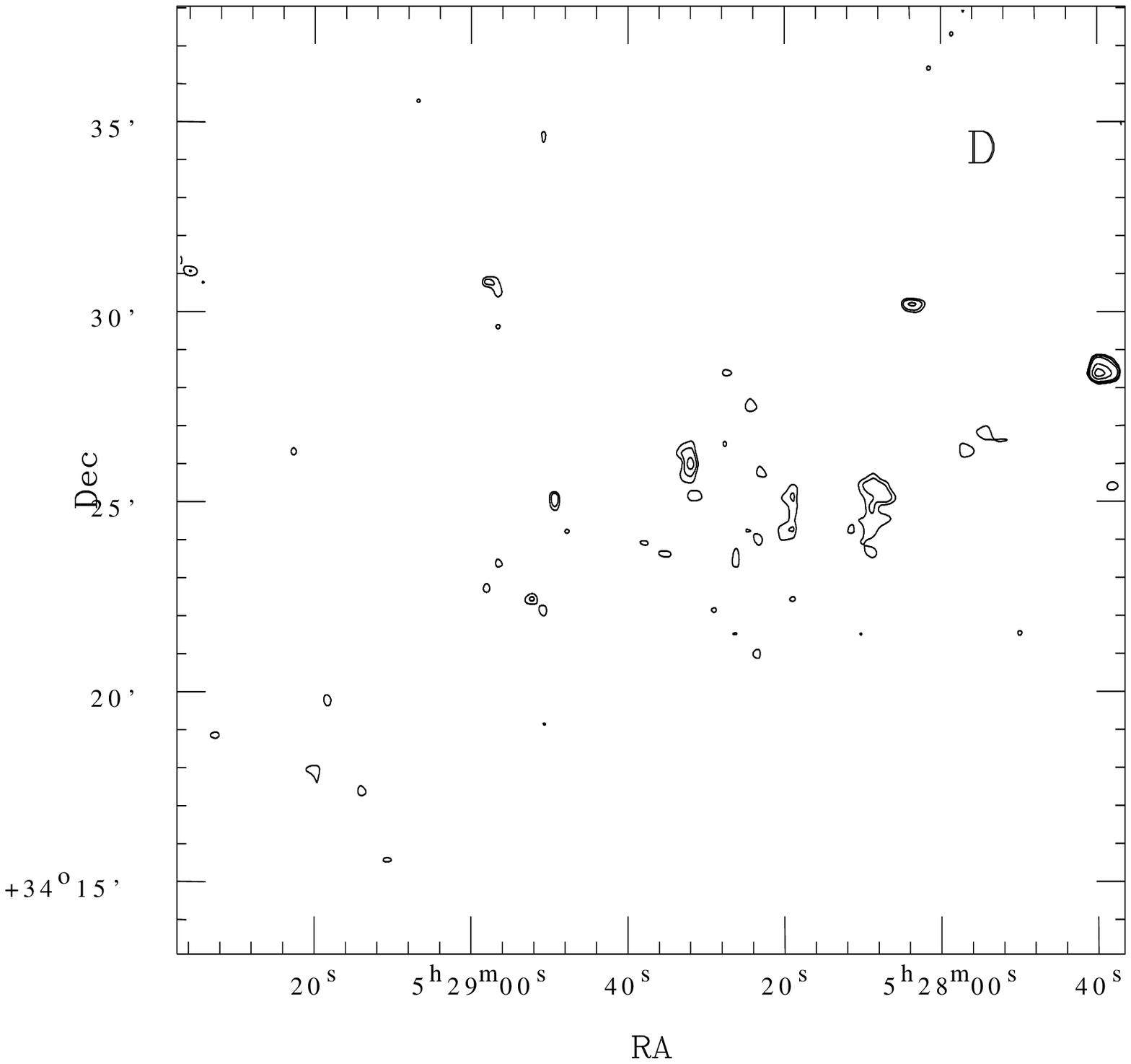}
\includegraphics[scale = .4, trim = 10 10 10 150, clip]{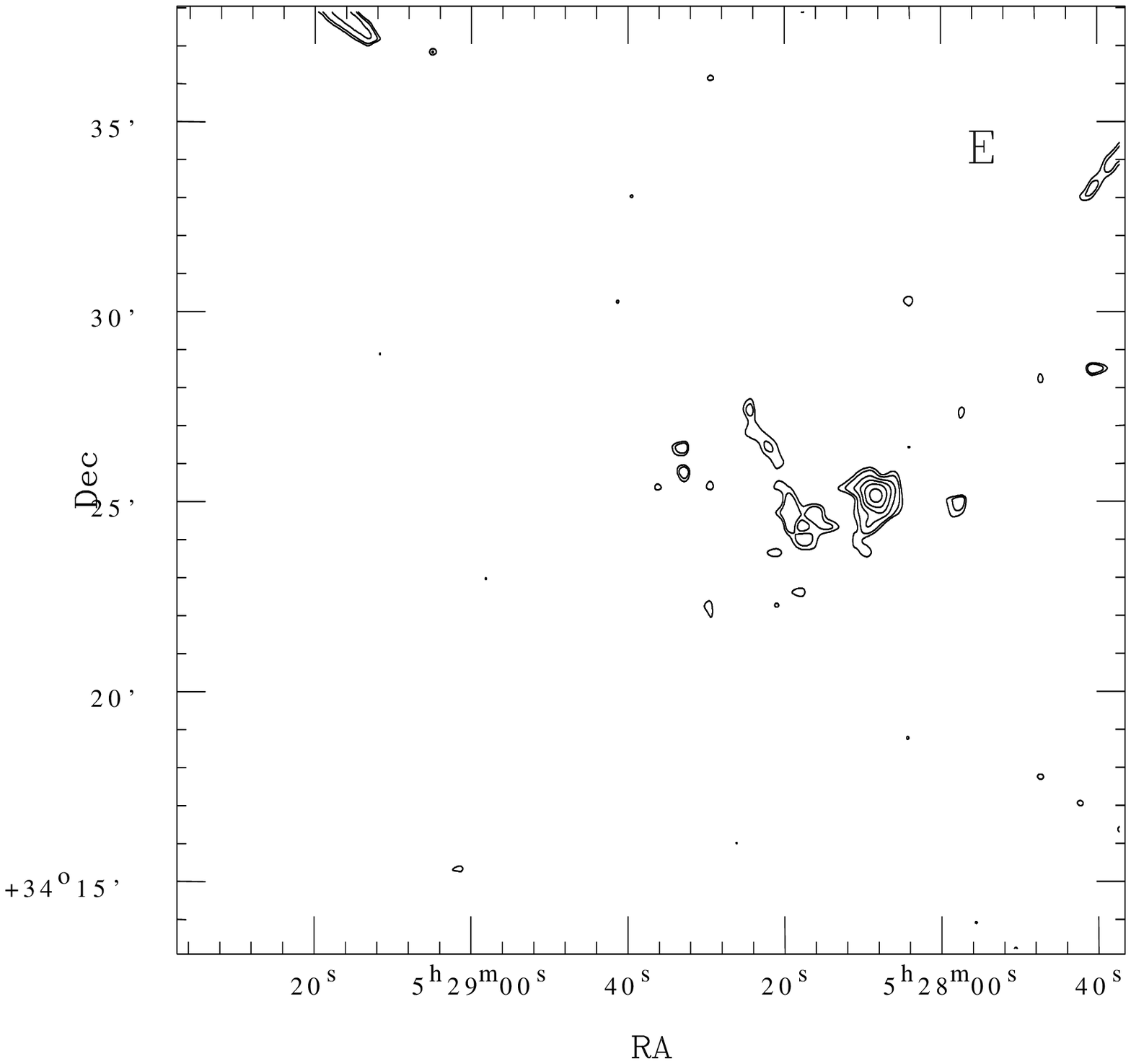}

\caption{$(top$ $left)$: MSX A-band intensity contours with levels as 
80, 60, 40, 20, 10, 5, 4, 3, 2, 1.5 \% of  the peak value 
$9.666 \times 10^{-5} Wm^{-2}Sr^{-1}$. $(top$ $right)$ and $(bottom$ $left)$: 
C and D-band intensity contours with levels as 90, 80, 60, 40, 20, 15, 12 \% 
of the peak value $1.218 \times 10^{-5} Wm^{-2}Sr^{-1}$ and 
$7.239 \times 10^{-6} Wm^{-2}Sr^{-1}$ respectively. $(bottom$ $right)$: 
E-band intensity contours with levels as 90, 80, 70, 60, 50, 40, 35 \% 
of the peak value $6.298 \times 10^{-6} Wm^{-2}Sr^{-1}$. 
The abscissa and ordinates are in  J2000 epoch.}
\label{msx}
\end{figure*}
%%%%%%%%%%%%%%%%%%%%%%%%%%%%%%%%%%%%%%%%%%%%%%%%%%%%%%%%%%%%%%%%%%%%%%%%%

The emission from MSX A-band is completely absent from the inner
region  ($r < 2^\prime$) as well as from  the western region of Stock
8. It is stronger and more extended towards the eastern side of the  cluster
and then  protrudes further to the east, nicely following the
Nebulous  Stream and showing a slight enhancement around CC 14. This
further  supports the physical connection between the H II region and
the Nebulous  Stream/CC 14. We show the MSX A, C, D and E-band contour maps of the 
Stock 8 region in Fig.  \ref{msx}. The contours of the MSX A and C-bands
show an elliptical ridge around $\alpha_{2000}\sim$
$05^{h}28^{m}26^{s}$,  $\delta_{2000}\sim$ $+34^{\circ}25^{\prime}
00^{\prime\prime}$.  The eastward  contours seen in A (8 $\mu$m)
and C (12 $\mu$m) bands lie beyond the  ionization front and get diluted 
as we move towards D (15 $\mu$m) and E (21 $\mu$m) bands, whereas the
emission towards the west remains present in all the four bands.  The
emission in the MIR bands traces presence of warm small dust
grains. The bands A and C also include several PAH features as
described in Sec. 3. The morphology of the emission shown in
Fig. \ref{msx} suggests that the westward emission may be due to warm
small dust grains, whereas eastward emission in A and C bands may be
due to PAHs because PAHs cannot survive in  ionized regions as the hard
radiation field  within HII regions destroys them (Cesarsky et
al. 1996).

%%%%%%%%%%%%%%%%%%%%%%%%%%%%%%%%%%%%%%%%%%%%%%%%%%%%%%%%%%%%%%%%%%%%%%%%%%
\begin{figure*}
%\centering
\includegraphics[scale = .4, trim = 0 10 10 150, clip]{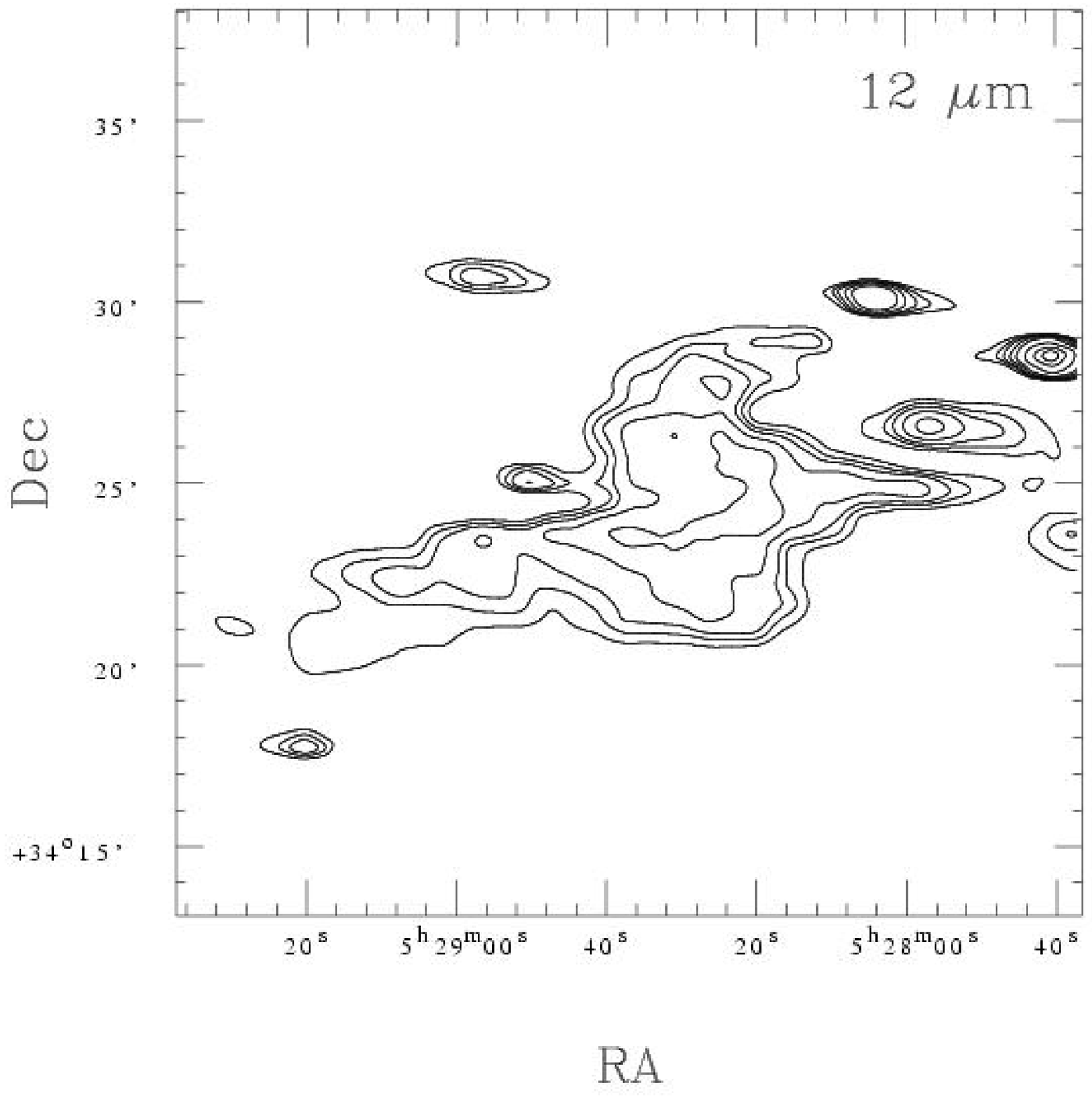}
\includegraphics[scale = .4, trim = 0 10 10 150, clip]{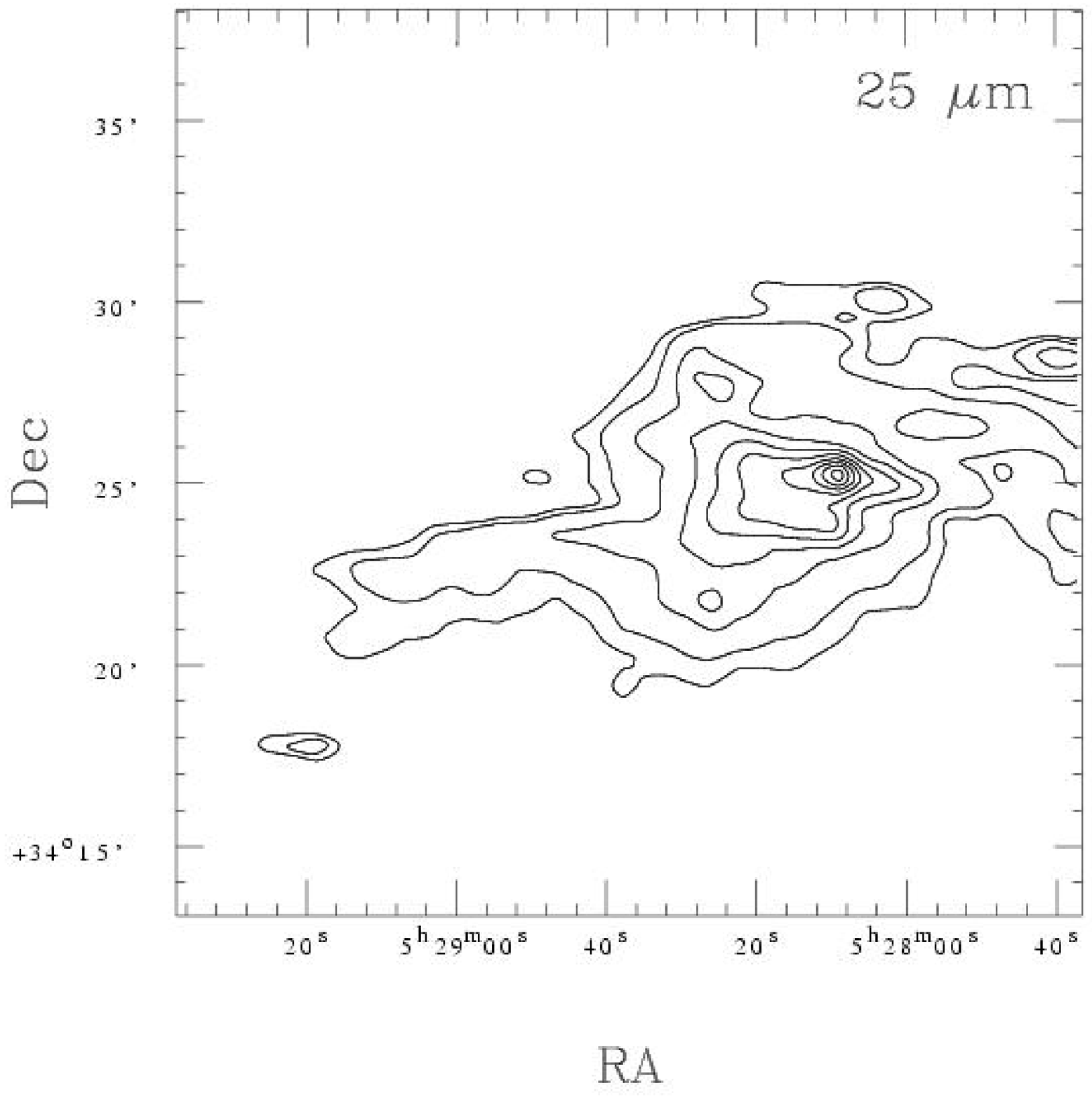}
\hspace{-.01cm}
\includegraphics[scale = .4, trim = 0 10 10 150, clip]{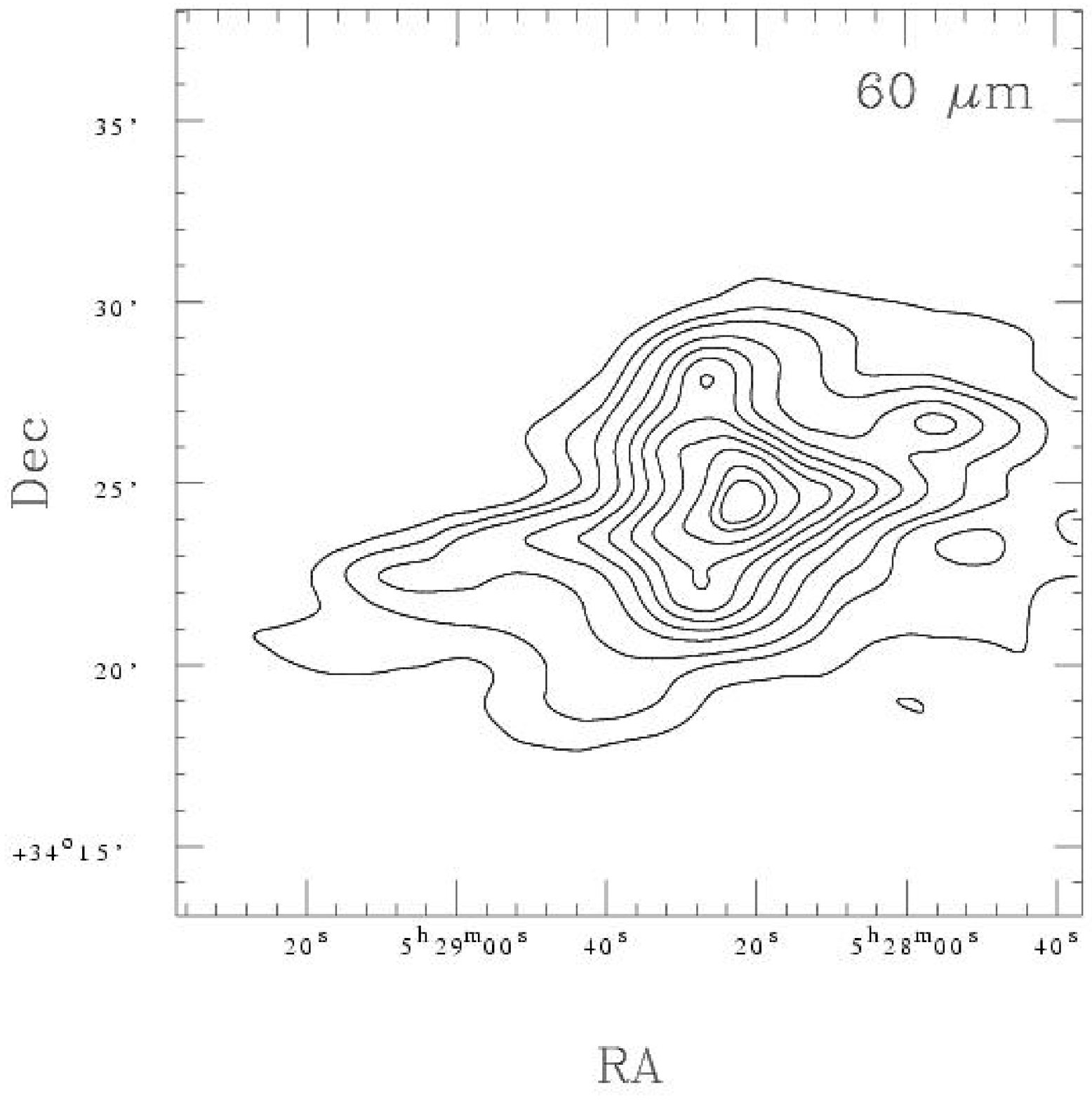}
\includegraphics[scale = .4, trim = 0 10 10 150, clip]{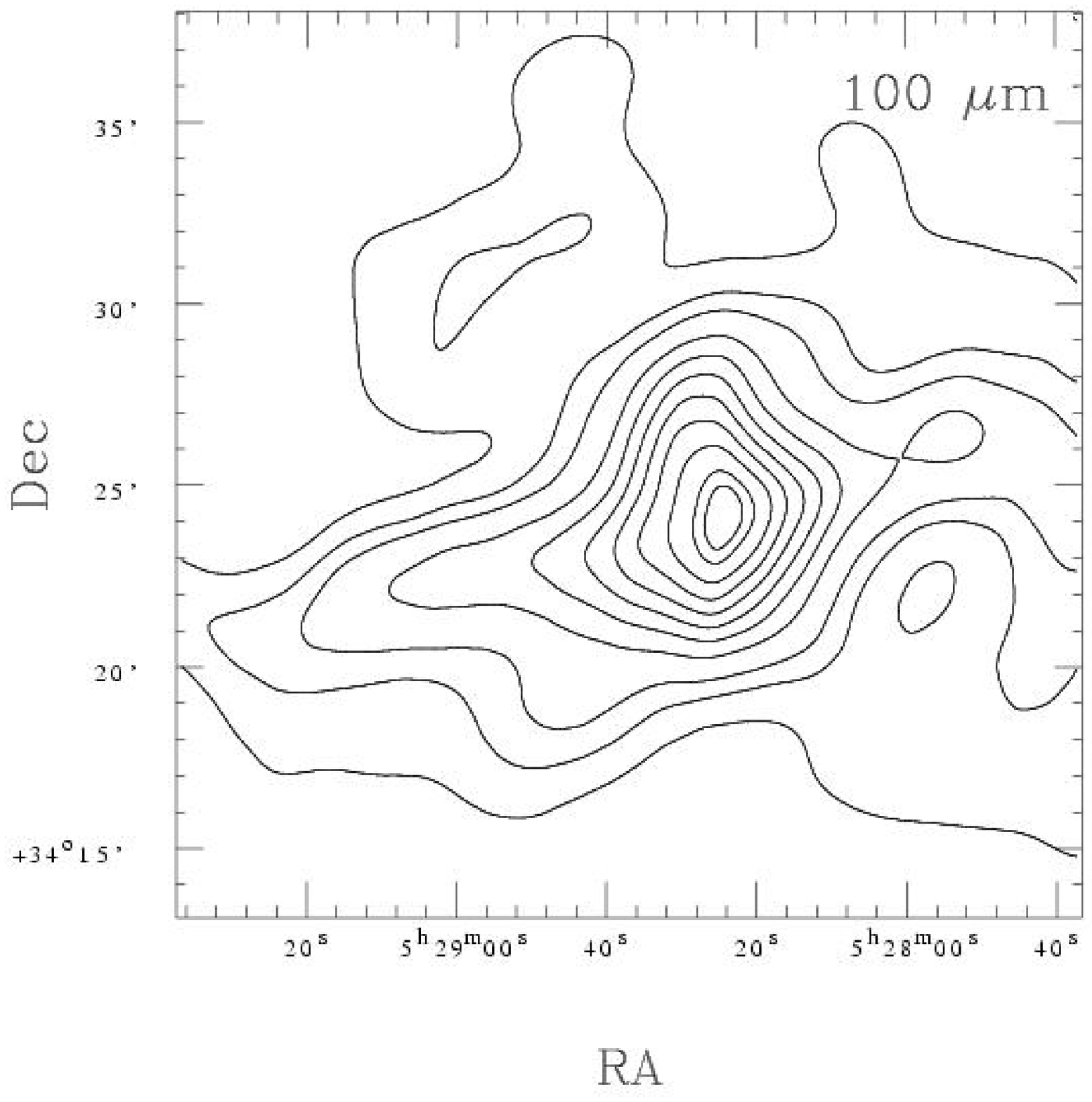}

\caption{IRAS-HIRES intensity maps of 12 $\mu$m ($top$ $left$), 25 $\mu$m ($top$ $right$),
60 $\mu$m ($bottom$ $left$) and 100 $\mu$m ($bottom$ $right$). The contours 
are at 3, 4, 5, 7, 10, 20, 40, 60, 80, 90 $\%$ of the peak value of 190.5 MJy/Sr 
at 12 $\mu$m; 7, 10, 20, 30, 40, 50, 60, 70, 80, 90 $\%$ of the peak value of  
97.7 MJy/Sr at  25 $\mu$m; 10, 15, 20, 30, 40, 50, 60, 70, 80, 90, 95 $\%$ of 
the peak values of 332.6 MJy/Sr and 515.4 MJy/Sr at 60 $\mu$m and 100 $\mu$m 
respectively. The abscissa and ordinates are in  J2000 epoch.}
\label{hires}
\end{figure*}
%%%%%%%%%%%%%%%%%%%%%%%%%%%%%%%%%%%%%%%%%%%%%%%%%%%%%%%%%%%%%%%%%%%%%%%%%%

%%%%%%%%%%%%%%%%%%%%%%%%%%%%%%%%%%%%%%%%%%%%%%%%%%%%%%%%%%%%%%%%%%%%%%%%%
\begin{figure*}
\includegraphics[scale = .4, trim = 10 10 10 10, clip]{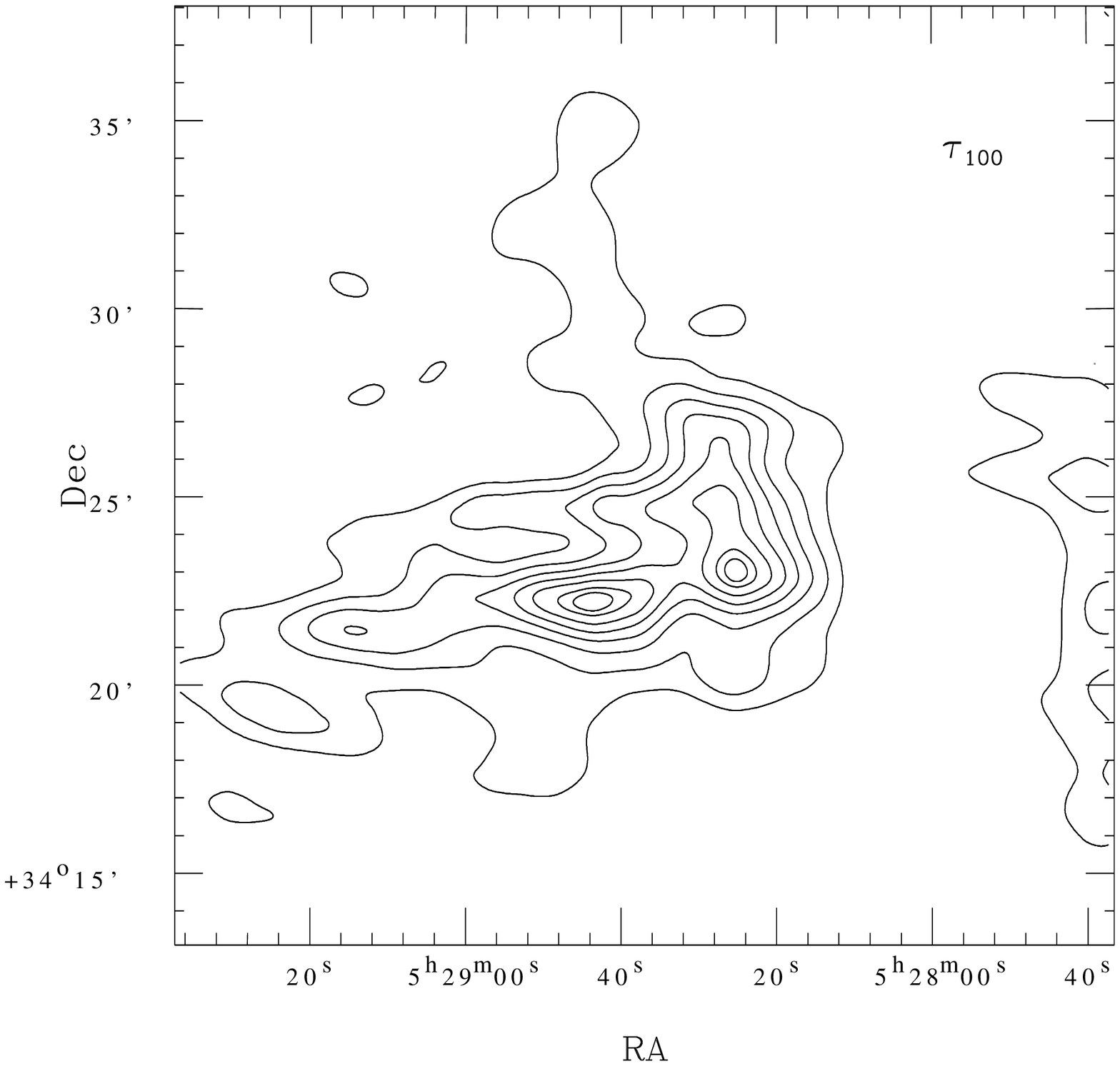}
\includegraphics[scale = .4, trim = 10 10 10 10, clip]{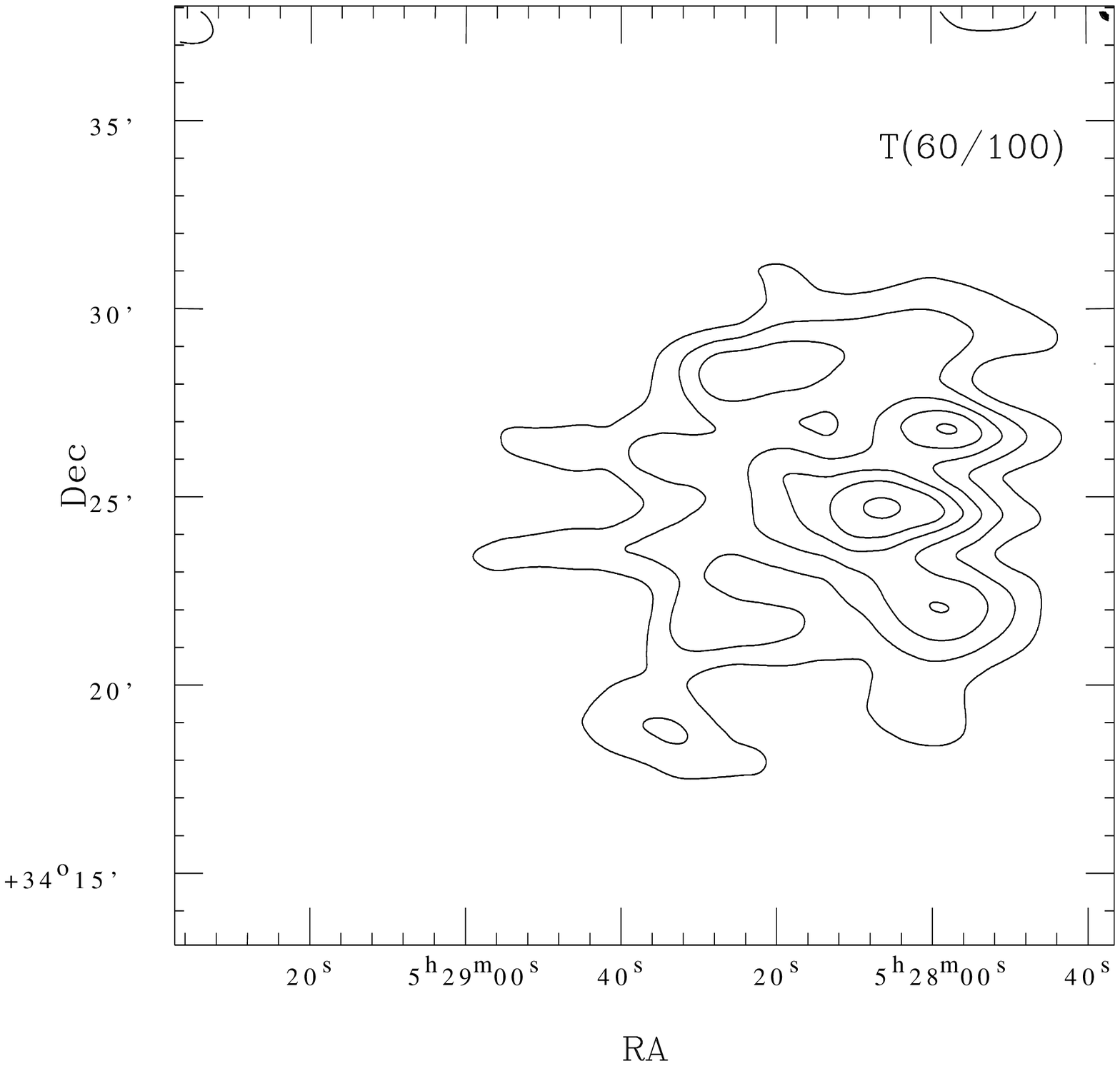}
\caption{($left$): The dust optical depth ($\tau_{100}$ at 100 $\mu$m ) distribution 
of the region around the cluster. The  contours are at 30, 40, 50, 60, 70, 80, 90, 
95 $\%$ of the peak value $6.85\times10^{-4}$. ($right$): Dust colour temperature 
map of the region around the cluster. The contour levels are at 34, 36, 39, 42, 44, 48 K.}
\label{tau_temp}
\end{figure*}
%%%%%%%%%%%%%%%%%%%%%%%%%%%%%%%%%%%%%%%%%%%%%%%%%%%%%%%%%%%%%%%%%%%%%%%%%%

In Fig. \ref{hires} we show  IRAS-HIRES intensity  maps for  
the cluster region at $12\mu$m, $25\mu$m, $60 \mu$m and $100\mu$m.  The
emission  at FIR  ($60 \mu$m and $100 \mu$m) peaks at $\alpha_{2000}
\sim$ $05^{h}28^{m}24^{s}$; $\delta_{2000} \sim$
$+34^{\circ}24^{\prime}00^{\prime\prime}$, whereas  at MIR  ($25\mu$m)
the emission peaks at $\alpha_{2000} \sim$ $05^{h}28^{m}10^{s}$;
$\delta_{2000} \sim$ $+34^{\circ}25^{\prime}30^{\prime\prime} $.  The
peaks of FIR and  MIR emissions are   $\sim$
2$^\prime$  away from  the peak  of the radio continuum emission,  which
indicates an  interaction between the  gas and dust around  the region.
IRAS-HIRES  maps at 60 and $100\mu$m were also used  to generate the 
spatial distribution of dust colour temperature (T(60/100)) and optical 
depth maps at 100 $\mu$m
($\tau_{100}$) using the  procedure given  by Ghosh  et al.  (1993). An
emissivity law of $\epsilon_\lambda \propto \lambda^{-1} $ was assumed
to generate the optical depth map. Fig. \ref{tau_temp} presents the contour 
maps of dust colour temperature ($T(60/100)$)
and optical depth ($\tau_{100}$). The dust colour temperature varies from 
34 to 48 $K$ 
as can be inferred from the map shown in Fig. \ref{tau_temp}.
The relatively high temperature around the cluster 
may probably be due to the radiation from massive star(s) in the region. 
The $\tau_{100}$ map represents low optical depth around the position of Stock 8,
whereas its peak matches well with the CO peak and the position of CC 14.

\section {Nature of the Nebulous Stream}

What is the nature of the Nebulous Stream? At first impression it 
appears like a giant Herbig-Haro jet and reminds us of HH 399 in Trifid
Nebula (Yusef-Zadeh et al. 2005). However this does not seem to be the case.
Fig. \ref{kimage} reveals that  it is closely associated with the 
distribution of the MSX A-band emission, as well as the distribution of 
H$\alpha$ emission stars
and IR-excess stars. The cluster CC 14 is also embedded within the Nebulous Stream. We 
therefore interpret the Nebulous Stream as a bright rim or an ionization front.

%%%%%%%%%%%%%%%%%%%%%%%%%%%%%%%%%%%%%%%%%%%%%%%%%%%%%%%%%%%%%%%%%%%%%%%%%%
\begin{figure}
\centering
\includegraphics[scale = 0.85, trim = 5 100 0 60, clip]{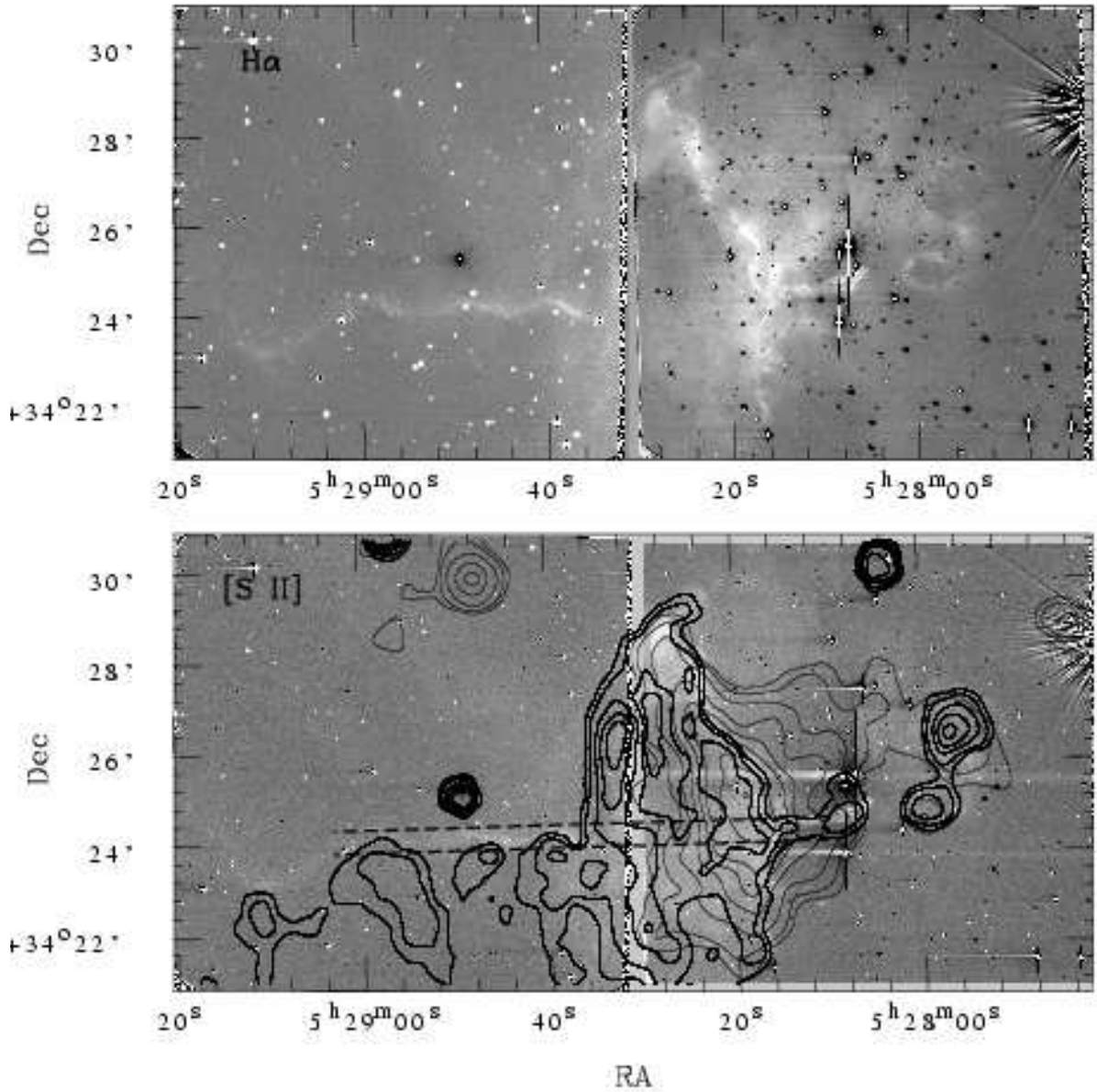}
\caption{ Continuum-subtracted ($20\times10$ arcmin$^2$) H$\alpha$ 
($upper$ $panel$) and [S II] ($lower$ $panel$) images of the regions 
around Stock 8  overlaid with MSX A-band contours (thick black) and NVSS radio 
continuum contours (light black). The contour intensity levels are the same as 
mentioned in Fig. \ref{kimage}. The abscissa and the 
ordinates are for the J2000 epoch. The discontinuity in the images at 
$\alpha_{2000}\sim$ $05^{h}28^{m}30^{s}$ is an artifact of observations as the 
region shown in the figure is mosaic of two non-overlapping frames. The area 
between dashed lines represent a 0.5 arc-min wide strip along the Nebulous Stream 
(see text). Spatial variation of [S II]/H$\alpha$ ratio along the strip is 
shown in Fig. \ref {s2overha_dust}.}
\label{has2} 
\end{figure}
%%%%%%%%%%%%%%%%%%%%%%%%%%%%%%%%%%%%%%%%%%%%%%%%%%%%%%%%%%%%%%%%%%%%%%%%%%
%%%%%%%%%%%%%%%%%%%%%%%%%%%%%%%%%%%%%%%%%%%%%%%%%%%%%%%%%%%%%%%%%%%%%%%%
%\clearpage
\begin{figure}
%centering
\includegraphics[scale = .7, trim = 10 10 10 100, clip]{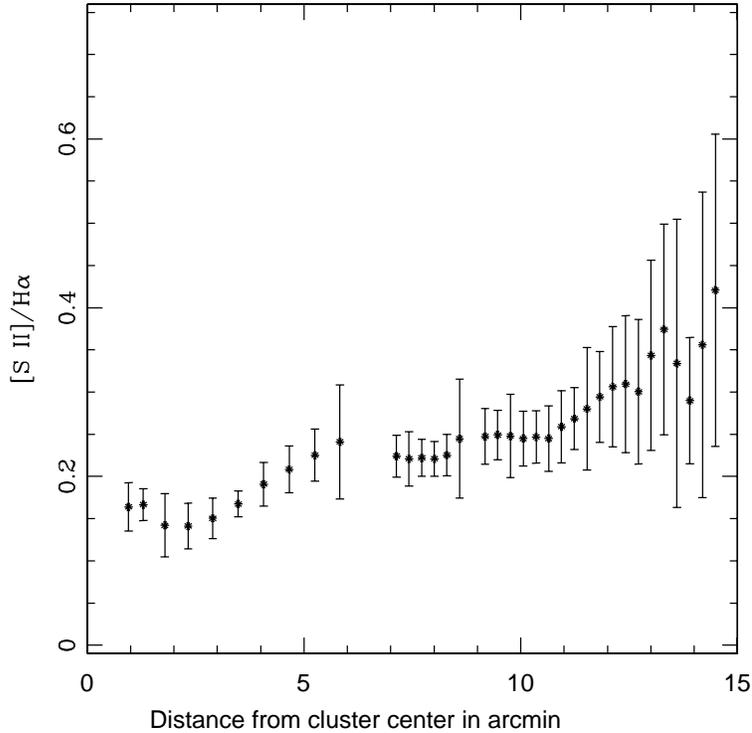}
\caption{Spatial variation of [S II]/H$\alpha$ ratio within a $0^\prime$.5 wide strip 
along the Nebulous Stream from Stock 8 as shown in Fig. \ref{has2}. The error bars 
represent the standard deviation.}
\label{s2overha_dust}
\end{figure}
%%%%%%%%%%%%%%%%%%%%%%%%%%%%%%%%%%%%%%%%%%%%%%%%%%%%%%%%%%%%%%%%%%%%%%%%%%

Fig. \ref{has2} shows continuum-subtracted H$\alpha$  and [S II]
images for  the region  along with contours for MSX A-band  emission
and radio continuum superimposed  on the [S II] image. 
The gap seen at the middle of  Fig. \ref{has2} is
because the field was observed in two frames without an
overlap. The H$\alpha$ emission is more extended compared 
to the [S II] emission. The  [S II] emission comes from low-excitation 
zones and is enhanced close to the ionization fronts (cf. 
Deharveng et al. 2003). Therefore the spatial distribution of the 
[S II]/H$\alpha$ ratio should  give information on the excitation conditions. The 
variation of the [S II]/H$\alpha$ ratio along  
a $0^\prime$.5 wide strip (indicated by black dashed lines 
in Fig. \ref{has2}) as a function of radial distance from  
Stock 8 is shown in Fig. \ref{s2overha_dust}. To improve the 
signal-to-noise ratio and to remove  faint  stars, the  images were 
median filtered  using a  3 $\times $ 3 pixel box. A  background was 
then  determined for these smoothed images by calculating the 
statistics inside a 3$\times$3 pixel box at various places where the 
H$\alpha$ and [S II] brightness  is at a  minimum. The average 
background value was then subtracted from each  pixel. As can be seen in 
Fig. \ref{s2overha_dust}
the ratio of  [S II]/H$\alpha$ is higher  eastwards of the cluster 
Stock 8. Note that the values in the eastern most region are not 
reliable because the error bars are very large. Mean value of the 
[S II]/H$\alpha$  ratio for the central region $r < 6^\prime$ of 
Stock 8 is $\sim$ 0.18$\pm$0.06, which is consistent with the ratio 
expected for an H II region associated with a late O-B0 type star (Reynolds 
1988). Whereas in the region between $7^\prime$ and 12 $^\prime$ the mean 
value of the ratio is higher, $\sim$ 0.24$\pm0.02$. This indicates that 
the Nebulous Stream is a low-excitation object suggestive of its 
bright-rim/ ionization front nature.

The question then arises; where is(are) the source(s) of its excitation?  
The O9.5/ B0 stars of  
Stock 8 or O8/ O9 stars further west do not seem to be the possible source  
because the morphology of the radio 
and MSX A-band emissions indicates an interface at $\alpha_{2000}\sim$ 
$05^{h}28.5^{m}$ and suggests that the ionization front has not reached 
the Nebulous Stream.  UV radiation from the west of  the Stream 
is most probably blocked by the molecular gas located at the interface 
at $\alpha_{2000}\sim$ $05^{h}28.5^{m}$. 
In addition, the Stream runs in the east-west direction, which suggests that the UV 
photons responsible for the Stream should arrive from the north. In a search
towards  the north of the Stream, we find an O8 star (star number 
13 at $\alpha_{2000} = $ $05^{h} 28^{m} 39^s, \delta_{2000}= +34^{\circ} 
40^{\prime} 09^{\prime\prime}$, distance $\sim$ 2.06 kpc: 
Cruz-Gonzalez et al. 1974; 2.7 kpc: Savage, 1985) located at a projected 
distance of 9 pc from the Nebulous Stream. We suspect that this 
star could be the possible source of UV photons which excite the Stream.

In order to verify whether this star emits enough UV photons to generate
the bright rim, we calculated the flux of Lyman continuum
photons which are expected to arrive at the location of the bright rim.
The flux associated with an O8 star is taken from Panagia (1973). 
We assumed in our calculations, that loss of photons due to absorption by interstellar
matter between the star and  bright rim is negligible. 
Using a projected distance of $\sim$ 9 pc
between the star and the  bright rim, the Lyman photon flux at the bright rim 
is estimated to be $\sim 4 \times 10^8$ $cm^{-2}$ $s^{-1}$. The Lyman photon
flux reaching  a radial distance of $6^{\prime} (\sim$
3.6 pc) from  B0 star  is $\sim 2.7 \times 10^8$ $cm^{-2}$ $s^{-1}$. The estimated
flux of $\sim 4 \times 10^8$ $cm^{-2}$ $s^{-1}$ in the case of O8
star is comparable to the ionizing flux associated with the bright
rimmed clouds (see Thompson et al. 2004, Morgan et al. 2004) which
supports an O8 star as responsible for the Nebulous Stream.

\section {Star formation scenario in the vicinity of Stock 8}

The energetic stellar UV radiation and winds from massive stars 
could disperse nearby clouds and consequently terminate further star
formation. Alternatively, they can enhance star formation activity in 
the region. Elmegreen \& Lada (1977) proposed that after formation of 
massive stars, the expanding ionization fronts play a constructive 
role in inducing a sequence of star-formation activities in the
neighbourhood. The  distribution of visible young stars and embedded 
YSOs and the morphological details of the environment around the cluster
containing OB stars can be used to probe the star formation history of
the region.

%%%%%%%%%%%%%%%%%%%%%%%%%%%%%%%%%%%%%%%%%%%%%%%%%%%%%%%%%%%%%%%%%%%%%%%%%
\begin{figure}
%\centering
\includegraphics[scale = .6, trim = 10 10 10 10, clip]{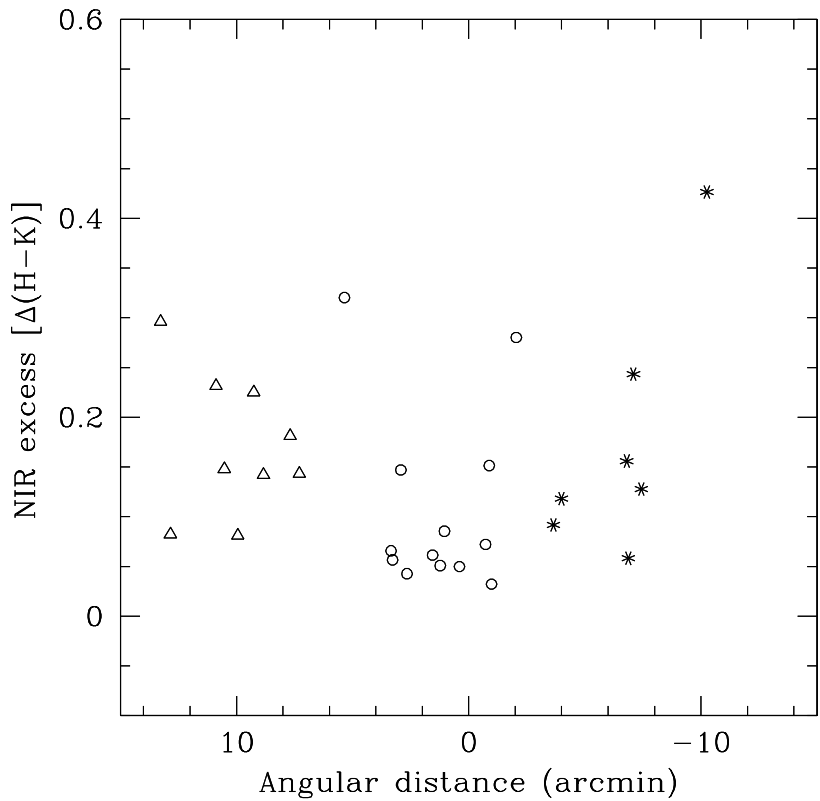}

\caption{Variation of NIR excess ($\Delta (H-K)$) as a function of radial
distance from the center of Stock 8 for three sub-regions as shown in  Fig. \ref{comap}.
The triangles, circles and asterisks are the sources falling in regions I, II 
and III respectively. }
\label{nirexc}
\end{figure}
%%%%%%%%%%%%%%%%%%%%%%%%%%%%%%%%%%%%%%%%%%%%%%%%%%%%%%%%%%%%%%%%%%%%%%%%%%

\begin{figure*}
%\centering
\includegraphics[scale = 0.8, trim = 0 250 0 10, clip]{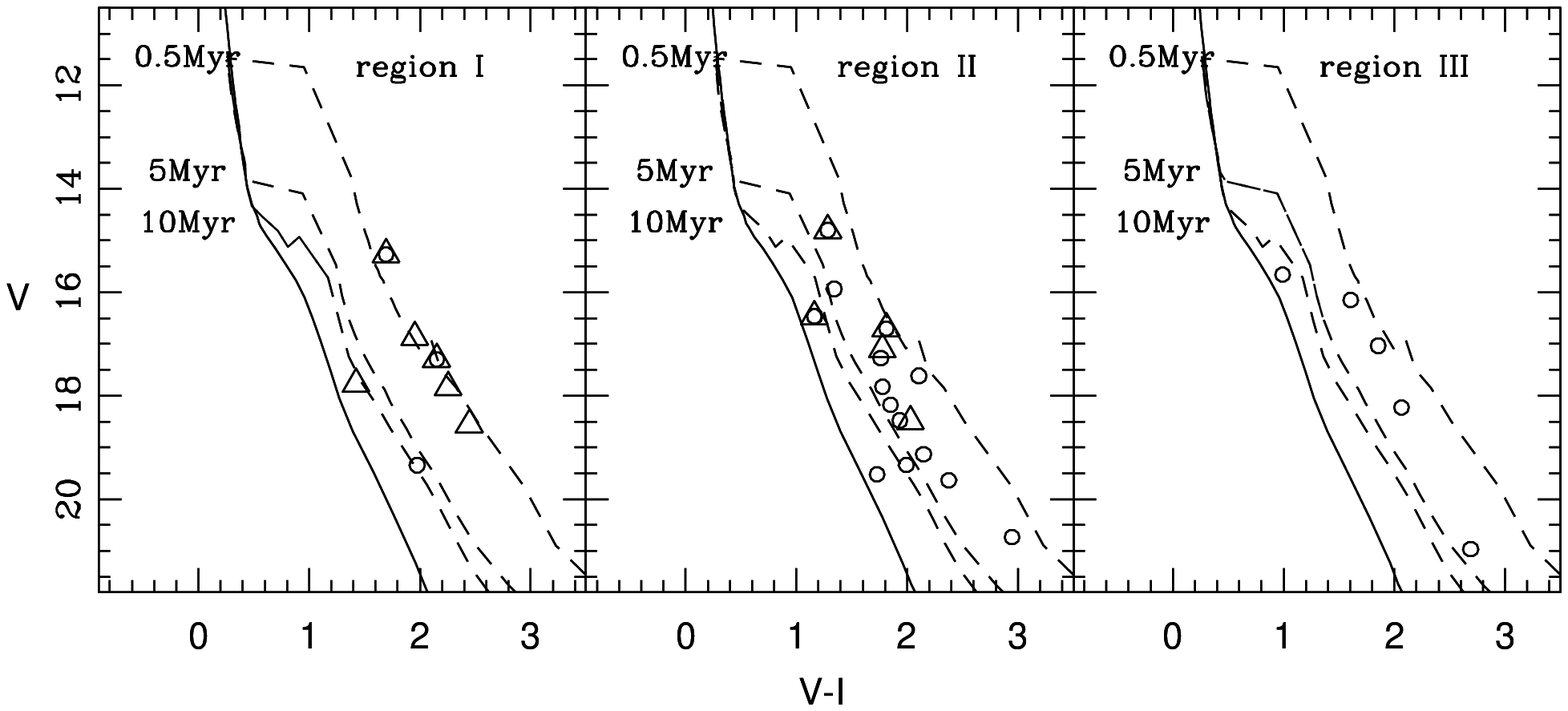}
\includegraphics[scale = 0.8, trim = 0 0 0 300, clip]{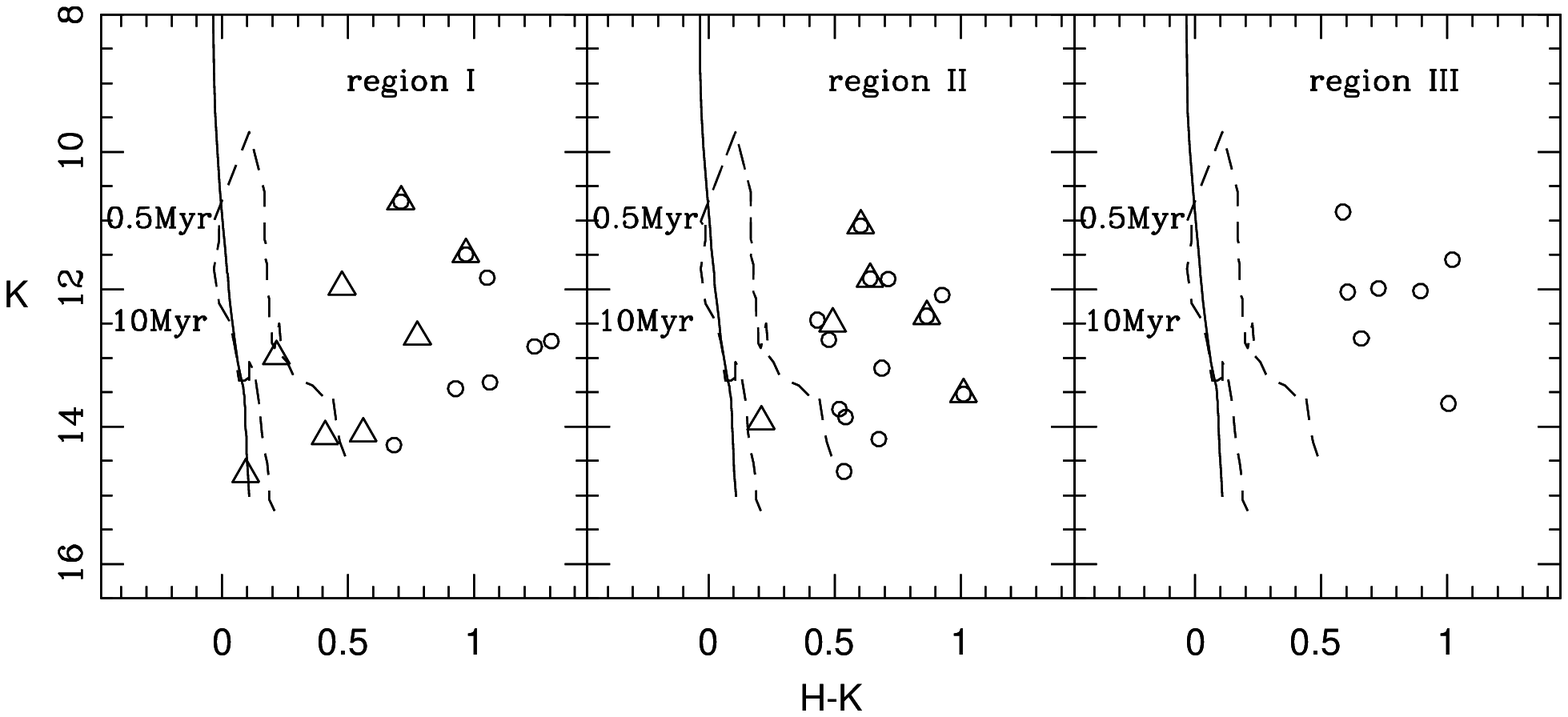}
 
\caption{$V/(V - I)$ and $K/(H-K)$ CMDs for the H$\alpha$ (triangles) and NIR excess 
sources (circles) lying in three sub-regions as shown in Fig. \ref{comap}. The 
isochrone of 2 Myr (continuous curve) by Girardi et al. (2004) and PMS isochrones 
(dashed curves) for ages 0.5, 5 and 10 Myr by Siess et al. (2000) corrected for 
cluster distance and reddening are also shown.}
\label{cmd_sel}
\end{figure*}

In addition to the $^{12}$CO contours and the locations of OB stars, 
Fig. \ref{comap} also shows spatial distribution of NIR-excess sources and
H$\alpha$ emission stars. 
In Fig. \ref{nirexc} we plot NIR excess, $\Delta (H-K)$,
for three sub-regions (regions I, II and III) shown 
in Fig. \ref{comap} as a function  of radial distance from the center 
of the cluster Stock 8. Fig. \ref{nirexc} indicates  that the stars 
lying in regions I and III show relatively higher NIR excess. This 
trend in NIR excess may be an indication of the youth of YSOs located in 
regions I and III as compared to those located near the center of Stock 
8. To verify further the age sequence, we plot $V/ (V-I)$ and 
$K/ (H-K)$ CMDs for the three sub-regions in Fig. \ref{cmd_sel}, which 
also suggests that the YSOs in regions I and III are relatively 
younger than those located near the cluster center. 

%%%%%%%%%%%%%%%%%%%%%%%%%%%%%%%%%%%%%%%%%%%%%%%%%%%%%%%%%%%%%%%%%%%%%%%%%%
%%%%%%%%%%%%%%%%%%%%%%%%%%%%%%%%%%%%%%%%%%%%%%%%%%%%%%%%%%%%%%%%%%%%%%%%%%

As mentioned earlier, several OB stars are located around the cluster Stock 8.
We suspect that these sources belong to the first generation of stars formed in 
the IC 417 region, although the evidence is rather  weak because of their 
scattered distribution. 
The age of the early-type stars in Stock 8 is estimated to be $\le$ 
2 Myr, whereas the YSOs in the Stock 8 region indicate an average age 
of $\sim$ 3 Myr (see Figs. \ref{excess} and \ref{cmd_sel}). However these two
age estimates probe different physical mechanisms, viz., the former being a nuclear 
age and the
latter a contraction age. So, we conclude that they are practically in agreement. We 
suspect that the formation of stars in Stock 8 region was triggered by the  
first-generation stars mentioned above. The UV radiation from the OB stars, in 
particular the O8/O9 stars (star numbers 2 and 3 of Table 3), seem 
to have swept  the pre-existing molecular cloud towards east of 
these stars. It formed a compressed gas layer which became unstable 
against self-gravity and collapsed to form second generation stars in 
the form of Stock 8 and surrounding YSOs in the  region II. 
The shock/ionization front seems to be moving further to the east into 
the low density gas and accumulating a denser layer of molecular gas, 
which we now observe as a hole in the MSX A-band emission region.

The younger age of the YSOs scattered in region III is very strange. 
We suspect that they  recently formed in scattered remnant clouds 
as seen in the Orion region (Ogura \& Sugitani 1998). A hint of the 
presence of such remnant clouds is noticed in the inset of 
Fig. \ref{comap}. In this sense they may well be considered as 
 stars of the third generation in the Stock 8 region.

The narrow strip along the Nebulous Stream appears to be another active 
region (region I) of recent star formation. Fig. \ref{comap} shows that a few NIR-excess 
sources (11 stars) and H$\alpha$ stars (13 stars) are distributed along 
the Nebulous Stream which also contains the CC 14 cluster. The peaks 
of $^{12}$CO and MSX A-band $\tau_{100}$ contours along the Nebulous Stream almost
coincide with the location of the embedded cluster. As we have discussed 
above, these stars show an average age of $\sim$ 1 Myr. The question 
is whether the star formation here has been triggered by the UV radiation 
from the O/B star in Stock 8 or from the O stars located further west. 
This is unlikely, because, as we have discussed in Sec. 10, the action 
of the UV radiation from these stars does not seem to reach this region. 
The morphology of the Nebulous Stream as well as the slightly-displaced 
distribution of the YSOs and CC 14 to the south with respect to the 
Stream suggest that the pre-existing molecular gas might have been 
compressed by the shock front associated with the ionization front caused by the O8 
star located northwards of the Stream (cf. Section 11). So, the 
star formation activity along the Nebulous Stream is probably independent 
of that in the Stock 8 region.

\section{Summary}

On  the basis  of a  comprehensive  multi-wavelength study  of the IC 417 (Sh2-234)
region we have made an attempt to determine the basic properties of the 
cluster Stock 8 as well as to study the star formation scenario in the 
region. Deep optical  $UBVI_c$ and narrow-band [S II], H$\alpha$
photometric data, slitless spectroscopy along with archival data from
the surveys such  as 2MASS, MSX, IRAS and NVSS  are used to understand
the star formation in and around the cluster Stock 8.

Reddening  ($E(B-V)$) in  the  direction  of the cluster  is  found to  be
varying  between 0.40  to  0.60 mag.  The  post-main-sequence age  and
distance of  the cluster are  found to be  $\leq 2$ Myr and  2.05 $\pm
0.10$ kpc respectively.  Using 2MASS NIR  two-colour  diagram  and  
grism survey for H$\alpha$ emission stars, we identified  candidate YSOs. 
A significant number  of YSOs (22 stars) are aligned along a Nebulous 
Stream   eastwards of the cluster and an embedded  cluster  is located  
along the  Stream at a  distance of  $\sim$13 arcmin (7.8 pc) from the 
center of Stock 8. The mass of the  YSOs lie in the range of $\sim$  
0.1 - 3.0 $M_\odot$.  The position of YSOs on the  CMDs indicates that 
the majority of these stars in cluster Stock 8 have ages $\sim$ 1 to  
5 Myr indicating star formation in the cluster may be non-coeval.

The embedded  cluster in  the Nebulous Stream  is found to  be physically
connected to  the H II region. The YSOs located in the Nebulous Stream, 
embedded cluster and in the western region of Stock 8 have larger $(H-K)$ excess
in comparison to those located in the central region of Stock 8. The
$(H-K)$ excess, $V/(V-I)$ and $K/(H-K)$ CMDs indicate that these YSOs are younger 
than those located in the central region of Stock 8. 
The  radio continuum, MSX, IRAS maps and the
ratio  of [S II]/H$\alpha$ intensities indicate that the eastern
region of the cluster is ionization bounded, whereas, the western region of
the Stock 8 is density bounded. The morphology of radio emission and MSX A-band
emission suggests that ionization/ shock front caused by the central ionization 
source of Stock 8 and by O type stars located in the western region of Stock 8, 
has not reached the Nebulous Stream. An O8 star located
$\sim$ 9 pc away towards the north of the Nebulous Stream may be a probable 
source of the bright-rimmed Nebulous Stream. 

The slope  of the  mass function  $\Gamma$ in the  mass range  $\sim 1.0
\le M/M_\odot < 13.4$ can be  represented by $-1.38\pm0.12$,  which agrees
well with Salpeter value (-1.35). In the mass range $0.3 \le M/M_\odot < 1.0$ 
the mass function
is found  to be  shallower with $\Gamma  = -0.58\pm 0.23$ indicating a break in 
the slope of the MF at $\sim$ $1  M_\odot$.  The slope of the $K$-band
luminosity function for the cluster  is found to be $0.31\pm0.02$ which
is smaller than the average value ($\sim$ 0.4) obtained for young 
star clusters (Lada et al. 1991; Lada \& Lada 1995; Lada \& Lada 2003).

\section{Acknowledgments}

Authors are thankful to the referee Dr. R. de Grijs for useful comments 
which improved contents and presentation of the paper.
The observations reported  in this paper were obtained  using the 1.05-m
Kiso Schmidt,  2-m  HCT at  IAO, Hanle,  the  high altitude
station  of Indian  Institute of  Astrophysics, Bangalore  and 1.04-m
Sampurnanad telescope of ARIES. We  are also  thankful to  the
Kiso Observatory, IAO and ARIES for alloting observation time. 
We thank the staff at Kiso Observatory
(Japan), IAO, Hanle and its  remote control station at CREST, Hosakote
and ARIES  (Naini Tal) for  their assistance during  observations. This
publication  makes use of  data from  the Two  Micron All  Sky Survey,
which is  a joint project of  the University of  Massachusetts and the
Infrared  Processing  and   Analysis  Center/California  Institute  of
Technology,   funded   by   the   National   Aeronautics   and   Space
Administration and the National  Science Foundation.  We also used MSX
data for which we have  used the NASA/IPAC Infrared Science Archive. We
thank IPAC  Caltech, for providing  us the HIRES-processed  IRAS maps.
This study  is a part of the DST (India) sponsored project and  JJ is 
thankful to DST for the support. AKP is thankful to the National Central  
University, Taiwan  and TIFR, Mumbai, India for the financial support 
during his visit to NCU and TIFR respectively. AKP and KO acknowledge 
the support given  by DST and JSPS (Japan) to carry out 
the CCD photometry  around open  clusters.

\section*{REFERENCES}
Adams F. C., Lada C. J., Shu F. H., 1987, ApJ, 312, 788\\
Aumann, H. H., Fowler, J. W., \& Melnyk, M., 1990, AJ, 99, 1674\\
Bertout, C., Basri, G., \& Bouvier, J., 1988, ApJ, 330, 350\\
Bessell, M. S., \& Brett, J. M., 1988, PASP, 100, 1134\\
Borissova, J., Pessev, P., Ivanov, V.D., Saviane, I., Kurtev, R., Ivanov  G.R., 2003, A\&A, 411, 83\\
Brandl, B., Brandner, W.,\& Eisenhauer, F., 1999, A\&A, 352, L69\\
Carraro G., Vzquez R. A., Moitinho A., Baume G., 2005, ApJ, 630, L153\\
Cesarsky, D., Lequeux, J., Abergel, A., Perault, M., Palazzi, E., et al., 1996, A\&A, 315, L305\\
Chen, H., Tafalla, M., \& Greene, T. P., 1997, ApJ, 475, 163\\
Chen L., de Grijs R., Zhao J. L., 2007, AJ, 134, 1368\\
Chini, R. \& Krugel, 1983, AA, 117, 289\\
Chini, R., \& Wargau, W. F., 1990, A\&A, 227, 213\\
Cohen, J. G., Frogel, J. A., Persson, S. E., \& Ellias, J. H., 1981, ApJ, 249, 481\\
Cruz-González, C., Recillas-Cruz, E., Costero, R., Peimbert, M., \& Torres-Peimbeert, S., 1974, RMxAA, 1, 211\\
Cutri R.M., Skrutskie M.F., Van Dyk S., et al., 2003, 2MASS All-Sky Catalog of Point Sources\\
Damiani, F., Micela, G., Sciortino, S., Huélamo, N., Moitinho, A., et al.,A\&A, 460, 133\\
de Grijs R., Gilmore G. F., Johnson R. A., Mackey A. D., 2002a, MNRAS, 331, 228\\
de Grijs, R., Gilmore, G. F., Johnson, R. A., \& Mackey, A. D., 2002b, MNRAS,  331, 245\\
de Grijs R., Gilmore G. F., Mackey A. D., Wilkinson M. I., Beaulieu S. F.,et al., 2002c, MNRAS, 337, 597\\
Deharveng, L., Zavagno, A., Salas, L., Porras, A., Caplan, J. \& Cruz-Gonzalez, I, 2003, A\&A, 399, 1135\\
Dolan, C. J., \& Mathieu, R. D., 2002, AJ, 123, 387\\
Draine, B. T.\& Lee, H. M., 1984, ApJ, 285, 89\\
Elmegreen, B. G. \& Lada, C. J., 1977, ApJ, 214, 725\\
Fischer, P., Pryor, C., Murray, S., Mateo, M. \& Richtler, T., 1998, AJ, 115, 592\\
Ghosh S. K., Verma R. P., Rengarajan T. N., Das B., Saraiya H. T., 1993, ApJS, 86, 401\\
Girardi, L., Bertelli, G., Bressan, A., Chiosi, C., Groenewegen, M. A. T., et al., 2002, A\&A, 391, 195\\
He, L., Whittet, D. C. B., Kilkenny, D., Spencer Jones, J. H., 1995, ApJS, 101, 335\\
Hillenbrand, L. A., Strom, S. E., Vrba, F. J. \& Keene, J., 1992, ApJ, 397, 613\\
Hillenbrand, L. A., Massey, P., Strom, S. E., Merrill, K. M., 1993, AJ, 106, 1906 \\
Hillenbrand, L. A., 1997, AJ, 113, 1733\\
Ivanov, V. D., Borissova, J., Bresolin, F. \& Pessev, P., 2005, A\&A, 435, 107\\
Johnson, H. L. \& Morgan, W. W., 1953, ApJ, 117, 313\\
Johnson, H. L., 1966, ARA\&A, 4, 193\\
King, I., 1962, AJ, 67, 471\\
Kroupa, P., 2001, MNRAS, 322, 231\\
Kroupa, P., 2002, Science, 295, 82\\
Kroupa, P., 2007, astro-ph 0708.1164\\
Lada, E. A., Evans, N. J., Depoy, D. L. \& Gatley, I., 1991, ApJ, 371, 171\\
Lada, E . A . \& Lada, C . J., 1995, AJ , 109, 1682\\
Lada, C. J. \& Lada E. A., 2003, ARA\&A, 41, 57\\
Lada, C. J., Muench, A. A., Rathborne, J. M., Alves, J. F., Lombardi, M., 2007, ApJ (in press), astro-ph 0709.1164\\
Landolt A.U., 1992, AJ, 104, 340\\
Larson, R. B., 1992, MNRAS, 256, 641\\
Lee, H. T., Chen, W. P., Zhang, Z. W., Hu, J.Y., 2005, ApJ, 624,808\\
Leisawitz, D., Bash, F. N. \& Thaddeus, P., 1989, ApJS, 70, 731\\
Leistra, A., Cotera, A. S., Leibert, J., \& Burton, M., 2005, AJ, 130, 1719\\
Luhman, K. L., Rieke, G. H., Young, E. T., Cotera, A. S., Chen, H., et al., 2000, ApJ, 540, 1016\\
Malysheva, L. K., 1990, Sov.Astron., 34, 122\\
Mart\'{i}n-Hern\'{a}ndez, N. L., van der Hulst, J. M., \& Tielens, A. G. G. M. 2003,  A\&A, 407, 957\\
Massey, P., Johnson, K. E., \& Degio-Eastwood, K., 1995, ApJ, 454, 151\\
Mayer, P., \& Macak, P., 1971,  Bull. Astron. Czech., 22, 46\\
Megeath, S. T., Herter, T., Beichman, C., Gautier, N., Hester, J. J., et al., 1996, A\&A, 307, 775\\
Meyer, M., Calvet, N., \& Hillenbrand, L. A., 1997, AJ, 114, 288\\
Morgan, L. K., Thompson, M. A., Urquhart, J. S., White, G. J., Miao, J., 2004, 426, 535
Muench, A. A., Lada, E. A., \& Lada, C. J., et al., 2002, ApJ, 573, 366\\
Muench, A. A., Lada, E. A., Lada, C. J., et al., 2003, AJ, 125, 2029\\
Neckel, Th. \& Chini, R., 1981, A\&AS, 45, 451\\
Ogura, K., \& Sugitani, K., 1998, Publ. Astron. Soc. Australia, 15, 91\\
Ojha, D. K., Tamura, M., Nakijama, Y., et al., 2004b, ApJ, 608, 797\\
Ojha, D. K., Tamura, M., Nakajima, Y., et al., 2004b, ApJ, 616, 1042\\
Panagia, N., 1973, AJ, 78, 929\\
Pandey, A.K., Mahra H.S., \& Sagar R., 1992, BASI, 20, 287\\
Pandey, A. K., Ogura, K., \& Sekiguchi, K., 2000, PASJ, 52, 847\\
Pandey A.K., Nilakshi, Ogura K., Sagar R., \& Tarusawa K., 2001, A\&A, 374, 504\\
Pandey, A. K., Upadhyay, K., Nakada, Y., \& Ogura, K., 2003, A\&A, 397, 191\\
Pandey, A. K., Upadhyay, K., Ogura, K., Sagar, R., Mohan, V., et al., 2005, MNRAS, 358, 1290\\
Pandey A. K., Sharma S., Ogura K., 2006, MNRAS, 373, 255\\
Pandey, A. K., Sharma, S., Ogura, K., Ojha, D. K., Chen, W. P., et al., 2007, MNRAS,  (arXiv:0710.5429)\\
Parker, R. J. \& Goodwin, S. P., 2007, MNRAS (in pres), astro-ph 0707.0605\\
Preibisch, T., \& Zinnecker, H., 1999, AJ, 117, 2381\\
Price, N. M., \& Podsiadlowski, Ph., 1995, MNRAS, 273, 1041\\
Price, S. D., Egan, M. P., Carey, S. J., Mizuno, D. R., \& Kuchar, T. A., 2001, AJ, 121, 2819\\
Prisinzano, L., Damiani, F., Micela, G., Sciortino, S., 2005, A\&A, 430, 941\\
Reynolds, R. J., 1988, ApJ, 333, 341\\
Robin, A. C., Reyle, C., Derriere, S., \& Picaud, S., 2003, A\&A, 409, 523\\
Salpeter, E.E., 1955, ApJ, 121, 161\\
Samal, M. R., Pandey, A. K., Ojha, D. K., Ghosh, S. K., Kulkarni, V. K., Bhatt, B. C.,ApJ (in press) astro-ph 0708.4137\\
Sanchawala, K., Chen, W.P., Ojha, D.K, Ghosh, S.K., Nakajima, Y., et al., 2007, arXiv:0706.1834S\\
Savage, B. D., Massa, D., Meade, M., \& Wesselius, P. R. 1985, ApJS, 59, 397\\
Scalo, J. 1998, in ASP Conf. Ser. 142, The Stellar Initial Mass Function, ed.
  G. Gilmore \& D. Howell (San Francisco: ASP), 201\\
Schaerer, D., \& de Koter, A. 1997, A\&A, 322, 598\\
Sharma, S., Pandey, A. K., Ojha, D. K., Chen, W. P., Ghosh, S. K., Bhatt, B. C., Maheswar, G., Sagar, Ram, 2007, MNRAS, 380, 1141\\
Siess, L., Dufour, E. \& Forestini, M., 2000, A\&A, 358, 593\\
Stetson, P. B., 1987, PASP, 99, 191\\
Tapia, M., Roth, M., Costero, R., Echevarria, J., \& Roth, M., 1991, MNRAS, 253, 649\\
Thompson, M. A., Urquhart, J. S., White, G. J., 2004, A\&A, 415, 627\\
Wegner, W. 1993, A\&A, 43, 209\\
Winkler, H. 1997, MNRAS, 287, 481\\
Yusef-Zadeh, F., Biretta, J., \& Wardle, M. 2005, ApJ, 624, 246\\
Zinnecker, H., 1986., IMF in starburst regions. In light on Dark Matter, ed. F.P.Israel, ApSS Library Vol. 124, pp.277-278\\
%\newpage

\bsp
%label{lastpage}

\end{document}